\newcommand{\dif}[1]{{\rm d}#1}

\newcommand{\rf}[1]{\mbox{(\ref{#1}\hspace{-.5pt})}}

\newcommand{\ppp}{{,...\,,\hspace{1.5pt}}}
\newcommand{\sdot}{\,{\scriptscriptstyle{}^{\bullet}}\,}

\newcommand{\pmu}{{\partial_{\mu}}}
\newcommand{\pno}{{\partial^{\nu}}}
\newcommand{\nablamu}{\nabla_{\!\!\mu\,}}
\newcommand{\nablanu}{\nabla_{\!\!\nu\,}}
\newcommand{\vnabla}{\vec{\nabla}}

\newcommand{\idr}{\int\!\!{\rm d}^3\vec{r}\;}
\newcommand{\idrs}{\int\!\!{\rm d}^3\vec{r}\!\;'\;}

\newcommand{\vr}{(\vec{r}\!\:)}
\newcommand{\vrs}{(\vec{r}\!\;'\!\!\;)}

\newcommand{\vsigma}{\vec{\sigma}}
\newcommand{\vgamma}{\vec{\gamma}}
\newcommand{\GGamma}{\mbox{\reflectbox{\rotatebox[origin=c]{180}{$\mathbbm{L}$}}}}
\newcommand{\Sigmunu}{\Sigma_{\mu\nu}}

\newcommand{\hho}{\mathbbm{h}_0}
\newcommand{\vhh}{\vec{\mathbbm{h}}}
\newcommand{\vbb}{\vec{\mathbbm{b}}}
\newcommand{\vbbS}{\vcenter{\offinterlineskip \hbox{$\scriptstyle\;\ast$}\vskip.0ex\hbox{$\vbb$}\vskip1.8ex}}
\newcommand{\vzz}{\vec{\mathbbm{z}}}
\newcommand{\vqqa}{\vec{\reflectbox{$\mathbbm{p}$}}_a}
\newcommand{\vqqe}{\vec{\reflectbox{$\mathbbm{p}$}}_1}
\newcommand{\vqqz}{\vec{\reflectbox{$\mathbbm{p}$}}_2}
\newcommand{\vqqd}{\vec{\reflectbox{$\mathbbm{p}$}}_3}
\newcommand{\akko}{{{}^{\alo\!\!\!\:}\mathbbm{k}_0}}
\newcommand{\vkka}{\vec{\mathbbm{k}}_a}

\newcommand{\mvZZ}{{{}^{\scriptscriptstyle\rm(\!\!\:m\!\!\:)\!\!\!\;}\vec{\mathbbm{Z}}}}

\newcommand{\hatHi}{\hat{H}_{\rm I}}
\newcommand{\hatHii}{\hat{H}_{\rm II}}
\newcommand{\hatHiii}{\hat{H}_{\rm III}}
\newcommand{\hathii}{\hat{h}_{\rm II}}
\newcommand{\hathiii}{\hat{h}_{\rm III}}

\newcommand{\als}{\alpha_{\rm S}}
\newcommand{\ab}{a_{\rm B}}
\newcommand{\am}{a_{\rm M}}
\newcommand{\ax}{a_{\rm X}}

\newcommand{\Dmu}{D_{\!\mu}}
\newcommand{\Dmo}{D^\mu}
\newcommand{\MDmu}{\mathcal{D}_{\!\mu}}
\newcommand{\MDnu}{\mathcal{D}_{\!\nu}}
\newcommand{\MDmo}{\mathcal{D}^\mu}

\newcommand{\ML}{\mathcal{L}}
\newcommand{\MLD}{\ML_{\rm D}}
\newcommand{\MLG}{\ML_{\rm G}}
\newcommand{\MLDe}{\ML_{\rm D(1)}}
\newcommand{\MLDz}{\ML_{\rm D(2)}}
\newcommand{\MLDd}{\ML_{\rm D(3)}}
\newcommand{\MLM}{\ML_{\rm M}}
\newcommand{\MLkin}{\ML_{\rm kin}}
\newcommand{\MLDem}{\ML_{\rm D}^{\rm(\!\!\:em\!\!\:)\!}}
\newcommand{\MLDhg}{\ML_{\rm D}^{\rm(\!\!\:hg\!\!\:)\!}}
\newcommand{\MLR}{\ML_{\rm R}}
\newcommand{\MLC}{\ML_{\rm C}}

\newcommand{\MM}{\mathcal{M}}

\newcommand{\Me}{M^{\rm(\!\!\:e\!\!\:)\!}}

\newcommand{\sMe}{{{}'\!\!\!\;M}^{\rm(\!\!\:e\!\!\:)\!}}
\newcommand{\tMe}{\tilde{M}^{\rm(\!\!\:e\!\!\:)\!}}
\newcommand{\Mm}{M^{\rm(\!\!\:m\!\!\:)\!}}
\newcommand{\Mh}{M^{\rm(\!\!\:h\!\!\:)\!}}
\newcommand{\Mg}{M^{\rm(\!\!\:g\!\!\:)\!}}
\newcommand{\Mhg}{M^{\rm(\!\!\:hg\!\!\:)\!}}
\newcommand{\Mfa}{M_{[a]}}
\newcommand{\Mfe}{M_{[1]}}
\newcommand{\Mfz}{M_{[2]}}
\newcommand{\Mfd}{M_{[3]}}
\newcommand{\tMfa}{\tilde{M}_{[a]}}
\newcommand{\Mefa}{M^{\rm(\!\!\:e\!\!\:)\!}_{[a]}}
\newcommand{\Mefe}{M^{\rm(\!\!\:e\!\!\:)\!}_{[1]}}
\newcommand{\Mefz}{M^{\rm(\!\!\:e\!\!\:)\!}_{[2]}}
\newcommand{\Mefd}{M^{\rm(\!\!\:e\!\!\:)\!}_{[3]}}
\newcommand{\Mmfa}{M^{\rm(\!\!\:m\!\!\:)\!}_{[a]}}
\newcommand{\Mmfe}{M^{\rm(\!\!\:m\!\!\:)\!}_{[1]}}
\newcommand{\Mmfz}{M^{\rm(\!\!\:m\!\!\:)\!}_{[2]}}
\newcommand{\Mmfd}{M^{\rm(\!\!\:m\!\!\:)\!}_{[3]}}
\newcommand{\Memfa}{M^{\rm(\!\!\:em\!\!\:)\!}_{[a]}}
\newcommand{\Memfe}{M^{\rm(\!\!\:em\!\!\:)\!}_{[1]}}
\newcommand{\Memfz}{M^{\rm(\!\!\:em\!\!\:)\!}_{[2]}}
\newcommand{\Memfd}{M^{\rm(\!\!\:em\!\!\:)\!}_{[3]}}
\newcommand{\Mgfa}{M^{\rm(\!\!\:g\!\!\:)\!}_{[a]}}

\newcommand{\Mgfz}{M^{\rm(\!\!\:g\!\!\:)\!}_{[2]}}
\newcommand{\Mgfd}{M^{\rm(\!\!\:g\!\!\:)\!}_{[3]}}
\newcommand{\Mhfa}{M^{\rm(\!\!\:h\!\!\:)\!}_{[a]}}

\newcommand{\Mhfz}{M^{\rm(\!\!\:h\!\!\:)\!}_{[2]}}
\newcommand{\Mhfd}{M^{\rm(\!\!\:h\!\!\:)\!}_{[3]}}
\newcommand{\Mhgfa}{M^{\rm(\!\!\:hg\!\!\:)\!}_{[a]}}

\newcommand{\Mhgfz}{M^{\rm(\!\!\:hg\!\!\:)\!}_{[2]}}
\newcommand{\Mhgfd}{M^{\rm(\!\!\:hg\!\!\:)\!}_{[3]}}
\newcommand{\Mqfe}{M^{\rm(\!\!\:q\!\!\:)\!}_{[1]}}
\newcommand{\Mqefe}{M^{\rm(\!\!\:qe\!\!\:)\!}_{[1]}}
\newcommand{\Mqmfe}{M^{\rm(\!\!\:qm\!\!\:)\!}_{[1]}}
\newcommand{\Msfe}{M^{\rm(\!\!\:s\!\!\:)\!}_{[1]}}
\newcommand{\MMe}{\mathbbm{M}^{\rm(\!\!\:e\!\!\:)\!}}
\newcommand{\dMMe}{\dot{\mathbbm{M}}^{\rm(\!\!\:e\!\!\:)\!}}
\newcommand{\ddMMe}{\ddot{\mathbbm{M}}^{\rm(\!\!\:e\!\!\:)\!}}
\newcommand{\MMm}{\mathbbm{M}^{\rm(\!\!\:m\!\!\:)\!}}
\newcommand{\MMh}{\mathbbm{M}^{\rm(\!\!\:h\!\!\:)\!}}
\newcommand{\MMg}{\mathbbm{M}^{\rm(\!\!\:g\!\!\:)\!}}
\newcommand{\MMefa}{\mathbbm{M}^{\rm(\!\!\:e\!\!\:)\!}_{[a]}}
\newcommand{\MMefe}{\mathbbm{M}^{\rm(\!\!\:e\!\!\:)\!}_{[1]}}
\newcommand{\MMefz}{\mathbbm{M}^{\rm(\!\!\:e\!\!\:)\!}_{[2]}}
\newcommand{\MMefd}{\mathbbm{M}^{\rm(\!\!\:e\!\!\:)\!}_{[3]}}
\newcommand{\MMemfe}{\mathbbm{M}^{\rm(\!\!\:em\!\!\:)\!}_{[1]}}
\newcommand{\MMemfz}{\mathbbm{M}^{\rm(\!\!\:em\!\!\:)\!}_{[2]}}
\newcommand{\MMemfd}{\mathbbm{M}^{\rm(\!\!\:em\!\!\:)\!}_{[3]}}
\newcommand{\MMmfa}{\mathbbm{M}^{\rm(\!\!\:m\!\!\:)\!}_{[a]}}
\newcommand{\MMmfe}{\mathbbm{M}^{\rm(\!\!\:m\!\!\:)\!}_{[1]}}
\newcommand{\MMmfz}{\mathbbm{M}^{\rm(\!\!\:m\!\!\:)\!}_{[2]}}
\newcommand{\MMmfd}{\mathbbm{M}^{\rm(\!\!\:m\!\!\:)\!}_{[3]}}
\newcommand{\MMqefe}{\mathbbm{M}^{\rm(\!\!\:qe\!\!\:)\!}_{[1]}}
\newcommand{\MMqmfe}{\mathbbm{M}^{\rm(\!\!\:qm\!\!\:)\!}_{[1]}}
\newcommand{\MMgfa}{\mathbbm{M}^{\rm(\!\!\:g\!\!\:)\!}_{[a]}}
\newcommand{\MMgfz}{\mathbbm{M}^{\rm(\!\!\:g\!\!\:)\!}_{[2]}}
\newcommand{\MMgfd}{\mathbbm{M}^{\rm(\!\!\:g\!\!\:)\!}_{[3]}}
\newcommand{\MMhfa}{\mathbbm{M}^{\rm(\!\!\:h\!\!\:)\!}_{[a]}}
\newcommand{\MMhfz}{\mathbbm{M}^{\rm(\!\!\:h\!\!\:)\!}_{[2]}}
\newcommand{\MMhfd}{\mathbbm{M}^{\rm(\!\!\:h\!\!\:)\!}_{[3]}}
\newcommand{\MMsfe}{\mathbbm{M}^{\rm(\!\!\:s\!\!\:)\!}_{[1]}}
\newcommand{\MfT}{M_{[\rm T]}}
\newcommand{\tMfT}{\tilde{M}_{[\rm T]}}

\newcommand{\NGe}{N_{\rm\!G}^{\rm(\!\!\:e\!\!\:)\!}}
\newcommand{\sNGe}{{{}'\!\!\!\;N}_{\rm\!G}^{\rm(\!\!\:e\!\!\:)\!}}
\newcommand{\tNGe}{\tilde{N}_{\rm\!G}^{\rm(\!\!\:e\!\!\:)\!}}
\newcommand{\NGm}{N_{\rm\!G}^{\rm(\!\!\:m\!\!\:)\!}}
\newcommand{\NGh}{N_{\rm\!G}^{\rm(\!\!\:h\!\!\:)\!}}
\newcommand{\NGg}{N_{\rm\!G}^{\rm(\!\!\:g\!\!\:)\!}}
\newcommand{\sND}{{{}'\!\!\!\;N}_{\!{\rm D}}}
\newcommand{\tND}{\tilde{N}_{\!{\rm D}}}
\newcommand{\NDa}{N_{\!{\rm D}(a)}}
\newcommand{\NDe}{N_{\!{\rm D}\!\!\:(\!\!\;1\!\!\;)}}
\newcommand{\NDz}{N_{\!{\rm D}\!\!\:(\!\!\;2\!\!\;)}}
\newcommand{\NNDa}{\mathbbm{N}_{\!{\rm D}\!\!\:(\!\!\;a\!\!\;)}}
\newcommand{\NNDe}{\mathbbm{N}_{\!{\rm D}\!\!\:(\!\!\;1\!\!\;)}}
\newcommand{\NNDz}{\mathbbm{N}_{\!{\rm D}\!\!\:(\!\!\;2\!\!\;)}}
\newcommand{\NNGe}{\mathbbm{N}_{\rm\!G}^{\rm(\!\!\:e\!\!\:)\!}}
\newcommand{\NNGm}{\mathbbm{N}_{\rm\!G}^{\rm(\!\!\:m\!\!\:)\!}}
\newcommand{\NNGh}{\mathbbm{N}_{\rm\!G}^{\rm(\!\!\:h\!\!\:)\!}}
\newcommand{\NNGg}{\mathbbm{N}_{\rm\!G}^{\rm(\!\!\:g\!\!\:)\!}}
\newcommand{\dNND}{\dot{\mathbbm{N}}_{\!{\rm D}}}
\newcommand{\ddNND}{\ddot{\mathbbm{N}}_{\!{\rm D}}}
\newcommand{\tNND}{\tilde{\mathbbm{N}}_{\!{\rm D}}}
\newcommand{\dNNGe}{\dot{\mathbbm{N}}_{\rm\!G}^{\rm(\!\!\:e\!\!\:)\!}}
\newcommand{\ddNNGe}{\ddot{\mathbbm{N}}_{\rm\!G}^{\rm(\!\!\:e\!\!\:)\!}}

\newcommand{\lambDa}{\lambda_{{\rm D}\!\!\:(\!\!\;a\!\!\;)}}
\newcommand{\lambDe}{\lambda_{{\rm D}\!\!\:(\!\!\;1\!\!\;)}}
\newcommand{\lambDz}{\lambda_{{\rm D}\!\!\:(\!\!\;2\!\!\;)}}
\newcommand{\lambDd}{\lambda_{{\rm D}\!\!\:(\!\!\;3\!\!\;)}}
\newcommand{\lambdapa}{\lambda_{{\rm P}\!\!\:(\!\!\;a\!\!\;)}}
\newcommand{\lambdape}{\lambda_{{\rm P}\!\!\:(\!\!\;1\!\!\;)}}
\newcommand{\lambdapz}{\lambda_{{\rm P}\!\!\:(\!\!\;2\!\!\;)}}
\newcommand{\lambdaGe}{\lambda_{\rm G}^{\rm(\!\!\:e\!\!\:)\!}}
\newcommand{\lambdaGm}{\lambda_{\rm G}^{\rm(\!\!\:m\!\!\:)\!}}
\newcommand{\lambdaGh}{\lambda_{\rm G}^{\rm(\!\!\:h\!\!\:)\!}}
\newcommand{\lambdaGg}{\lambda_{\rm G}^{\rm(\!\!\:g\!\!\:)\!}}

\newcommand{\dlambdaS}{\dot{\lambda}_{\rm S}}
\newcommand{\ddlambdaS}{\ddot{\lambda}_{\rm S}}

\newcommand{\MZ}{\mathcal{Z}}
\newcommand{\sMZ}{{\!\:{}'\!\!\!\;\mathcal{Z}}}
\newcommand{\tMZ}{\tilde{\mathcal{Z}}}
\newcommand{\MZa}{\mathcal{Z}_{(a)}}
\newcommand{\MZe}{\mathcal{Z}_{(1)}}
\newcommand{\MZz}{\mathcal{Z}_{(2)}}
\newcommand{\MZd}{\mathcal{Z}_{(3)}}

\newcommand{\WD}{W_{\!\rm D}}
\newcommand{\WG}{W_{\!\rm G}}

\newcommand{\RRST}{_{\!\rm RST}}
\newcommand{\WRST}{W_{\!\rm RST}}
\newcommand{\MLRST}{\ML_{\rm RST}}

\newcommand{\Tmunu}{T_{\!\mu\nu}}
\newcommand{\TTmunu}{{{}^{\scriptscriptstyle\rm(T)\!}T_{\!\mu\nu}}}
\newcommand{\DTmunu}{{{}^{\scriptscriptstyle\rm(D)\!}T_{\!\mu\nu}}}
\newcommand{\GTmunu}{{{}^{\scriptscriptstyle\rm(G)\!}T_{\!\mu\nu}}}
\newcommand{\Too}{T_{\!00}}
\newcommand{\TToo}{{{}^{\scriptscriptstyle\rm(T)\!}T_{\!00}}}
\newcommand{\DToo}{{{}^{\scriptscriptstyle\rm(D)\!}T_{\!00}}}
\newcommand{\GToo}{{{}^{\scriptscriptstyle\rm(G)\!}T_{\!00}}}

\newcommand{\ES}{E_{\rm S}}
\newcommand{\ET}{E_{\rm T}}
\newcommand{\tET}{\tilde{E}_{\rm T}}

\newcommand{\EET}{\mathbbm{E}_{\rm T}}
\newcommand{\tEET}{\tilde{\mathbbm{E}}_{\rm T}}

\newcommand{\EivT}{E^{\rm(\raisebox{.4pt}{\rotatebox[origin=c]{270}{$\scriptscriptstyle\geq$}}\!\!\;)\!}_{\rm T}}
\newcommand{\EifT}{E^{\rm(\!\raisebox{.3pt}{\rotatebox[origin=c]{270}{$\scriptscriptstyle -$}}\!)\!}_{\rm[T]}}
\newcommand{\EiifT}{E^{\rm(\!\!\;\raisebox{.8pt}{\rotatebox[origin=c]{270}{$\scriptscriptstyle =$}})\!}_{\rm[T]}}
\newcommand{\EiiifT}{E^{\rm(\raisebox{.5pt}{\rotatebox[origin=c]{270}{$\scriptscriptstyle\equiv$}})\!}_{\rm[T]}}
\newcommand{\EivfT}{E^{\rm(\raisebox{.4pt}{\rotatebox[origin=c]{270}{$\scriptscriptstyle\geq$}}\!\!\;)\!}_{\rm[T]}}
\newcommand{\ED}{E_{\rm D}}
\newcommand{\EG}{E_{\rm G}}
\newcommand{\EfD}{E_{\rm[D]}}
\newcommand{\EfG}{E_{\rm[G]}}
\newcommand{\ER}{E_{\rm R}}
\newcommand{\EC}{E_{\rm C}}
\newcommand{\ERe}{E_{\rm R}^{\rm(\!\!\:e\!\!\:)\!}}
\newcommand{\ERm}{E_{\rm R}^{\rm(\!\!\:m\!\!\:)\!}}
\newcommand{\ECh}{E_{\rm C}^{\rm(\!\!\:h\!\!\:)\!}}
\newcommand{\ECg}{E_{\rm C}^{\rm(\!\!\:g\!\!\:)\!}}

\newcommand{\Epa}{E_{{\rm P}\!\!\:(\!\!\;a\!\!\;)}}
\newcommand{\Epe}{E_{{\rm P}\!\!\:(\!\!\:1\!\!\:)}}
\newcommand{\Epz}{E_{{\rm P}\!\!\:(\!\!\:2\!\!\:)}}
\newcommand{\Epd}{E_{{\rm P}\!\!\:(\!\!\:3\!\!\:)}}
\newcommand{\Efp}{E_{[{\rm P}]}}
\newcommand{\tEfp}{\tilde{E}_{[{\rm P}]}}
\newcommand{\Epfa}{E_{{\rm P}\!\!\:[\!\!\;a\!\!\;]}}
\newcommand{\Epfe}{E_{{\rm P}\!\!\:[\!\!\:1\!\!\;]}}
\newcommand{\Epfz}{E_{{\rm P}\!\!\:[\!\!\:2\!\!\;]}}
\newcommand{\Epfd}{E_{{\rm P}\!\!\:[\!\!\:3\!\!\;]}}
\newcommand{\EfT}{E_{[{\rm T}]}}
\newcommand{\tEpfa}{\tilde{E}_{{\rm P}\!\!\:[\!\!\:a\!\!\;]}}
\newcommand{\tEpfe}{\tilde{E}_{{\rm P}\!\!\:[\!\!\:1\!\!\;]}}

\newcommand{\EDfa}{E_{{\rm D}\!\!\:[\!\!\;a\!\!\;]}}
\newcommand{\EDfe}{E_{{\rm D}\!\!\:[\!\!\:1\!\!\;]}}
\newcommand{\EDfz}{E_{{\rm D}\!\!\:[\!\!\:2\!\!\;]}}
\newcommand{\EDfd}{E_{{\rm D}\!\!\:[\!\!\:3\!\!\;]}}
\newcommand{\tEfT}{\tilde{E}_{[{\rm T}]}}
\newcommand{\tEEfT}{\tilde{\mathbbm{E}}_{[{\rm T}]}}
\newcommand{\tEEfTo}{\tilde{\mathbbm{E}}_{[{\rm T}]}^{(\!\!\:\circ\!\!\:)}}
\newcommand{\tEETo}{\tilde{\mathbbm{E}}_{{\rm T}}^{(\!\!\:\circ\!\!\:)}}
\newcommand{\stEEfT}{{\!\:{}'\!\!\;\tilde{\mathbbm{E}}}_{[{\rm T}]}}
\newcommand{\stEET}{{\!\:{}'\!\!\;\tilde{\mathbbm{E}}}_{{\rm T}}}

\newcommand{\stEfT}{{{}'\!\tilde{E}}_{[\rm T]}}
\newcommand{\dEkin}{\dot{E}_{\rm kin}}
\newcommand{\ddEkin}{\ddot{E}_{\rm kin}}
\newcommand{\dERe}{\dot{E}_{\rm R}^{\rm(\!\!\:e\!\!\:)\!}}

\newcommand{\sERe}{{{}'\!\!\!\;E}_{\rm R}^{\rm(\!\!\:e\!\!\:)\!}}
\newcommand{\Tkin}{T_{\!\!\:\rm kin}}
\newcommand{\sTkin}{{{}'\!\!\;T}_{\!\!\:\rm kin}}
\newcommand{\tTkin}{\tilde{T}_{\!\!\:\rm kin}}
\newcommand{\Tkina}{T_{\!\!\:{\rm kin}\!\!\:(\!\!\;a\!\!\;)}}
\newcommand{\Tkine}{T_{\!\!\:{\rm kin}\!\!\:(\!\!\:1\!\!\:)}}
\newcommand{\Tkinz}{T_{\!\!\:{\rm kin}\!\!\:(\!\!\:2\!\!\:)}}
\newcommand{\Tkind}{T_{\!\!\:{\rm kin}\!\!\:(\!\!\:3\!\!\:)}}
\newcommand{\Ekin}{E_{\!\!\:{\rm kin}}}
\newcommand{\Ekina}{E_{\!\!\:{\rm kin}\!\!\:(\!\!\;a\!\!\;)}}
\newcommand{\Ekine}{E_{\!\!\:{\rm kin}\!\!\:(\!\!\:1\!\!\:)}}
\newcommand{\Ekinz}{E_{\!\!\:{\rm kin}\!\!\:(\!\!\:2\!\!\:)}}
\newcommand{\Ekind}{E_{\!\!\:{\rm kin}\!\!\:(\!\!\:3\!\!\:)}}
\newcommand{\Eskine}{E'_{\!\!\:{\rm kin}\!\!\:(\!\!\:1\!\!\:)}}

\newcommand{\Econv}{E_{\!\!\:{\rm conv}}}
\newcommand{\Es}{E_{\!\!\:{\rm S}}}
\newcommand{\Esz}{E_{\!\!\:{\rm S}\!\!\:(\!\!\:2\!\!\:)}}
\newcommand{\epskin}{\varepsilon_{\!\!\;{\rm kin}}}
\newcommand{\epspot}{\varepsilon_{\!\!\:{\rm pot}}}

\newcommand{\MAmu}{\mathcal{A}_\mu}
\newcommand{\MAnu}{\mathcal{A}_\nu}

\newcommand{\MFmunu}{\mathcal{F}_{\!\mu\nu}}

\newcommand{\MJmu}{\mathcal{J}_{\!\mu}}
\newcommand{\MJnu}{\mathcal{J}_{\!\nu}}

\newcommand{\Bmu}{B_\mu}
\newcommand{\Bmo}{B^\mu}
\newcommand{\Bnu}{B_\nu}
\newcommand{\Bno}{B^\nu}
\newcommand{\BSmu}{\vcenter{\offinterlineskip \hbox{$\scriptstyle\;\:\ast$}\vskip.2ex\hbox{$\Bmu$}\vskip.7ex}}
\newcommand{\BSmo}{\vcenter{\offinterlineskip \hbox{$\scriptstyle\;\:\ast$}\vskip.2ex\hbox{$\Bmo$}\vskip1.35ex}}
\newcommand{\BSnu}{\vcenter{\offinterlineskip \hbox{$\scriptstyle\;\:\ast$}\vskip.2ex\hbox{$\Bnu$}\vskip1ex}}
\newcommand{\BSno}{\vcenter{\offinterlineskip \hbox{$\scriptstyle\;\:\ast$}\vskip.2ex\hbox{$\Bno$}\vskip1.35ex}}

\newcommand{\Gmunu}{G_{\!\mu\nu}}
\newcommand{\Gmono}{G^{\mu\nu}}
\newcommand{\Gmonu}{G^\mu_{\;\nu}}
\newcommand{\GSmunu}{\vcenter{\offinterlineskip \hbox{$\scriptstyle\;\:\ast$}\vskip.2ex\hbox{$\Gmunu$}\vskip.7ex}}
\newcommand{\GSmono}{\vcenter{\offinterlineskip \hbox{$\scriptstyle\;\:\ast$}\vskip.2ex\hbox{$\Gmono$}\vskip1.35ex}}

\newcommand{\Zmunu}{Z_{\mu\nu}}

\newcommand{\Kalubeu}{K_{\alpha\beta}}
\newcommand{\Kalobeo}{K^{\alpha\beta}}

\newcommand{\Aaomu}{A^a_{\;\mu}}
\newcommand{\Aeomu}{A^1_{\;\mu}}
\newcommand{\Azomu}{A^2_{\;\mu}}
\newcommand{\Adomu}{A^3_{\;\mu}}
\newcommand{\Aaonu}{A^a_{\;\nu}}
\newcommand{\Aeonu}{A^1_{\;\nu}}
\newcommand{\Azonu}{A^2_{\;\nu}}
\newcommand{\Adonu}{A^3_{\;\nu}}

\newcommand{\Azomo}{A^{2\mu}}
\newcommand{\Adomo}{A^{3\mu}}

\newcommand{\Azono}{A^{2\nu}}
\newcommand{\Adono}{A^{3\nu}}
\newcommand{\Bkomu}{B^k_{\;\mu}}

\newcommand{\alo}{{\scriptscriptstyle(\!\!\;}a{\scriptscriptstyle\!\!\;)}}
\newcommand{\plo}{{\scriptscriptstyle(}p{\scriptscriptstyle\!\!\;)}}
\newcommand{\exAmu}{{{}^{\scriptscriptstyle\rm(\!\!\:ex\!\!\:)\!\!\!}A_\mu}}
\newcommand{\aAo}{{{}^{\alo\!\!\!}A_0}}
\newcommand{\dotAo}{\dot{A}_0}
\newcommand{\ddotAo}{\ddot{A}_0}
\newcommand{\pAo}{{{}^{\plo\!\!\!}A_0}}
\newcommand{\sAo}{{\:{}'\!\!\!\!\:A}_0}
\newcommand{\eAo}{{{}^{\rm(\!\!\:1\!\!\:)\!\!\!}A_0}}
\newcommand{\zAo}{{{}^{\rm(\!\!\:2\!\!\:)\!\!\!}A_0}}
\newcommand{\dAo}{{{}^{\rm(\!\!\:3\!\!\:)\!\!\!}A_0}}
\newcommand{\iAo}{{{}^{\rm(\!\!\:I\!\!\:)\!\!\!}A_0}}
\newcommand{\iiAo}{{{}^{\rm(\!\!\:II\!\!\:)\!\!\!}A_0}}
\newcommand{\iiiAo}{{{}^{\rm(\!\!\:III\!\!\:)\!\!\!}A_0}}
\newcommand{\vAa}{\vec{A}_a}
\newcommand{\vAp}{\vec{A}_p}
\newcommand{\vAe}{\vec{A}_1}
\newcommand{\vAz}{\vec{A}_2}
\newcommand{\vAd}{\vec{A}_3}
\newcommand{\vAi}{\vec{A}_{\rm I}}
\newcommand{\vAii}{\vec{A}_{\rm II}}
\newcommand{\vAiii}{\vec{A}_{\rm III}}
\newcommand{\aAr}{{{}^{\alo\!\!\!}A_r}}
\newcommand{\aAt}{{{}^{\alo\!\!\!}A_\vartheta}}
\newcommand{\aAp}{{{}^{\alo\!\!\!}A_\phi}}
\newcommand{\eAr}{{{}^{\rm(\!\!\:1\!\!\:)\!\!\!}A_r}}
\newcommand{\eAt}{{{}^{\rm(\!\!\:1\!\!\:)\!\!\!}A_\vartheta}}
\newcommand{\eAp}{{{}^{\rm(\!\!\:1\!\!\:)\!\!\!}A_\phi}}
\newcommand{\zAr}{{{}^{\rm(\!\!\:2\!\!\:)\!\!\!}A_r}}
\newcommand{\zAt}{{{}^{\rm(\!\!\:2\!\!\:)\!\!\!}A_\vartheta}}
\newcommand{\zAp}{{{}^{\rm(\!\!\:2\!\!\:)\!\!\!}A_\phi}}

\newcommand{\eAmo}{{{}^{\rm(\!\!\:1\!\!\:)\!\!\!}A^\mu}}
\newcommand{\zAmo}{{{}^{\rm(\!\!\:2\!\!\:)\!\!\!}A^\mu}}
\newcommand{\ajo}{{{}^{\alo\!\!}j_0}}
\newcommand{\pjo}{{{}^{\plo\!\!}j_0}}
\newcommand{\ejo}{{{}^{\rm(\!\!\:1\!\!\:)\!\!}j_0}}
\newcommand{\zjo}{{{}^{\rm(\!\!\:2\!\!\:)\!\!}j_0}}
\newcommand{\djo}{{{}^{\rm(\!\!\:3\!\!\:)\!\!}j_0}}
\newcommand{\vja}{\vec{j}_a}
\newcommand{\vjp}{\vec{j}_p}
\newcommand{\vje}{\vec{j}_1}
\newcommand{\vjz}{\vec{j}_2}
\newcommand{\vjd}{\vec{j}_3}
\newcommand{\ako}{{{}^{\alo\!\!\!\:}k_0}}
\newcommand{\pko}{{{}^{\plo\!\!\!\:}k_0}}
\newcommand{\eko}{{{}^{\rm(\!\!\:1\!\!\:)\!\!\!\:}k_0}}
\newcommand{\zko}{{{}^{\rm(\!\!\:2\!\!\:)\!\!\!\:}k_0}}
\newcommand{\dko}{{{}^{\rm(\!\!\:3\!\!\:)\!\!\!\:}k_0}}
\newcommand{\vka}{\vec{k}_a}
\newcommand{\vkp}{\vec{k}_p}
\newcommand{\vke}{\vec{k}_1}
\newcommand{\vkz}{\vec{k}_2}
\newcommand{\vkd}{\vec{k}_3}

\newcommand{\vqe}{\vec{q}_1}

\newcommand{\vsa}{\vec{s}_a}
\newcommand{\Bo}{B_0}
\newcommand{\vB}{\vec{B}}
\newcommand{\BSo}{\vcenter{\offinterlineskip \hbox{$\scriptstyle\;\:\ast$}\vskip.2ex\hbox{$\Bo$}\vskip1ex}}
\newcommand{\vBS}{\vcenter{\offinterlineskip \hbox{$\scriptstyle\;\:\ast$}\vskip.0ex\hbox{$\vB$}\vskip1.8ex}}
\newcommand{\ho}{h_0}
\newcommand{\vh}{\vec{h}}
\newcommand{\hSo}{\vcenter{\offinterlineskip \hbox{$\scriptstyle\;\ast$}\vskip.2ex\hbox{$\ho$}\vskip1ex}}
\newcommand{\vhS}{\vcenter{\offinterlineskip \hbox{$\scriptstyle\;\ast$}\vskip.0ex\hbox{$\vh$}\vskip1.8ex}}
\newcommand{\aEjo}{{{}^{\alo\!\!}E^j}}
\newcommand{\vE}{\vec{E}}
\newcommand{\vEa}{\vec{E}_a}
\newcommand{\vEe}{\vec{E}_1}
\newcommand{\vEz}{\vec{E}_2}
\newcommand{\vEd}{\vec{E}_3}
\newcommand{\aHjo}{{{}^{\alo\!\!}H^j}}
\newcommand{\vH}{\vec{H}}
\newcommand{\vHa}{\vec{H}_a}
\newcommand{\vHp}{\vec{H}_p}
\newcommand{\vHe}{\vec{H}_1}
\newcommand{\vHz}{\vec{H}_2}
\newcommand{\vHd}{\vec{H}_3}
\newcommand{\vHi}{\vec{H}_{\rm I}}
\newcommand{\vHii}{\vec{H}_{\rm II}}
\newcommand{\vHiii}{\vec{H}_{\rm III}}
\newcommand{\vHiiiiii}{\vec{H}_{\rm I,II,III}}
\newcommand{\vX}{\vec{X}}
\newcommand{\vXS}{\vcenter{\offinterlineskip \hbox{$\scriptstyle\;\:\ast$}\vskip.0ex\hbox{$\vX$}\vskip1.8ex}}
\newcommand{\vY}{\vec{Y}}
\newcommand{\vYS}{\vcenter{\offinterlineskip \hbox{$\scriptstyle\;\:\ast$}\vskip.0ex\hbox{$\vY$}\vskip1.8ex}}
\newcommand{\aphipm}{{{}^{\alo\!\!\!\;}\varphi_{\!\pm}}}
\newcommand{\pphipm}{{{}^{\plo\!\!\!\;}\varphi_{\!\pm}}}
\newcommand{\ephipm}{{{}^{\rm(\!\!\:1\!\!\:)\!\!}\varphi_{\!\pm}}}
\newcommand{\zphipm}{{{}^{\rm(\!\!\:2\!\!\:)\!\!}\varphi_{\!\pm}}}
\newcommand{\dphipm}{{{}^{\rm(\!\!\:3\!\!\:)\!\!}\varphi_{\!\pm}}}

\newcommand{\pphimp}{{{}^{\plo\!\!\!\;}\varphi_{\!\mp}}}
\newcommand{\ephimp}{{{}^{\rm(\!\!\:1\!\!\:)\!\!}\varphi_{\!\mp}}}
\newcommand{\zphimp}{{{}^{\rm(\!\!\:2\!\!\:)\!\!}\varphi_{\!\mp}}}
\newcommand{\dphimp}{{{}^{\rm(\!\!\:3\!\!\:)\!\!}\varphi_{\!\mp}}}
\newcommand{\aphip}{{{}^{\alo\!\!\!\;}\varphi_{\!+}}}
\newcommand{\pphip}{{{}^{\plo\!\!\!\;}\varphi_{\!+}}}
\newcommand{\ephip}{{{}^{\rm(\!\!\:1\!\!\:)\!\!}\varphi_{\!+}}}
\newcommand{\zphip}{{{}^{\rm(\!\!\:2\!\!\:)\!\!}\varphi_{\!+}}}
\newcommand{\dphip}{{{}^{\rm(\!\!\:3\!\!\:)\!\!}\varphi_{\!+}}}
\newcommand{\aphim}{{{}^{\alo\!\!\!\;}\varphi_{\!-}}}
\newcommand{\pphim}{{{}^{\plo\!\!\!\;}\varphi_{\!-}}}
\newcommand{\ephim}{{{}^{\rm(\!\!\:1\!\!\:)\!\!}\varphi_{\!-}}}
\newcommand{\zphim}{{{}^{\rm(\!\!\:2\!\!\:)\!\!}\varphi_{\!-}}}
\newcommand{\dphim}{{{}^{\rm(\!\!\:3\!\!\:)\!\!}\varphi_{\!-}}}
\newcommand{\aphipmk}{{{}^{\alo\!\!\!\;}\varphi_{\!\pm}^\dagger}}

\newcommand{\aphipk}{{{}^{\alo\!\!\!\;}\varphi_{\!+}^\dagger}}
\newcommand{\ephipk}{{{}^{\rm(\!\!\:1\!\!\:)\!\!}\varphi_{\!+}^\dagger}}
\newcommand{\zphipk}{{{}^{\rm(\!\!\:2\!\!\:)\!\!}\varphi_{\!+}^\dagger}}
\newcommand{\dphipk}{{{}^{\rm(\!\!\:3\!\!\:)\!\!}\varphi_{\!+}^\dagger}}
\newcommand{\aphimk}{{{}^{\alo\!\!\!\;}\varphi_{\!-}^\dagger}}
\newcommand{\ephimk}{{{}^{\rm(\!\!\:1\!\!\:)\!\!}\varphi_{\!-}^\dagger}}
\newcommand{\zphimk}{{{}^{\rm(\!\!\:2\!\!\:)\!\!}\varphi_{\!-}^\dagger}}

\newcommand{\evSe}{{{}^{\scriptscriptstyle\rm(\!\!\:e\!\!\:)\!\!}\vec{S}_1}}
\newcommand{\mvSa}{{{}^{\scriptscriptstyle\rm(\!\!\:m\!\!\:)\!\!}\vec{S}_a}}
\newcommand{\mvSe}{{{}^{\scriptscriptstyle\rm(\!\!\:m\!\!\:)\!\!}\vec{S}_1}}
\newcommand{\mvSz}{{{}^{\scriptscriptstyle\rm(\!\!\:m\!\!\:)\!\!}\vec{S}_2}}
\newcommand{\mvSd}{{{}^{\scriptscriptstyle\rm(\!\!\:m\!\!\:)\!\!}\vec{S}_3}}
\newcommand{\eSejo}{{{}^{\scriptscriptstyle\rm(\!\!\:e\!\!\:)\!\!}S_1^{\;j}}}
\newcommand{\mSejo}{{{}^{\scriptscriptstyle\rm(\!\!\:m\!\!\:)\!\!}S_1^{\;j}}}
\newcommand{\eso}{{{}^{\rm(\!\!\:1\!\!\:)\!\!}s_0}}
\newcommand{\vse}{\vec{s}_1}
\newcommand{\eqo}{{{}^{\rm(\!\!\:1\!\!\:)\!\!}q_0}}

\newcommand{\Faomunu}{F^a_{\;\,\mu\nu}}
\newcommand{\Feomunu}{F^1_{\;\,\mu\nu}}
\newcommand{\Fzomunu}{F^2_{\;\,\mu\nu}}
\newcommand{\Fdomunu}{F^3_{\;\,\mu\nu}}
\newcommand{\Falomunu}{F^\alpha_{\;\,\mu\nu}}
\newcommand{\Fbeomunu}{F^\beta_{\;\,\mu\nu}}
\newcommand{\Falomono}{F^{\alpha\mu\nu}}
\newcommand{\Faoouju}{F^a_{\;\,0j}}
\newcommand{\Faokulo}{F^{a\,\,l}_{\;\,k}}

\newcommand{\Seomunu}{S^{(1)}_{\;\,\mu\nu}}
\newcommand{\Seonumu}{S^{(1)}_{\;\,\nu\mu}}
\newcommand{\Szomunu}{S^{(2)}_{\;\,\mu\nu}}
\newcommand{\Sdomunu}{S^{(3)}_{\;\,\mu\nu}}
\newcommand{\Seoouju}{S^{(1)}_{\;\,0j}}
\newcommand{\Seokulu}{S^{(1)}_{\;\,kl}}

\newcommand{\Gouju}{G_{\!\!\:0j}}
\newcommand{\Gkulo}{G_{\!k}^{\:\:l}}

\newcommand{\kaumu}{k_{a\mu}}
\newcommand{\keumu}{k_{1\mu}}
\newcommand{\kzumu}{k_{2\mu}}
\newcommand{\kdumu}{k_{3\mu}}
\newcommand{\keumo}{k_1^{\;\mu}}
\newcommand{\kzumo}{k_2^{\;\mu}}
\newcommand{\kdumo}{k_3^{\;\mu}}
\newcommand{\qaumu}{q_{a\mu}}
\newcommand{\qeumu}{q_{1\mu}}

\newcommand{\saumu}{s_{a\mu}}
\newcommand{\seumu}{s_{1\mu}}

\newcommand{\seujo}{s_1^{\;j}}

\newcommand{\hmu}{h_\mu}
\newcommand{\hnu}{h_\nu}
\newcommand{\hSmu}{\vcenter{\offinterlineskip \hbox{$\scriptstyle\;\ast$}\vskip.2ex\hbox{$\hmu$}\vskip.7ex}}
\newcommand{\hSnu}{\vcenter{\offinterlineskip \hbox{$\scriptstyle\,\,\ast$}\vskip.2ex\hbox{$\hnu$}\vskip1ex}}
\newcommand{\hmo}{h^\mu}

\newcommand{\jalomu}{j^\alpha_{\;\mu}}
\newcommand{\jalomo}{j^{\alpha\mu}}
\newcommand{\jalumu}{j_{\alpha\mu}}

\newcommand{\jbeumu}{j_{\beta\mu}}
\newcommand{\jaomu}{j^a_{\;\mu}}
\newcommand{\jaonu}{j^a_{\;\nu}}
\newcommand{\jeomu}{j^1_{\;\mu}}
\newcommand{\jzomu}{j^2_{\;\mu}}
\newcommand{\jdomu}{j^3_{\;\mu}}
\newcommand{\jvomu}{j^4_{\;\mu}}
\newcommand{\jfomu}{j^5_{\;\mu}}
\newcommand{\jeonu}{j^1_{\;\nu}}
\newcommand{\jzonu}{j^2_{\;\nu}}
\newcommand{\jdonu}{j^3_{\;\nu}}
\newcommand{\jeumu}{j_{1\mu}}
\newcommand{\jzumu}{j_{2\mu}}
\newcommand{\jdumu}{j_{3\mu}}
\newcommand{\jvumu}{j_{4\mu}}
\newcommand{\jfumu}{j_{5\mu}}

\newcommand{\gmu}{g_\mu}
\newcommand{\gnu}{g_\nu}
\newcommand{\gSmu}{\vcenter{\offinterlineskip \hbox{$\scriptstyle\;\ast$}\vskip.2ex\hbox{$\gmu$}\vskip.1ex}}

\newcommand{\bmu}{b_\mu}
\newcommand{\zmu}{z_\mu}

\newcommand{\hvJ}{\hat{\vec{J}}}
\newcommand{\hJz}{\hat{J}_z}
\newcommand{\hvL}{\hat{\vec{L}}}
\newcommand{\hvS}{\hat{\vec{S}}}

\newcommand{\zjml}{\zeta^{j,m}_l}
\newcommand{\zopp}{\zeta^{\frac{1}{2},\frac{1}{2}}_0}
\newcommand{\zopm}{\zeta^{\frac{1}{2},-\frac{1}{2}}_0}
\newcommand{\zepp}{\zeta^{\frac{1}{2},\frac{1}{2}}_1}
\newcommand{\zepm}{\zeta^{\frac{1}{2},-\frac{1}{2}}_1}
\newcommand{\zdee}{\zeta^{\frac{3}{2},\frac{1}{2}}_1}
\newcommand{\zdez}{\zeta^{\frac{3}{2},\frac{1}{2}}_2}

\newcommand{\woepm}{\omega^{(\pm)}_{0,1}}
\newcommand{\wop}{\omega^{(+)}_0}
\newcommand{\wom}{\omega^{(-)}_0}
\newcommand{\wep}{\omega^{(+)}_1}
\newcommand{\wem}{\omega^{(-)}_1}

\newcommand{\akr}{{{}^{\alo\!\!\!\:}k_r}}
\newcommand{\akt}{{{}^{\alo\!\!\!\:}k_\vartheta}}
\newcommand{\akp}{{{}^{\alo\!\!\!\:}k_\phi}}

\newcommand{\tRpm}{\tilde{R}_\pm}
\newcommand{\tRp}{\tilde{R}_+}
\newcommand{\tRm}{\tilde{R}_-}
\newcommand{\tSpm}{\tilde{S}_\pm}
\newcommand{\tSp}{\tilde{S}_+}
\newcommand{\tSm}{\tilde{S}_-}
\newcommand{\dRpm}{\dot{R}_\pm}
\newcommand{\dRp}{\dot{R}_+}

\newcommand{\dSpm}{\dot{S}_\pm}
\newcommand{\dSp}{\dot{S}_+}

\newcommand{\ddRp}{\ddot{R}_+}

\newcommand{\ddSp}{\ddot{S}_+}

\newcommand{\nRpm}{\vcenter{\offinterlineskip \hbox{$\scriptscriptstyle\,\:\scoh$}\vskip-.1ex\hbox{$R_\pm$}\vskip.9ex}{}}
\newcommand{\nRp}{\vcenter{\offinterlineskip \hbox{$\scriptscriptstyle\,\:\scoh$}\vskip-.1ex\hbox{$R_+$}\vskip.9ex}{}}
\newcommand{\nRm}{\vcenter{\offinterlineskip \hbox{$\scriptscriptstyle\,\:\scoh$}\vskip-.1ex\hbox{$R_-$}\vskip.9ex}{}}

\newcommand{\sRpm}{{{}'\!\!\!\;R}_\pm}
\newcommand{\sRp}{{{}'\!\!\!\;R}_+}
\newcommand{\sRm}{{{}'\!\!\!\;R}_-}
\newcommand{\sSpm}{{{}'\!\!\!\:S}_\pm}
\newcommand{\sSp}{{{}'\!\!\!\:S}_+}
\newcommand{\sSm}{{{}'\!\!\!\:S}_-}
\newcommand{\zRpm}{{{}^{\rm(\!\!\:2\!\!\:)\!\!\!\!\;}R_\pm}}
\newcommand{\zRp}{{{}^{\rm(\!\!\:2\!\!\:)\!\!\!\!\;}R_+}}
\newcommand{\zRm}{{{}^{\rm(\!\!\:2\!\!\:)\!\!\!\!\;}R_-}}
\newcommand{\pRpm}{{{}^{\plo\!\!\!\!\;}R_\pm}}
\newcommand{\pRp}{{{}^{\plo\!\!\!\!\;}R_+}}
\newcommand{\pRm}{{{}^{\plo\!\!\!\!\;}R_-}}
\newcommand{\nSpm}{\vcenter{\offinterlineskip \hbox{$\scriptscriptstyle\;\scoh$}\vskip-.1ex\hbox{$S_\pm$}\vskip.9ex}{}}
\newcommand{\nSp}{\vcenter{\offinterlineskip \hbox{$\scriptscriptstyle\;\scoh$}\vskip-.1ex\hbox{$S_+$}\vskip.9ex}{}}
\newcommand{\nSm}{\vcenter{\offinterlineskip \hbox{$\scriptscriptstyle\;\scoh$}\vskip-.1ex\hbox{$S_-$}\vskip.9ex}{}}

\newcommand{\oRp}{\vcenter{\offinterlineskip \hbox{$\scriptscriptstyle\;\;\circ$}\vskip.1ex\hbox{$R_+$}\vskip.6ex}{}}

\newcommand{\oSp}{\vcenter{\offinterlineskip \hbox{$\scriptscriptstyle\;\,\circ$}\vskip.1ex\hbox{$S_+$}\vskip.6ex}{}}

\newcommand{\po}{p_\circ}
\newcommand{\fo}{f_{\!\!\:\circ}}
\newcommand{\nuo}{{\nu_\circ}}
\newcommand{\beto}{\beta_\circ}
\newcommand{\zSpm}{{{}^{\rm(\!\!\:2\!\!\:)\!\!}S_\pm}}
\newcommand{\zSp}{{{}^{\rm(\!\!\:2\!\!\:)\!\!}S_+}}
\newcommand{\zSm}{{{}^{\rm(\!\!\:2\!\!\:)\!\!}S_-}}
\newcommand{\pSpm}{{{}^{\plo\!\!}S_\pm}}
\newcommand{\pSp}{{{}^{\plo\!\!}S_+}}
\newcommand{\pSm}{{{}^{\plo\!\!}S_-}}
\newcommand{\aMRpm}{{{}^{\alo\!\!\:}\mathcal{R}_\pm}}

\newcommand{\aMRp}{{{}^{\alo\!\!\:}\mathcal{R}_+}}

\newcommand{\aMRm}{{{}^{\alo\!\!\:}\mathcal{R}_-}}

\newcommand{\eMRpm}{{{}^{\rm(\!\!\:1\!\!\:)\!}\mathcal{R}_\pm}}
\newcommand{\eMRp}{{{}^{\rm(\!\!\:1\!\!\:)\!}\mathcal{R}_+}}
\newcommand{\eMRm}{{{}^{\rm(\!\!\:1\!\!\:)\!}\mathcal{R}_-}}
\newcommand{\zMRpm}{{{}^{\rm(\!\!\:2\!\!\:)\!}\mathcal{R}_\pm}}
\newcommand{\zMRp}{{{}^{\rm(\!\!\:2\!\!\:)\!}\mathcal{R}_+}}
\newcommand{\zMRm}{{{}^{\rm(\!\!\:2\!\!\:)\!}\mathcal{R}_-}}
\newcommand{\aMSpm}{{{}^{\alo\!\!}\mathcal{S}_\pm}}

\newcommand{\aMSp}{{{}^{\alo\!\!}\mathcal{S}_+}}

\newcommand{\aMSm}{{{}^{\alo\!\!}\mathcal{S}_-}}

\newcommand{\eMSpm}{{{}^{\rm(\!\!\:1\!\!\:)\!\!}\mathcal{S}_\pm}}
\newcommand{\eMSp}{{{}^{\rm(\!\!\:1\!\!\:)\!\!}\mathcal{S}_+}}
\newcommand{\eMSm}{{{}^{\rm(\!\!\:1\!\!\:)\!\!}\mathcal{S}_-}}
\newcommand{\zMSpm}{{{}^{\rm(\!\!\:2\!\!\:)\!\!}\mathcal{S}_\pm}}
\newcommand{\zMSp}{{{}^{\rm(\!\!\:2\!\!\:)\!\!}\mathcal{S}_+}}
\newcommand{\zMSm}{{{}^{\rm(\!\!\:2\!\!\:)\!\!}\mathcal{S}_-}}
\newcommand{\zMRpmS}{{{}^{\rm(\!\!\:2\!\!\:)\!}{\vcenter{\offinterlineskip\hbox{$\scriptstyle\:\,\ast$}\vskip.2ex\hbox{$\mathcal{R}_\pm$}\vskip.7ex}}}}
\newcommand{\zMRpS}{{{}^{\rm(\!\!\:2\!\!\:)\!}{\vcenter{\offinterlineskip\hbox{$\scriptstyle\:\,\ast$}\vskip.2ex\hbox{$\mathcal{R}_+$}\vskip.7ex}}}}
\newcommand{\zMRmS}{{{}^{\rm(\!\!\:2\!\!\:)\!}{\vcenter{\offinterlineskip\hbox{$\scriptstyle\:\,\ast$}\vskip.2ex\hbox{$\mathcal{R}_-$}\vskip.7ex}}}}
\newcommand{\aMRpS}{{{}^{\alo\!\!\:}{\vcenter{\offinterlineskip\hbox{$\scriptstyle\:\,\ast$}\vskip.2ex\hbox{$\mathcal{R}_+$}\vskip.7ex}}}}
\newcommand{\aMRmS}{{{}^{\alo\!\!\:}{\vcenter{\offinterlineskip\hbox{$\scriptstyle\:\,\ast$}\vskip.2ex\hbox{$\mathcal{R}_-$}\vskip.7ex}}}}

\newcommand{\zMSpS}{{{}^{\rm(\!\!\:2\!\!\:)\!\!}{\vcenter{\offinterlineskip\hbox{$\scriptstyle\:\,\ast$}\vskip.2ex\hbox{$\mathcal{S}_+$}\vskip.7ex}}}}
\newcommand{\zMSmS}{{{}^{\rm(\!\!\:2\!\!\:)\!\!}{\vcenter{\offinterlineskip\hbox{$\scriptstyle\:\,\ast$}\vskip.2ex\hbox{$\mathcal{S}_-$}\vskip.7ex}}}}
\newcommand{\aMSpS}{{{}^{\alo\!\!}{\vcenter{\offinterlineskip\hbox{$\scriptstyle\:\,\ast$}\vskip.2ex\hbox{$\mathcal{S}_+$}\vskip.7ex}}}}
\newcommand{\aMSmS}{{{}^{\alo\!\!}{\vcenter{\offinterlineskip\hbox{$\scriptstyle\:\,\ast$}\vskip.2ex\hbox{$\mathcal{S}_-$}\vskip.7ex}}}}
\newcommand{\MCa}{{\mathcal{C}_{\!\!\:(\!\!\:a\!\!\:)}}}
\newcommand{\MCaS}{{{\vcenter{\offinterlineskip\hbox{$\scriptstyle\;\ast$}\vskip.2ex\hbox{$\mathcal{C}_{\!\!\:(\!\!\:a\!\!\:)}$}\vskip.7ex}}}}

\newcommand{\zpdh}{{\sf 2p_{\!\!\:\frac{3}{2}}}}
\newcommand{\zePe}{{\sf 2^{\!\:1\!\!\:}P_{\!1}}}
\newcommand{\eSo}{{\sf {}^{\,1\!}S_{\!\!\;0}}}
\newcommand{\eeSo}{{\sf 1^{\!\!\;1\!}S_{\!\!\;0}}}

\newcommand{\nN}{\vcenter{\offinterlineskip \hbox{$\scriptscriptstyle\;\:\scoh$}\vskip0ex\hbox{$N$}\vskip1.4ex}{}}

\renewcommand{\d}{\displaystyle}

\newcommand{\bbar}{{\mathchoice{\mbox{\it\textblank}}{\mbox{\it\textblank}}{\mbox{\scriptsize\it\textblank}}{\mbox{\tiny\it\textblank}}}}

\documentclass[12pt,letter]{article}
\usepackage{amsfonts,amssymb,amsmath,epsfig}
\usepackage{bbm,cmll,enumerate,graphics,textcomp,titling,tocloft}

\numberwithin{equation}{section}
\makeatletter
\def\subsection{\@startsection{subsection}{2}{\z@}{3.25ex plus 1ex minus .2ex}{1.5ex plus .2ex}{\centering\large\bf\itshape}}
\def\subsubsection{\@startsection{subsubsection}{3}{\z@}{3.25ex plus 1ex minus .2ex}{1.5ex plus .2ex}{\centering\large\em}}
\def\paragraph{\@startsection{paragraph}{4}{\z@}{3.25ex plus 1ex minus .2ex}{1.5ex plus .2ex}{\centering\large\bf}}
\makeatother

\renewcommand{\cfttoctitlefont}{\hfill\Large\bfseries}
\renewcommand{\cftaftertoctitle}{\hfill}

\renewcommand{\cftsecfont}{\large\bf}
\renewcommand{\cftsubsecfont}{\bf\itshape}
\renewcommand{\cftsubsubsecfont}{\itshape}

\begin{document}

\title{\bf Exchange Interactions\\ and\\ Principle of Minimal Energy\\ in\\ Relativistic Schr\"odinger Theory}
\preauthor{\begin{center}}
\author{M.\ Mattes and M.\ Sorg\\[1cm] II.\ Institut f\"ur Theoretische Physik der
Universit\"at Stuttgart\\ Pfaffenwaldring 57 \\ D 70550 Stuttgart, Germany\\Email: sorg@theo2.physik.uni-stuttgart.de\\http://www.theo2.physik.uni-stuttgart.de/institut/sorg/publika.html}
\postauthor{\end{center}}
\date{ }
\maketitle
\begin{abstract}
   The {\em principle of minimal energy}\/, which has been set up in the preceding papers for systems of non-identical particles (e.\,g. positronium), is now generalized to include also identical particles. Since the latter kind of particles feels also the exchange forces (besides the usual electromagnetic forces), one has to deal with non-zero exchange potentials which render the theory nonlinear, according to the non-Abelian character of Relativistic Schr\"odinger Theory (RST). However, the present extension of the variational principle refers only to the linearized version of RST in order to keep the calculations sufficiently simple. It is also demonstrated that in RST the Dirac particles can occur in fermionic and bosonic quantum states; and the mathematical and physical consistency of the variational principle is validated for both types of states (concretely the fermionic hydrogen state $\zpdh$ and the bosonic positronium state $\zePe$).
\vspace{4cm}
\noindent

\end{abstract}

\newpage
\tableofcontents

\newpage
\section{Introduction and Survey of Results}\label{s1}
\indent

The present paper is intended to promote the fluid-dynamic approach to the quantum phenomena. The conventional quantum theory is generally conceived to be a probabilistic framework for the description of microscopic matter (preferrably atoms and molecules). In order that this probabilistic description be most successful, it is also thought that it must be equipped with some counterintuitive elements, e.\,g. ``... {\em the so-called quantum postulate, which attributes to any atomic process an essential discontinuity, or rather individuality, completely foreign to the classical theories and symbolized by Planck's quantum of action. This postulate implies a renunciation as regards the causal space-time co-ordination of atomic processes}\/'' \cite{c1,c2}. It seems that this viewpoint of the early Bohr has been generally accepted in the meantime, because it seems to be safely supported by the acausal features of those observations which violate Bell's inequalities~\cite{c3}. Indeed, Bohr even anticipated the occurence of such acausal effects by saying, ``{\em The very nature of the quantum theory thus forces us to regard the space-time co-ordination and the claim of causality, the union of which characterizes the classical theories, as complementary but exclusive features of the description}\/ ...'' \cite{c1,c2}. Bohr's original idea of the impossibility of describing the quantum phenomena along the lines of a classical causal field theory (such as, e.\,g., classical electrodynamics) is held true up to the present days, as the following quotation may perhaps demonstrate most clearly: ``{\em If the classical concept of the space-time continuum were accepted, then quantum theory could not be considered complete, i.\,e., if it were accepted that the persisting objects of nature literally reside in a space-time continuum, ... then a complete scientific account of atomic phenomena would ... be required to describe whatever it was that is located at the points or infinitesimal regions of that continuum. Quantum theory does not do this, and hence a claim of completeness would be an abuse of language.}\/''~\cite{c4}

It seems to us that such assertions of Bohr, Stapp (and others) can display only half of the truth. The point here is that, for the description of the quantum phenomena, we have to resort to a certain kind of dichotomic thinking which is usually termed ``{\em wave-particle duality}\/'' \cite{c5}. By properly regarding this fact, one should feel being forced to admit that those claims about the  unfeasibility of an ordinary space-time description for the quantum phenomena can refer only to the probabilistic (i.\,e. particle) approach, {\em but by no means to the fluid-dynamic (i.\,e. wave) approach}\/. Indeed, the existence of a very successful fluid-dynamic approach (i.\,e. {\em density functional theory}\/~\cite{c6,c7}) demonstrates that the fluid-dynamic picture of quantum matter is at least as useful in atomic and molecular physics as the probabilistic viewpoint. Clearly, such an epistemological situation should bring forth sufficient motivation in order to further promote the fluid-dynamic approach which sometimes seems to have been overrun by the probabilistic way of thinking.

In this sense, the subsequent discussion aims at the advancement of an alternative fluid-dynamic approach, i.\,e. Relativistic Schr\"odinger Theory (RST) \cite{c8,c9,c10}. In comparison to its competitors (i.\,e. the probabilistic approach and density functional theory), RST has obviously no problems at all with a completely relativistic formulation which, on the other hand, causes certain problems for both the conventional probabilistic approach and density functional theory. Indeed it has already been demonstrated at various occasions that the Whitney sum construction (as opposed to the conventional tensor product construction) is perfectly compatible with the relativity principle (both special and general). This pleasant feature of RST is further worked out subsequently at many places, namely by passing over from any relativistic result to its non-relativistic approximation. Furthermore it will also be very instructive to explicitly verify the numerical coincidence of the RST and conventional predictions for some standard problem (i.\,e. the hydrogen atom) where both the conventional theory and RST admit exact analytic solutions. The meaning of this is to demonstrate that RST owns at least the same degree of mathematical and physical consistency as the conventional quantum theory or density functional theory.

However, the central point of the present investigation refers to the RST {\em principle of minimal energy}\/ \cite{c9,c10,c11}. This is a variational principle on the basis of an appropriately defined energy functional $\EfT$ (both relativistic and non-relativistic) so that for the stationary bound systems the coupled set of mass eigenvalue and gauge field equations turns out as the extremal equations due to that energy functional $\EfT$. In some of the preceding papers, the {\em principle of minimal energy}\/ has been set up for systems of non-identical particles (e.\,g. positronium) where the exchange fields do trivially vanish. But of course one wishes to apply such a variational principle also to systems of {\em identical particles}\/ which in RST are subjected to the action of the exchange forces. Since the latter are described in RST by additional gauge potentials, one would like to see the {\em principle of minimal energy}\/ being validated also for this situation where the additional exchange potentials ($\Bmu$) are non-trivially present besides the usual electromagnetic potentials ($A_\mu$). The main result of the paper is now that the inclusion of the exchange interactions is possible, but here the desired energy functional $\EfT$ must receive a slight modification and thus is no longer immediately equal to the spatial integral of the energy density $\Too$ (as the time component of the energy-momentum density). Because of the considerable complications caused by the non-Abelian (and therefore nonlinear) character of RST, we present the proof only for the linearized version of the theory; but there should be no doubt that the non-Abelian generalization of the constructed energy functional $\EfT$ will exist also for the original non-Abelian theory.

As a concrete demonstration of the usefulness and consistency of the proposed RST energy functional $\EfT$, we explicitly compute with its help the energy of both a fermionic and a bosonic field configuration (i.\,e. the hydrogen state $\zpdh$ and the excited positronium state $\zePe$).

These results are worked out in the following order:

\paragraph{Linearization}

In {\bf sect.\,\ref{s2}}, the basic concepts of RST are briefly sketched in order to have the subsequent discussions sufficiently self-contained: the kinematics of the gauge fields is presented as the first part of the RST action principle whose second part consists then of the field equation for the pure states $\Psi$ (for the mixtures see ref.~\cite{c12}). As in any field theory of the elemental particles, the field equations form the central part also for RST and they consist of the coupled system of matter equations and (non-Abelian) gauge field equations, see equations \rf{231}-\rf{233} below. Since these field equations emerge as the Euler-Lagrange equations due to the RST action principle, the corresponding Noether theorems automatically ensure the validity of the most important conservation laws, such as for current $\MJmu$~\rf{235} and for energy-momentum $T_{\!\mu\nu}$~\rf{245}. The field-theoretic concept of energy-momentum is the starting point for the proper goal of this paper; namely to set up an energy functional ($\EfT$) for the bound few-particle and many-particle systems in just such a way that the stationary form of the original Euler-Lagrange equations (due to the action principle) do appear simultaneously as the variational equations (``{\em extremal equations}\/'') due to the desired energy functional $\EfT$!

Naturally, one thinks that the existence of an energy-momentum density $\Tmunu$ could provide a nearby basis for the definition of ``energy'', i.\,e. one will first resort to the spatial integral of its time component $\Too\vr$, see equation~\rf{251} below. Indeed, this procedure has been shown to be viable for a system of {\em non-identical}\/ particles (such as positronium~\cite{c9,c10,c11}). But for a system of {\em identical particles}\/ the feasibility of an analogous construction must be proven separately, because here the set of RST fields must be complemented by an additional field degree of freedom, i.\,e. the exchange potential $\Bmu$ which is responsible for the exchange interactions occurring exclusively among the identical particles. On the other hand, the inclusion of the exchange interactions equips the theory with a truly non-Abelian character and this is a considerable complication. For this reason, we restrict ourselves in this paper to the {\em linear approximation}\/ of the gauge field equations. Indeed this linearization process keeps the exchange fields present in the theory but on the other hand linearizes the field equations for the gauge potentials, see the d'Alembert equations \rf{259a}-\rf{259b} below, so that the residual degree of complexity remains manageable. But clearly, there can be no doubt that an energy functional $\EfT$ of the desired kind would exist also for the original non-Abelian theory. The present treatment of the linearized theory, however, will show that for the inclusion of the exchange interactions the original idea of ``energy'' as the spatial integral~\rf{251} of the time component $\Too\vr$ must be slightly modified. The nature of this modification is just one of the main results of the present paper. 

\paragraph{Electromagnetic and Exchange Fields}

Surely, the linearization of a non-Abelian gauge theory represents a considerable truncation of the original mathematical framework, and therefore it becomes necessary to reassure that the residual linear formalism is still logically consistent. Therefore in {\bf sect.\,\ref{s3}} the linearized gauge field equations are analysed in detail. Since the goal of the paper aims at the stationary bound systems, this analysis is carried out in three-vector notation which presumes a certain space-time splitting so that the wave functions appear as the usual products of an exponential time factor and a time-independent wave function $\psi_a\vr$, see equation~\rf{39} below. Accordingly, the electromagnetic fields $A_\mu$ become completely time-independent, whereas the exchange fields $B_\mu$ still appear as the product of an exponential time factor and a time-independent spatial function, see equation~\rf{36a} below. This difference, however, does not disturb the time-independent curl and source equations which couple the three-vector fields to the charge and current densities as in ordinary Maxwellian electromagnetism, see equations \rf{313a}-\rf{314b} below. It is merely the appearance of additional charge and current densities, i.\,e. the exchange densities $\ho,\,\vh$, which complete the ordinary densities of the Maxwellian theory. But what is important here, this refers to the {\em Poisson and exchange identities}\/ which guarantee the numerical equality between the interaction energy $M^{\rm(\!\!\:e,m;h,g\!\!\:)\!}c^2$ of the gauge field modes with the matter modes on the one hand and of the gauge field energies $E_{\rm R}^{\rm(\!\!\:e,m\!\!\:)\!},\,E_{\rm C}^{\rm(\!\!\:h,g\!\!\:)\!}$ on the other hand; see equations \rf{333}-\rf{336} below. This formal equivalence of the electromagnetic and exchange interaction energies once more points at the fact that in RST the exchange phenomena among identical particles are treated on the basis of real forces, quite analogously as their electromagnetic counterparts.

\paragraph{Exchange Polarization}

The latter feature of RST is also responsible for a certain complication of the matter subsystem, in comparison to conventional quantum theory. The point here is that in RST all the ordinary densities (of charge, current, polarization, etc.) are accompanied by their exchange counterparts which generate the corresponding field strengths according to the gauge field equations, see below the equations \rf{33}-\rf{34} for the ordinary electromagnetic case and \rf{310a}-\rf{310b} for the exchange case. This forces us to take into account also the exchange polarization effects which then entails a somewhat more intricate form of the mass eigenvalue equations. This complication becomes especially evident for the non-relativistiv form of the eigenvalue equations, cf. the simple Hamiltonian form \rf{373}-\rf{374} for the {\em different}\/ (i.\,e. positively charged) particle vs. the extended form \rf{375a}-\rf{377b} of both {\em identical}\/ (negatively charged) particles.

\paragraph{Mass and Energy Functionals}

Concerning now the search for an appropriate energy functional $\EfT$ for the stationary bound systems, one can deduce a kind of precursor directly from the (relativistic or non-relativistic) eigenvalue equations: this is the (relativistic) mass functional $\MfT c^2$~\rf{343}, or the (non-relativistic) Pauli energy functional $\Efp$ \rf{3128a}-\rf{3128b} as the non-relativistic limit of $\MfT c^2$. The point with these mass (or energy, resp.) functionals is that they do generate the corresponding eigenvalue equations (both relativistic and non-relativistic) via their variational (i.\,e. ``extremal'') equations. Indeed, the non-relativistic version may be viewed as the RST counterpart of the well-known Ritz variational principle \cite{c11,c13,c14} if this is combined with the Hartree and Hartree-Fock approximations (see the discussion of this in ref.~\cite{c11}). However, with respect to our original intention these relativistic mass and non-relativistic energy functionals $\MfT c^2$ and $\Efp$ suffer from a serious deficiency, namely it is not possible to deduce from them the gauge field equations! Therefore one has to undertake a fresh attempt from a somewhat different side.

\paragraph{Exchange Interactions}

The new point of departure in {\bf sect.\,\ref{s4}} refers to the total energy $\ET$~\rf{251} as the integral of the total energy density $\TToo\vr$ over all three-space. As it originally stands, this quanity $\ET$ cannot be adopted as the desired energy functional because the stationary field equations do not appear as the corresponding variational equations. However, when the Poisson and exchange identities are added as constraints, one arrives at a modified form of $\ET$, i.\,e. $\EiiifT$~\rf{412}, which {\em almost}\/ meets with the expectations. Namely, the numerical value of that preliminary proposal $\EiiifT$ upon the solutions of the RST field equations is the same as for the original $\EfT$~\rf{251} because both functionals $\EiiifT$ and $\EfT$ differ only by the Poisson and exchange identities and these are automatically satisfied by any solution of the field equations.

But despite this pleasant feature of numerical equality of the latter proposal $\EiiifT$~\rf{412} and the original $\EfT$~\rf{251}, the functional $\EiiifT$ cannot be accepted as the true solution of the problem {\em if some of the particles are identical and thus undergo the exchange interactions}\/. The reason for this rejection is two-fold:
   \begin{enumerate}[\bf (i)]
   \item The proposal $\EiiifT$~\rf{412} contains not only the gauge {\em field strengths}\/ but also the exchange {\em potential}\/ $\Bmu$. The usual expectation in field theory is that the energy concentrated in some field configuration should depend exclusively upon the field strengths (curvature components) and not upon the gauge potentials (connection components).
   \item The RST field equations do appear as the variational equations due to the proposed functional $\EiiifT$~\rf{412} only if no identical particles are present and thus the exchange potential $\Bmu$ is zero. Consequently, in the presence of identical particles ($\Bmu\not\equiv0$) the tentative proposal $\EiiifT$ is not extremalized by the solutions of the RST eigenvalue problem!
   \end{enumerate}

In view of these problems with the right definition of the RST energy $\EfT$ we obviously have to abandon the energy concept~\rf{251} which is essentially based upon the existence of an energy-momentum tensor $\Tmunu$. But fortunately, this is to be done in a minimal way: it is merely necessary to drop that questionable potential term from the proposal~\rf{412} so that the final result is represented by the energy functional $\EivfT$~\rf{413}. Indeed, this functional in both its relativistic form $\tEfT$~\rf{414} and non-relativistic version $\tEEfT$~\rf{421} has the corresponding RST field equations (matter plus gauge field) as its extremal equations and can therefore be viewed as the proper physical energy to be associated with any solution of the RST eigenvalue problem. But unfortunately, this reasonable definition of energy does then no longer coincide with the original RST energy $\ET$~\rf{251} which rests on the basis of the energy-momentum tensor $\Tmunu$. Notice that the numerical equivalence of our final result $\EivfT$ and of the starting point $\ET$~\rf{251} is secured for all those situations where the exchange interactions are inactive ($\leadsto$ non-identical particles).

\paragraph{Application: Hydrogen State $\zpdh$}

As a concrete application of the constructed energy functional, we consider in {\bf sect.\,\ref{s5}} a one-particle system (the hydrogen state $\zpdh$) and a two-particle system (the positronium state $\zePe$). Here it should not come as a surprise that for the one-particle states the RST predictions for the binding energy {\em exactly}\/ coincide with their conventional counterparts. In fact, it has already been demonstrated for the one-particle systems by means of rather general arguments \cite{c15,c16} that the conventional one-particle energy ($\Econv$) always agrees with the corresponding total energy ($\tET$) of RST. Furthermore, the latter always equals the mass eigenvalue ($M_2c^2$) so that for the one-particle systems one always has $\Econv=\tET=M_2c^2$, see equation~\rf{553} below. But despite this somewhat simple result it is nevertheless very instructive to see in detail how the relatively complicated RST energy functional $\tEfT$ manages to exactly coincide with the mass eigenvalue $M_2c^2$ on the one hand and with the conventional prediction $\Econv$ on the other hand. But beyond this numerical coincidence, the RST functional $\tEfT$ provides a much more detailed picture of the intra-atomic situation.

\paragraph{Positronium State $\zePe$}

However, a more rigorous scrutiny of the RST energy functional $\EfT$ concerns the few-particle (or even many-particle) systems, because their treatment is on principle different in RST and the conventional theory (i.\,e. Whitney sum of one-particle bundles vs. tensor product of one-particle Hilbert spaces). As a test case we choose here positronium although its constituents (i.\,e. electron and positron) are non-identical particles and therefore the exchange potential $\Bmu$ is zero. But nevertheless this is an absolutely non-trivial situation for testing the RST energy functional, since the bosonic character of positronium is in RST transferred to its constituents so that both the positron and the electron do occur in {\em bosonic states}\/ (though being described by Dirac four-spinors). Restricting ourselves here again to te non-relativistic limit, we can demonstrate that the RST prediction for the excited state $\zePe$ approaches the conventional prediction (and the experimental value) up to a magnitude of (roughly) 10\%, although we resorted to the spherically symmetric approximation and to the simplest possible trial function with only two variational parameters, see equation~\rf{5125} below and the diagram on p.~\pageref{f1}. Since this magnitude of deviation is the same as for the analogous calculation of the groundstate ($\eeSo$) energy in the preceding paper \cite{c10}, we conclude that the use of the spherically symmetric approximation in principle generates such a 10-percent deviation which, however, would vanish through the use of better approximation techniques.

Therefore it seems to us that the essential logical elements of RST (i.\,e. Whitney sum, energy functional and bosonic quantum states for Dirac particles) must have something to do with physical reality. In this sense it would also be very interesting to develop more accurate (relativistic and non-relativistic) approximation techniques in order to check how close the RST predictions do really approach their conventional counterparts (and the corresponding experimental values). The outcome of such a competition between a fluid-dynamic and a probabilistic view of the quantum phenomena would surely shed new light on the microscopic world.

\section{Action Principle}\label{s2}
\indent

Once the basic field equations for RST had been set up, it was oberved that they may also be deduced from an action principle \cite{c12}. Indeed, the existence of an action principle is always a very nice feature of a field theory because the relationship between the conservation laws and both the external and internal symmetries of the theory become elucidated in an elegant way (Noether theorem). But for the present case of RST, the consideration of the action principle is especially instructive also from another point of view, namely concerning its relationship with the \/{\em principle of minimal energy}\/. The latter principle is not only a pleasant feature of the general logical structure of the theory, but rather it is also of considerable practical relevance when one wants to develop variational techniques for the approximate calculation of the energy levels of bound systems~\cite{c10}.

\subsection{Relativistic Field Kinematics}

The crucial element of any action principle, i.\,e. more concretely for RST
   \begin{eqnarray}
   \label{21}
   \delta \WRST&=&0\;,
   \end{eqnarray}
is the Lagrangean density ($\MLRST$) which builds up the action integral $W\RRST$ in the following way
   \begin{eqnarray}
   \label{22}
   \WRST&=&\int\!\!{\rm d}^4x\;\MLRST[\Psi,\MAmu]\;.
   \end{eqnarray}
Since the RST field system is a coupled set of matter ($\Psi$) and gauge fields ($\MAmu$), the corresponding Lagrangean appears naturally as the sum of a matter part ($\MLD$) and a gauge field part ($\MLG$):
   \begin{eqnarray}
   \label{23}
   \MLRST[\Psi,\MAmu]&=&\MLD[\Psi]+\MLG[\MAmu]\;.
   \end{eqnarray}
This splitting then induces a corresponding partitioning of the action integral $\WRST$~\rf{22} into two contributions, i.\,e.
   \begin{eqnarray}
   \label{24}
   \WRST&=&\WD+\WG\;,
   \end{eqnarray}
with the self-evident arrangement
   \begin{subequations}
   \begin{align}
   \label{25a}
   \WD[\Psi]&=\int\!\!{\rm d}^4x\;\MLD[\Psi]\\
   \label{25b}
   \WG[\MAmu]&=\int\!\!{\rm d}^4x\;\MLG[\MAmu]\;.
   \end{align}
   \end{subequations}
\indent Here, the matter Lagrangean $\MLD$ looks as follows:
   \begin{eqnarray}
   \label{26}
   \MLD[\Psi]&=&\frac{i\hbar c}{2}\,\Big[\bar{\Psi}\GGamma^\mu(\MDmu\Psi)-(\MDmu\bar{\Psi})\GGamma^\mu\Psi\Big]-\bar{\Psi}\MM c^2\Psi\;,
   \end{eqnarray}   
where the $N$-particle wave function $\Psi$ is the direct (i.\,e. Whitney) sum of the single-particle wave functions $\psi_a$ ($a=1{\ppp}N$)
   \begin{eqnarray}
   \label{27}
   \Psi&=&\psi_1\oplus\psi_2\oplus...\oplus\psi_N\;.
   \end{eqnarray}
Thus, from the mathematical point of view, the wave function $\Psi(x)$ is a section of a complex vector bundle over space-time with typical fibre $\mathbbm{C}^{4N}$ where the individual wave functions $\psi_a$ are chosen as Dirac's four-spinors. Furthermore, the \/{\em total velocity operator}\/ $\GGamma^\mu$ is the direct sum of the ordinary Dirac matrices $\gamma^\mu$,
   \begin{eqnarray}
   \label{28}
   \GGamma^\mu&=&(-\gamma^\mu)\oplus\gamma^\mu\oplus\gamma^\mu\oplus...\;,
   \end{eqnarray}
where the explicitly displayed combination of Dirac matrices would then refer to a three-particle system ($N=3$), with the first particle ($a=1$) carrying a positive charge unit ($\leadsto$ positron, proton, etc.) and the second ($a=2$) and third particle ($a=3$) being negatively charged ($\leadsto$ electrons), see ref.\,\cite{c17}.\\
The gauge covariant derivative of the wave function $\Psi$ is as usual
   \begin{eqnarray}
   \label{29}
   \MDmu\Psi&=&\pmu\Psi+\MAmu\Psi\;,
   \end{eqnarray}
with the gauge potential $\MAmu$ (\/{\em bundle connection}\/) taking its value in the Lie algebra $\mathfrak{u}(N)$ of the {\em structure group} $U(N)$. This structure group becomes reduced to its subgroup \mbox{$U(N,N')\subset U(N)$}, which is of dimension \mbox{${N'}^2+(N-N')=N+N'(N'-1)$} if $N'$ particles are identical and the other \mbox{$(N-N')$} particles are different. For instance, for the three-particle system of {\em two identical} particles (\mbox{$N'=2$}, electrons) and {\em one different} particle (\mbox{$N-N'=1$}, positron) mentioned below equation~\rf{28} the original nine-dimensional structure group $U(3)$ for three identical particles \cite{c17} becomes reduced to the five-dimensional product group $U(1)\times U(2)$, where the Abelian factor $U(1)$ refers to the positively charged particle; and the remaining four-dimensional factor group $U(2)$ refers to the two (identical) electrons. The point with this bundle reduction is that the {\em identical} particles do not only undergo the Abelian electromagnetic interactions (described by $U(1)$\/) but are additionally subjected to the {\em non-Abelian exchange interactions} (described by the non-Abelian $U(2)$, see ref.\,\cite{c17} for a more detailed discussion). Accordingly, the $N+N'(N'-1)$ generators $\tau_\alpha$ ($\alpha=1{\ppp}N+N'(N'-1)$) of the reduced Lie algebra $\mathfrak{u}(N,N')$ are subdivided into the set of {\em commuting electromagnetic generators} $\tau_a$ ($a=1{\ppp}N$) and the remaining $N'(N'-1)$ {\em exchange generators} $\chi_k$ ($k=1{\ppp}N'(N'-1)$) so that the decomposition of the bundle comnection $\MAmu$ with respect to such a special Lie algebra basis appears in the following form:
   \begin{eqnarray}
   \label{210}
   \MAmu&=&\sum_{a=1}^N\Aaomu\tau_a+\sum_{k=1}^{N'(N'-1)}\Bkomu\chi_k\;.
   \end{eqnarray}
\indent In particular, for the three-particle system~\rf{28} considered here one has
   \begin{eqnarray}
   \label{211}
   \MAmu&=&\sum_{a=1}^3\Aaomu\tau_a+\Bmu\chi-\BSmu\bar{\chi}\;,
   \end{eqnarray}
where the two exchange generators $\chi_k$ ($k=1,2$) are taken as anti-Hermitean conjugates. Of course, the curvature $\MFmunu$ of $\MAmu$ ($\leadsto$ {\em field strength}\/)
   \begin{eqnarray}
   \label{212}
   \MFmunu&=&\nabla_{\!\!\mu\,}\MAnu-\nabla_{\!\!\nu\,}\MAmu+[\MAmu,\MAnu]
   \end{eqnarray}
does then admit a similar decomposition
   \begin{eqnarray}
   \label{213}
   \MFmunu&=&\sum_{a=1}^3\Faomunu\tau_a+\Gmunu\chi-\GSmunu\bar{\chi}
   \end{eqnarray}
where the $\Faomunu$ are the (real-valued) electromagnetic field strengths and the (complex-valued) $\Gmunu$ plays the role of the exchange field strength. The proper (continuous) gauge group for such an arrangement is now taken as the $N$-fold Abelian product $U(1)\times U(1)\times...\times U(1)$, which is the maximal Abelian subgroup of the original structure group $U(N)$. The permutation of identical particles is an exact symmetry of the theory, with the corresponding discrete transformations to be considered as elements of the reduced structure group $U(N,N')$ \cite{c17}.\\
\indent The (Whitney) sum structure of the $N$-particle wave function $\Psi$~\rf{27} lets now appear the matter Lagrangean $\MLD$~\rf{26} as an (ordinary) sum of $N$ single-particle contributions, i.\,e. for $N=3$\/:
   \begin{eqnarray}
   \label{214}
   \MLD[\Psi]&=&\MLDe+\MLDz+\MLDd\;,
   \end{eqnarray}
with the individual terms being given by
   \begin{subequations}
   \begin{align}
   \label{215a}
   \MLDe&=-\frac{i\hbar c}{2}\,\Big[\bar{\psi}_1\gamma^\mu(\Dmu\psi_1)
   -(\Dmu\bar{\psi}_1)\gamma^\mu\psi_1\Big]
   -M_pc^2\,\bar{\psi}_1\psi_1\\
   \label{215b}
   \MLDz&={}\hspace{.8em}\frac{i\hbar c}{2}\,\Big[\bar{\psi}_2\gamma^\mu(\Dmu\psi_2)
   -(\Dmu\bar{\psi}_2)\gamma^\mu\psi_2\Big]
   -M_ec^2\,\bar{\psi}_2\psi_2\\
   \label{215c}
   \MLDd&={}\hspace{.8em}\frac{i\hbar c}{2}\,\Big[\bar{\psi}_3\gamma^\mu(\Dmu\psi_3)
   -(\Dmu\bar{\psi}_3)\gamma^\mu\psi_3\Big]
   -M_ec^2\,\bar{\psi}_3\psi_3\;.
   \end{align}
   \end{subequations}
Here, the gauge-covariant derivatives $(\Dmu\psi_a)$ of the single-particle wave functions $\psi_a$ ($a=1,2,3$) arise from the three-particle derivative $\MDmu\Psi$~\rf{29} via the sum structure of $\Psi$~\rf{27} as
   \begin{eqnarray}
   \label{216}
   \MDmu\Psi&=&(\Dmu\psi_1)\oplus(\Dmu\psi_2)\oplus(\Dmu\psi_3)\;,
   \end{eqnarray}
with the individual derivatives looking as follows
   \begin{subequations}
   \begin{align}
   \label{217a}
   \Dmu\psi_1&=\pmu\psi_1-i\,[\Azomu+\Adomu]\,\psi_1\\
   \label{217b}
   \Dmu\psi_2&=\pmu\psi_2-i\,[\Aeomu+\Adomu]\,\psi_2-i\Bmu\psi_3\\
   \label{217c}
   \Dmu\psi_3&=\pmu\psi_3-i\,[\Aeomu+\Azomu]\,\psi_3-i\BSmu\psi_2\;.
   \end{align}
   \end{subequations}
Consequently, the matter Lagrangean $\MLD[\Psi]$~\rf{214} is composed of four parts
   \begin{eqnarray}
   \label{218}
   \MLD[\Psi]&=&\MLM[\Psi]+\MLkin[\Psi]+\MLDem[\Psi]+\MLDhg[\Psi]\;,
   \end{eqnarray}
and any single one of these parts describes an important feature of the {\em matter subsystem}\/:
   \begin{enumerate}[\bf (i)]
   \item The rest-mass term $\MLM$ equips the RST field configurations with a rest-mass density which implies a rest-mass renormalization (see below)
   \begin{eqnarray}
   \label{219}
   \MLM[\Psi]&=&-M_pc^2\,\bar{\psi}_1\psi_1-M_ec^2\Big[\bar{\psi}_2\psi_2+\bar{\psi}_3\psi_3\Big]\;.
   \end{eqnarray}
   \item The kinetic term $\MLkin$ takes into account the four-dimensional kinetic energy of the particles and differs in sign for differently charged particles
   \begin{eqnarray}
   \label{220}
   \MLkin[\Psi]&=&-\frac{i\hbar c}{2}\,\Big[\bar{\psi}_1\gamma^\mu(\pmu\psi_1)
   -(\pmu\bar{\psi}_1)\gamma^\mu\psi_1\Big]\nonumber\\
   &&\hspace{-5.5em}{}+\frac{i\hbar c}{2}\,\Big[\bar{\psi}_2\gamma^\mu(\pmu\psi_2)
   -(\pmu\bar{\psi}_2)\gamma^\mu\psi_2\Big]
   +\frac{i\hbar c}{2}\,\Big[\bar{\psi}_3\gamma^\mu(\pmu\psi_3)
   -(\pmu\bar{\psi}_3)\gamma^\mu\psi_3\Big]\;.\nonumber\\
   &&
   \end{eqnarray}
   \item The third term $\MLDem$ in~\rf{218} describes the electromagnetic interactions of the particles and is given by
   \begin{eqnarray}
   \label{221}
   \MLDem[\Psi]&=&\hbar c\,\Big[\Aeomu(\kzumo+\kdumo)+\Azomu(-\keumo+\kdumo)+\Adomu(-\keumo+\kzumo)\Big]\nonumber\\
   &&
   \end{eqnarray}
   where the {\em Dirac currents} $\kaumu$ ($a=1,2,3$) are defined as usual, i.\,e.
   \begin{eqnarray}
   \label{222}
   \kaumu&=&\bar{\psi}_a\gamma_\mu\psi_a\;.
   \end{eqnarray}
   \item The last contribution $\MLDhg$ to the matter Lagrangean $\MLD$~\rf{218} is a peculiarity of RST because it describes the exchange interactions among the {\em identical} particles (i.\,e. electrons; $a=2,3$) by means of the non-Abelian part of the gauge field theory:
   \begin{eqnarray}
   \label{223}
   \MLDhg[\Psi]&=&\hbar c\,\Big[\Bmo\hmu+\BSmo\hSmu\Big]\;.
   \end{eqnarray}
   Observe here that the exchange generators $\chi,\,\bar{\chi}$~\rf{211} do not commute, in contrast to the electromagnetic generators $\tau_a$ ($a=1,2,3$), and thus they give rise to the non-Abelian character of the theory which is encoded in the {\em exchange current} $\hmu$ of the identical particles ($a=2,3$):
   \begin{eqnarray}
   \label{224}
   \hmu&\doteqdot&\bar{\psi}_2\gamma_\mu\psi_3\;.
   \end{eqnarray}
   \end{enumerate}
   
\indent Once the general structure of the matter Lagrangean $\MLD$ is clarified, one may inspect the gauge field Lagrangean $\MLG$ in a similar way:
   \begin{eqnarray}
   \label{225}
   \MLG[\MAmu]&=&\frac{\hbar c}{16\pi\als}\,\Kalubeu\Falomono\Fbeomunu\\\nonumber
   &&{}\hspace{-1.8em}(\alpha,\beta=1\ppp5)\;.
   \end{eqnarray}
Here, $\Kalubeu$ is a covariantly constant fibre metric of the Lie algebra bundle and invariant under the reduced structure group $U(3,2)$. Therefore, if the self-interactions of the particles are neglected (see ref.s\,\cite{c8,c12}), this symmetric fibre metric ($\Kalubeu=K_{\beta\alpha}$) adopts a very simple form, namely
   \begin{eqnarray}
   \label{226}
   K_{12}\;=\;K_{13}\;=&K_{23}&=\;K_{45}\;=\;-1
   \end{eqnarray}
with the other (independent) elements being zero. Accordingly, the gauge field part $\MLG[\MAmu]$ of the RST Lagrangean $\MLRST$~\rf{23} appears as the sum of the electromagnetic and exchange contributions
   \begin{eqnarray}
   \label{227}
   \MLG[\MAmu]&=&\MLR+\MLC\;,
   \end{eqnarray}   
where the electromagnetic part ($\MLR$) refers to the (real-valued) gauge field modes $\Faomunu$~\rf{213}, i.\,e.
   \begin{eqnarray}
   \label{228}
   \MLR&=&-\frac{\hbar c}{8\pi\als}\,\Big[F^1_{\;\mu\nu}F^{2\mu\nu}+F^2_{\;\mu\nu}F^{3\mu\nu}+F^3_{\;\mu\nu}F^{1\mu\nu}\Big]\;,
   \end{eqnarray}
and similarly the exchange contribution ($\MLC$) is related to the (complex-valued) exchange modes $\Gmunu$
   \begin{eqnarray}
   \label{229}
   \MLC&=&\frac{\hbar c}{8\pi\als}\,\GSmono\Gmunu\;.
   \end{eqnarray}
\indent For the subsequent variational procedure it is important to note that the electromagnetic field strengths $\Faomunu$ and the exchange field strengths $\Gmunu$ ($\leadsto$ {\em curvature components}\/) are linked to the {\em connection components} $\Aaomu,\,\Bmu$ through the following relations, to be deduced from the abstract equation~\rf{212}:
   \begin{subequations}
   \begin{align}
   \label{230a}
   F^1_{\;\mu\nu}&=\nablamu\Aeonu-\nablanu\Aeomu\\
   \label{230b}
   F^2_{\;\mu\nu}&=\nablamu\Azonu-\nablanu\Azomu+i\left[\Bmu\BSnu-\Bnu\BSmu\right]\\
   \label{230c}
   F^3_{\;\mu\nu}&=\nablamu\Adonu-\nablanu\Adomu-i\left[\Bmu\BSnu-\Bnu\BSmu\right]\\
   \label{230d}
   \Gmunu&=\nablamu\Bnu-\nablanu\Bmu+i\left[\Azomu-\Adomu\right]\Bnu-i\left[\Azonu-\Adonu\right]\Bmu\;.
   \end{align}
   \end{subequations}
Naturally, the first field strength $\Feomunu$~\rf{230a} does not couple to the exchange potential $\Bmu$ since the first electromagnetic mode $(\Aeomu,\Feomunu)$ is due to the first (differently charged) particle. On the other hand, the second and third modes $\Fzomunu,\,\Fdomunu$ are linked also to the exchange potential $\Bmu$ since the second ($a=2$) and third ($a=3$) particle do interact also via the exchange force. The exchange mode $(\Bmu,\Gmunu)$~\rf{230d} does of course exclusively couple to both {\em identical}\/ electrons ($a=2,3$), but not to the positively charged particle ($a=1$)!

\subsection{Relativistic Field Equations}

After the Lagrangean is fixed now in full detail, the variational procedure~\rf{21} can be carried out in order to yield the complete set of coupled field equations. More concretely, the extremalization of the action integral $\WRST$~\rf{22} with respect to the wave function $\Psi$ produces the $N$-particle Dirac equation \cite{c17}
   \begin{eqnarray}
   \label{231}
   i\hbar\,\GGamma^\mu\MDmu\Psi&=&\MM c\,\Psi\;,
   \end{eqnarray}
or, resp., in component form for the presently considered three-particle system ($N=3$)
   \begin{subequations}
   \begin{align}
   \label{232a}
   i\hbar\,\gamma^\mu\Dmu\psi_1&=-M_pc\,\psi_1\\
   \label{232b}
   i\hbar\,\gamma^\mu\Dmu\psi_2&=M_ec\,\psi_2\\
   \label{232c}
   i\hbar\,\gamma^\mu\Dmu\psi_3&=M_ec\,\psi_3\;.
   \end{align}
   \end{subequations}
Furthermore, the extremalization of the action integral with respect to the bundle connection $\MAmu$ yields the non-Abelian Maxwell equations for the bundle curvature $\MFmunu$
   \begin{eqnarray}
   \label{233}
   \MDmo\MFmunu\hspace{-.5em}&=&\hspace{-.4em}-4\pi i\als\,\MJnu\\\nonumber
   &&\hspace{-2.8em}(\als=\frac{e^2}{\hbar c})\;,
   \end{eqnarray}
i.\,e. in component form for the three-particle systems
   \begin{subequations}
   \begin{align}
   \label{234a}
   \nabla^\mu\Feomunu&=4\pi\als\,\jeonu\\
   \label{234b}
   \nabla^\mu\Fzomunu+i\left[\Bmo\GSmunu-\BSmo\Gmunu\right]&=4\pi\als\,\jzonu\\
   \label{234c}
   \nabla^\mu\Fdomunu-i\left[\Bmo\GSmunu-\BSmo\Gmunu\right]&=4\pi\als\,\jdonu\\
   \label{234d}
   \nabla^\mu\Gmunu+i\left[\Azomu-\Adomu\right]\Gmonu-i\left[\Fzomunu-\Fdomunu\right]\Bmo&=4\pi\als\,\gnu\;.
   \end{align}
   \end{subequations}
   
Here, the emergence of the {\em Maxwell currents} $\jalomu$ ($\alpha=1\ppp5$) needs an explanation because, up to now, the theory contains only the electromagnetic Dirac currents $\kaumu$~\rf{222} and the exchange current $\hmu$~\rf{224}. Namely, the point with the Maxwell four-currents $\jalomu$ is that they should obey a certain conservation law, i.\,e. in operator form
   \begin{eqnarray}
   \label{235}
   \MDmo\MJmu\hspace{-.5em}&=&\hspace{-.4em}0\;,
   \end{eqnarray}
in order that the Maxwell equations~\rf{233} be consistent. Decomposing here the {\em current operator} $\MJmu$ with respect to the chosen basis $\{\tau_a,\chi,\bar{\chi}\}$ lets appear just the Maxwell currents $\jalomu$, i.\,e.
   \begin{eqnarray}
   \label{236}
   \MJmu&\doteqdot&i\jalomu\tau_\alpha\;=\;i\left[\sum_{a=1}^3\jaomu\tau_a+\gmu\chi-\gSmu\bar{\chi}\right]\,,
   \end{eqnarray}
so that the conservation law~\rf{235} reads in component form
   \begin{eqnarray}
   \label{237}
   \Dmu\jalomo&\equiv&0\;.
   \end{eqnarray}
Thus, one becomes forced to define the Maxwell currents $\jalomu$ in terms of the matter fields $\psi_a$ in such a way that these source equations~\rf{237} do emerge as an implication of the matter dynamics~\rf{231} itself! But this {\em compatibility requirement}\/ for the matter and gauge field dynamics can be met just by means of the fibre metric $\Kalubeu$~\rf{225} whose covariant constancy
   \begin{eqnarray}
   \label{238}
   \Dmu\Kalubeu&\equiv&0
   \end{eqnarray}
helps solving the problem: First, define the {\em RST currents} $\jalumu$ through
   \begin{eqnarray}
   \label{239}
   \jalumu&\doteqdot&\bar{\Psi}\tau_\alpha\GGamma_{\!\!\mu}\Psi\\\nonumber
   &&{}\hspace{-3em}(\,[\tau_\alpha,\GGamma_{\!\!\mu}]\;=\;0\,)\;,
   \end{eqnarray}
which by a suitable choice of the generators $\tau_\alpha$ turn out as nothing else than a combination of Dirac and exchange currents $\kaumu,\,\hmu$ \cite{c8}
   \begin{subequations}
   \begin{align}
   \label{240a}
   \jeumu&={}\hspace{.8em}\kzumu+\kdumu\\
   \label{240b}
   \jzumu&=-\keumu+\kdumu\\
   \label{240c}
   \jdumu&=-\keumu+\kzumu\\
   \label{240d}
   \jvumu&={}\hspace{.8em}\hmu\\
   \label{240e}
   \jfumu&=-\hSmu\;.
   \end{align}
   \end{subequations}
\indent Next, observe that these RST currents $\jalumu$~\rf{239} do obey the following source equation:
   \begin{eqnarray}
   \label{241}
   \Dmo\jalumu&\equiv&0\;,
   \end{eqnarray}
namely as an immediate consequence of the matter dynamics~\rf{231}. And finally, define the Maxwell currents $\jalomu$~\rf{236} in terms of the RST currents $\jalumu$~\rf{239} just with the help of the fibre metric $\Kalubeu$, namely through
   \begin{eqnarray}
   \label{242}
   \jalomu&\doteqdot&\Kalobeo\jbeumu\;,
   \end{eqnarray}
so that the covariant constancy~\rf{238} of the fibre metric ensures the compatibility of the desired source equations~\rf{237} for the Maxwell currents $\jalomu$ with those for the RST currents $\jalumu$~\rf{241}. In this sense, the fibre metric $\Kalubeu$ works also as a ``{\em compatibility tensor}\/''; and the Maxwell currents can appear now in terms of the Dirac and exchange currents as
   \begin{subequations}
   \begin{align}
   \label{243a}
   \jeomu\;=\;\keumu&\doteqdot\bar{\psi}_1\gamma_\mu\psi_1\\
   \label{243b}
   \jzomu\;=\;-\kzumu&\doteqdot-\bar{\psi}_2\gamma_\mu\psi_2\\
   \label{243c}
   \jdomu\;=\;-\kdumu&\doteqdot-\bar{\psi}_3\gamma_\mu\psi_3\\
   \label{243d}
   \jvomu\;\equiv\;\gmu\;=\;\hSmu\;&\doteqdot\bar{\psi}_3\gamma_\mu\psi_2\\
   \label{243e}
   \jfomu\;\equiv\;-\gSmu\;=\;-\hmu\;&\doteqdot-\bar{\psi}_2\gamma_\mu\psi_3\;.
   \end{align}
   \end{subequations}
\indent But once a logical path from the Dirac and exchange currents to the Maxwell currents has been established, the source equations~\rf{237} for the latter can be transcribed to the coresponding source equations for the first kind of currents:
   \begin{subequations}
   \begin{align}
   \label{244a}
   \nabla^\mu\keumu&\equiv0\\
   \label{244b}
   \nabla^\mu\kzumu-i\left[\Bmo\hmu-\BSmo\hSmu\right]&\equiv0\\
   \label{244c}
   \nabla^\mu\kdumu+i\left[\Bmo\hmu-\BSmo\hSmu\right]&\equiv0\\
   \label{244d}
   \nabla^\mu\hmu-i\left[\Azomu-\Adomu\right]\hmo+i\BSmo\left[\kzumu-\kdumu\right]&\equiv0\;.
   \end{align}
   \end{subequations}
Observe here again that it is eclusively the electromagnetic current $\keumu$~\rf{244a} of the first (positively charged) particle which does obey a rigorous conservation law, whereas the other two (identical) particles \rf{244b}-\rf{244c} must satisfy a more complicated source equation because of their exchange interactions.\\
\indent Besides charge conservation there is a further important conservation law which appears here as the source equation for energy-momentum:
   \begin{eqnarray}
   \label{245}
   \nabla^\mu\TTmunu&\equiv&0\;.
   \end{eqnarray}
Since the RST field configurations represent a coupled system of matter and gauge fields, it is natural to assume that the {\em total energy-momentum density}\/ $\TTmunu$ will be built up by a matter part $\DTmunu$ and a gauge field part $\GTmunu$
   \begin{eqnarray}
   \label{246}
   \TTmunu&=&\DTmunu+\GTmunu\;.
   \end{eqnarray}
Indeed, the matter part has been identified as \cite{c17}
   \begin{eqnarray}
   \label{248}
   \hspace{-3em}\DTmunu&=&\frac{i\hbar c}{4}\,\Big[\bar{\Psi}\GGamma_{\!\!\mu}(\MDnu\Psi)-(\MDnu\bar{\Psi})\GGamma_{\!\!\mu}\Psi+\bar{\Psi}\GGamma_{\!\!\nu}(\MDmu\Psi)-(\MDmu\bar{\Psi})\GGamma_{\!\!\nu}\Psi\Big]
   \end{eqnarray}
and the gauge field part by
   \begin{eqnarray}
   \label{249}
   \GTmunu&=&\frac{\hbar c}{4\pi\als}\,\Kalubeu\Big(F^\alpha_{\;\,\mu\lambda}F^{\beta\;\,\lambda}_{\;\,\nu}-\frac{1}{4}\,g_{\mu\nu}F^{\alpha}_{\;\,\sigma\lambda}F^{\beta\sigma\lambda}\Big)\;.
   \end{eqnarray}
The (local) conservation law~\rf{245} comes now about through the mutual annihilation of the sources of both energy-momentum densities, i.\,e.
   \begin{eqnarray}
   \label{250}
   \nabla^\mu\DTmunu\;=\;-\nabla^\mu\GTmunu&=&\hbar c\,\Falomunu j_\alpha^{\;\mu}\;.
   \end{eqnarray}
Obviously the sources of the partial densities $\DTmunu$ and $\GTmunu$ are just the well-known {\em Lorentz forces}\/ in non-Abelian form.\\
\indent It should now appear self-suggesting that the definition of the total energy ($\ET$) of an RST field configuration is to be based upon the time component $\TToo$ of the energy-momentum density $\TTmunu$, i.\,e.
   \begin{eqnarray}
   \label{251}
   \ET&\doteqdot&\int\!\!{\rm d}^3\vec{r}\;\TToo\;.
   \end{eqnarray}
But since the total density $\TTmunu$ is the sum of a matter part and a gauge field part, cf.\,\rf{246}, the total energy $\ET$~\rf{251} naturally breaks up in an analogous way
   \begin{eqnarray}
   \label{252}
   \ET&=&\ED+\EG\;,
   \end{eqnarray}
with the self-evident definitions
   \begin{subequations}
   \begin{align}
   \label{253a}
   \ED&\doteqdot\idr\DToo\\
   \label{253b}
   \EG&\doteqdot\idr\GToo\;.
   \end{align}
   \end{subequations}
However, such an arrangement obviously requires the splitting up of space-time into space and time, which will readily be carried through for all the RST objects in the context of the bound systems. But the key question for the subsequent discussion will refer to the proposed definition~\rf{251} for the energy concept in RST. Actually, we will readily show how this definition~\rf{251} must be somewhat modified when the exchange interactions are taken into account.

\subsection{Linear Approximation}

The non-Abelian character of the $N$-particle theory clearly implies a considerable complication when one wishes to look for concrete solutions of the field equations, i.\,e. the Dirac equation~\rf{231} and the Maxwell equation~\rf{233}. On the other hand, it seems reasonable to assume that some aspects of the RST solutions can be discussed (at least in a qualitative way) by neglecting the nonlinearities which invade the theory through just its non-Abelian character. Therefore it is a nearby idea to consider first the {\em linearized theory}\/, namely by calculating certain energy levels in the linear approximation and comparing the corresponding results to the observational data. Such a linearization does not mean that one has to dispense with the exchange effects, because the exchange potential $\Bmu$ persists as part of the theory and it is only its field equation which becomes linearized. In this way one may deal with the exchange effects in a first approximation.\\
\indent The procedure of linearization refers mainly to the gauge field subsystem and consists in neglecting certain nonlinear terms emerging in that subsystem. In this sense, one neglects the commutators for the curvature definition~\rf{212} and puts in a simplifying way
   \begin{eqnarray}
   \label{254}
   \MFmunu&\Rightarrow&\nablamu\MAnu-\nablanu\MAmu\;,
   \end{eqnarray}
which truncates the component version \rf{230a}-\rf{230d} to ($a=1,2,3$)
   \begin{subequations}
   \begin{align}
   \label{255a}
   \Faomunu&=\nablamu\Aaonu-\nablanu\Aaomu\\
   \label{255b}
   \Gmunu&=\nablamu\Bnu-\nablanu\Bmu\;.
   \end{align}
   \end{subequations}
Correspondingly, the component form \rf{234a}-\rf{234d} of the Maxwell equation~\rf{233} becomes truncated to ($a=1,2,3$)
   \begin{subequations}
   \begin{align}
   \label{256a}
   \nabla^\mu\Faomunu&=4\pi\als\,\jaonu\\
   \label{256b}
   \nabla^\mu\Gmunu&=4\pi\als\,\hSnu\;.
   \end{align}
   \end{subequations}
But this implies immediately that the non-Abelian imprints of the source system \rf{244a}-\rf{244d} become omitted and all the currents do obey local conservation laws in the strict sense:
   \begin{subequations}
   \begin{align}
   \label{257a}
   \nabla^\mu\kaumu&\equiv0\\
   \label{257b}
   \nabla^\mu\hmu&\equiv0\;.
   \end{align}
   \end{subequations}
Thus the linearized gauge field system adopts a shape very similar to ordinary electrodynamics; and if one takes over also the Lorentz gauge conditions
   \begin{subequations}
   \begin{align}
   \label{258a}
   \nabla^\mu\Aaomu&=0\\
   \label{258b}
   \nabla^\mu\Bmu&=0\;,
   \end{align}
   \end{subequations}
then the ordinary Maxwell equations \rf{256a}-\rf{256b} are transcribed to the four-potentials in form of the well-known {\em d'Alembert equations}\/ ($a=1,2,3$)
   \begin{subequations}
   \begin{align}
   \label{259a}
   \square\Aaomu&=4\pi\als\,\jaomu\\
   \label{259b}
   \square\Bmu&=4\pi\als\,\hSmu\;.
   \end{align}
   \end{subequations}

Moreover, the fact that the gauge field equations can be deduced from an action principle is not affected at all by the linearization procedure. Indeed, it is merely necessary to substitute the truncated curvature components \rf{255a}-\rf{255b} into the gauge field Lagrangean~\rf{225}; and then the variational procedure yields the linearized field equations \rf{259a}-\rf{259b} in place of the original nonlinear ones \rf{234a}-\rf{234d}. On the other hand, the matter dynamics~\rf{231}, or its component form \rf{232a}-\rf{232c}, resp., is left completely unaffected by the linearization process.

Subsequently we will generalize the {\em RST principle of minimal energy} \cite{c9,c11} in order to include also the exchange interactions, albeit only in the linear approximation; and thereby we will also have to modify somewhat the na\"{\i}ve definition of total energy $\ET$~\rf{251}.

\section{Stationary Bound Systems}\label{s3}
\indent 

\textrm One of the most important applications of RST will surely be in the field of atomic and molecular physics, where the calculation of the relativistic energy levels must naturally receive the main interest. Therefore it becomes necessary to subject the {\em stationary}\/ RST states to a more thorough inspection. But this requires to break up the Lorentz covariant objects into their space and time components, which presumes the selection of some special Lorentz frame so that the manifest Lorentz covariance of the theory is lost. But the advantage of this reformulation of the {\em relativistic}\/ theory in terms of three-tensors (of various rank) is that one can more conveniently obtain the stationary solutions of the RST field equations, both for the relativistic and non-relativistic situations.

\subsection{Gauge Field Subsystem}

The simplest space-time splitting refers to the four-potentials $\Aaomu$, which for the stationary states become time-independent and thus appear in the following form:
   \begin{eqnarray}
   \label{31}
   \Aaomu(x)&\Rightarrow&\left\{\aAo\vr\,;-\vAa\vr\right\}\\
   &&\hspace{-1.6em}(a=1,2,3)\;.\nonumber
   \end{eqnarray}
A similar arrangement does apply also to the Maxwell four-currents $\jaomu$
   \begin{eqnarray}
   \label{32}
   \jaomu&\Rightarrow&\left\{\ajo\vr\,;-\vja\vr\right\}\,,
   \end{eqnarray}
so that the d'Alembert equations~\rf{259a} become split up into the Poisson equations, for both the scalar potentials $\aAo\vr$
   \begin{eqnarray}
   \label{33}
   \Delta\aAo\vr\!\!\!&=&\!\!\!-4\pi\als\ajo\vr
   \end{eqnarray}
and the three-vector potentials $\vAa\vr$
   \begin{eqnarray}
   \label{34}
   \Delta\vAa\vr\!\!\!&=&\!\!\!-4\pi\als\,\vja\vr\;.
   \end{eqnarray}
Recall here that the standard solutions of these equations are formally given by
   \begin{subequations}
   \begin{align}
   \label{35a}
   \aAo\vr&=\als\idrs\frac{\ajo\vrs}{\|\vec{r}-\vec{r}\!\;'\|}\\
   \label{35b}
   \vAa\vr&=\als\idrs\frac{\vAa\vrs}{\|\vec{r}-\vec{r}\!\;'\|}\;.
   \end{align}
   \end{subequations}

But somewhat more complicated is the space-time splitting of the exchange potentials $\Bmu$ because they are left time-dependent, even for the stationary states:
   \begin{subequations}
   \begin{align}
   \label{36a}
   \Bmu(x)&=\exp\Big[\!-i\,\frac{ct}{\am}\Big]\cdot\Bmu\vr\\
   \Bmu\vr&=\left\{\Bo\vr\,;-\vB\vr\right\}\,,
   \end{align}
   \end{subequations}
where the ``{\em exchange radius}\/'' $\am$ is defined as
   \begin{eqnarray}
   \label{37}
   \am&\doteqdot&\frac{\hbar}{(M_2-M_3)c}\;.
   \end{eqnarray}
The reason for such a time-dependence is that the exchange potential $\Bmu(x)$ must be a solution of the d'Alembert equation~\rf{259b} with the space-time splitting of the exchange current $\hmu(x)$~\rf{224} appearing as
   \begin{subequations}
   \begin{align}
   \label{38a}
   \hmu(x)&=\exp\Big[\!+i\,\frac{ct}{\am}\Big]\cdot\hmu\vr\\
   \label{38b}
   \hmu\vr&=\left\{\ho\vr\,;-\vh\vr\right\}\;.
   \end{align}
   \end{subequations}
This specific time-dependence of the exchange current is directly induced by the single-particle wave functions $\psi_a(x)$, which adopt their stationary forms in terms of the {\em mass eigenvalues}\/ $M_a$ as
   \begin{eqnarray}
   \label{39}
   \psi_a(x)&=&\exp\Big[\!-i\,\frac{M_ac^2}{\hbar}\,t\Big]\cdot\psi_a\vr\;.
   \end{eqnarray}
Accordingly, the d'Alembert equation~\rf{259b} for the exchange potential $\Bmu$ becomes not reduced to the ordinary Poisson equation as for the electromagnetic potentials $\Aaomu$, cf. \mbox{\rf{33}-\rf{34}}, but rather does appear in the following form:
   \begin{subequations}
   \begin{align}
   \label{310a}
   \Delta\Bo\vr+\frac{1}{\am^2}\,\Bo\vr&=-4\pi\als\,\hSo\vr\\
   \label{310b}
   \Delta\vB\vr+\frac{1}{\am^2}\,\vB\vr&=-4\pi\als\,\vhS\vr\;.
   \end{align}
   \end{subequations}
The formal solution of this type of equation is easily found to be the following:
   \begin{subequations}
   \begin{align}
   \label{311a}
   \Bo\vr&=\als\idrs\frac{\hSo\vrs\cdot\cos\!\left(\!\frac{\|\vec{r}-\vec{r}\!\;'\|}{ \am}\!\right)}{\|\vec{r}-\vec{r}\!\;'\|}\\
   \label{311b}
   \vB\vr&=\als\idrs\frac{\vhS\vrs\cdot\cos\!\left(\!\frac{\|\vec{r}-\vec{r}\!\;'\|}{ \am}\!\right)}{\|\vec{r}-\vec{r}\!\;'\|}\;,
   \end{align}
   \end{subequations}
which for infinite exchange radius (\mbox{$\am\rightarrow\infty$}, i.\,e. \mbox{$M_2\rightarrow M_3$}) tends to the massless form of the electromagnetic potentials $\aAo\vr,\,\vAa\vr$ \mbox{\rf{35a}-\rf{35b}}.

It is true, the particle interactions are organized via the (electromagnetic and exchange) {\em potentials}\/ which, according to the {\em principle of minimal coupling}\/, are entering the covariant derivatives $\Dmu\psi_a$ of the wave functions $\psi_a$ as shown by equations \mbox{\rf{217a}-\rf{217c}}. But nevertheless it is very instructive to glimpse also at the {\em field strengths}\/ $\Faomunu,\,\Gmunu$. Their space-time splitting is given by
   \begin{subequations}
   \begin{align}
   \label{312a}
   \vEa&=\left\{\aEjo\right\}\;\doteqdot\;\left\{\Faoouju\right\}\\
   \label{312b}
   \vHa&=\left\{\aHjo\right\}\;\doteqdot\;\left\{\frac{1}{2}\varepsilon^{jk}_{\;\;\;l}\Faokulo\right\}\\
   \label{312c}
   \vX&=\left\{X^j\right\}\;\doteqdot\;\left\{\Gouju\right\}\\
   \label{312d}
   \vY&=\left\{Y^j\right\}\;\doteqdot\;\left\{\frac{1}{2}\varepsilon^{jk}_{\;\;\;l}\Gkulo\right\}\;,
   \end{align}
   \end{subequations}
and thus the {\em linearized}\/ Maxwell equations \mbox{\rf{256a}-\rf{256b}} do split up in three-vector form $(a=1,2,3)$ to the scalar equations for the electric fields
   \begin{subequations}
   \begin{align}
   \label{313a}
   \vnabla\sdot\vEa&=4\pi\als\ajo\\
   \label{313b}
   \vnabla\sdot\vX&=4\pi\als\,\hSo\;,
   \end{align}
   \end{subequations}
and to the curl equations for the magnetic fields
   \begin{subequations}
   \begin{align}
   \label{314a}
   \vnabla\times\vHa&=4\pi\als\vja\\
   \label{314b}
   \vnabla\times\vY+\frac{i}{\am}\,\vX&=4\pi\als\,\vhS\;.
   \end{align}
   \end{subequations}
There is a pleasant consistency check for these linearized (but still relativistic) field equations in three-tensor form; namely one may first link the field strengths to the potentials in three-vector notation (cf.~\rf{255a}-\rf{255b} for the corresponding relativistic link):
   \begin{subequations}
   \begin{align}
   \label{315a}
   \vEa\vr&=-\vnabla\aAo\vr\\
   \label{315b}
   \vX\vr&=-\vnabla\Bo\vr+\frac{i}{\am}\,\vB\vr\\
   \label{315c}
   \vHa\vr&=\vnabla\times\vAa\vr\\
   \label{315d}
   \vY\vr&=\vnabla\times\vB\vr\;,
   \end{align}
   \end{subequations}
and then one substitutes these three-vector field strengths into their source and curl equations \mbox{\rf{313a}-\rf{314b}}. In this way one actually recovers the Poisson equations \mbox{\rf{33}-\rf{34}} for the electromagnetic potentials $\aAo,\,\vAa$; but the exchange potentials $\Bo,\,\vB$ require an extra discussion.

The point here is that the Lorentz gauge condition~\rf{258b} must in three-tensor form be consistent with the linearized source and curl equations for those three-vector field strengths. In order to become convinced of this, substitute first the magnetic exchange field strength $\vY$~\rf{315d} into the curl equation~\rf{314b} and find
   \begin{eqnarray}
   \label{316}
   \hspace{-3em}\vnabla\big(\vnabla\sdot\vB\vr\big)-\Delta\vB\vr-\frac{i}{\am}\,\vnabla\Bo\vr-\frac{1}{\am^2}\,\vB\vr &=&4\pi\als\,\vhS\vr\;.
   \end{eqnarray}
Obviously it becomes necessary here to select some gauge condition for the magnetic exchange potential $\vB$ in order that the latter equation~\rf{316} can coincide with the already fixed equation~\rf{310b} for $\vB$. This requirement implies that one has to put
   \begin{eqnarray}
   \label{317}
   \vnabla\left[\big(\vnabla\sdot\vB\vr\big)-\frac{i}{\am}\,\Bo\vr\right]&=&0\;,
   \end{eqnarray}
whose simplest solution is of course
   \begin{eqnarray}
   \label{318}
   \vnabla\sdot\vB-\frac{i}{\am}\,\Bo\vr&\equiv&0\;.
   \end{eqnarray}
But this is nothing else than the Lorentz gauge condition~\rf{258b} for the four-potential $\Bmu$ when its general time-dependence~\rf{36a} is respected.

Next, contract the curl equation~\rf{314b} by the gradient operator ($\vnabla$), observe also the generally valid identity
   \begin{eqnarray}
   \label{319}
   \vnabla\sdot\big(\vnabla\times\vY\vr\big)&\equiv&0
   \end{eqnarray}
and thus arrive at
   \begin{eqnarray}
   \label{320}
   \frac{i}{\am}\,\vnabla\sdot\vX\vr&=&4\pi\als\,\vnabla\sdot\vhS\vr\;,
   \end{eqnarray}
which by means of the source equation~\rf{313b} leads to
   \begin{eqnarray}
   \label{321}
   \frac{i}{\am}\,\hSo\vr&=&\vnabla\sdot\vhS\vr\;.
   \end{eqnarray}
But this is again nothing else than the conservation law~\rf{257b} for the four-current $\hmu$ when its general time-dependence~\rf{38a} for the stationary states is respected. In this way it is evident that the gauge part of the RST field equations remains a completely consistent framework for the stationary systems after the process of {\em linearization}\/ has been carried through.

Moreover, the exchange subsystem of the gauge part admits an important conclusion for the matter subsystem. Namely, observing the general definition~\rf{224} for the exchange current $\hmu$, one finds for its time component $\ho\vr$
   \begin{eqnarray}
   \label{322}
   \ho\vr&\doteqdot&\bar{\psi}_2\vr\gamma_0\psi_3\vr\;=\;\psi_2^\dagger\vr\psi_3\vr\;.
   \end{eqnarray}
Therefore integrating equation~\rf{321} over all three-space, with observation of the exchange radius $\am$~\rf{37}, yields the following important result:
   \begin{eqnarray}
   \label{323}
   (M_2-M_3)\idr\psi_2^\dagger\vr\psi_3\vr&=&-i\,\frac{\hbar}{c}\oint\limits_{(\!\!\:S_\infty\!)}\!\vhS\vr\sdot\dif{\vec{S}}\;,
   \end{eqnarray}
where $S_\infty$ is some infinitely distant two-surface. Assuming now that the matter fields $\psi_a\vr$ ($a=2,3$) decay rapidly enough (e.\,g. exponentially) at spatial infinity ($r\rightarrow\infty$), the exchange flux through this infinitely distant surface will vanish,
   \begin{eqnarray}
   \label{324}
   \oint\limits_{(\!\!\:S_\infty\!)}\!\vhS\vr\sdot\dif{\vec{S}}&=&0\;,
   \end{eqnarray}
and thus we are left with the {\em orthogonality condition}
   \begin{eqnarray}
   \label{325}
   \idr\psi_2^\dagger\vr\psi_3\vr&=&0\;.
   \end{eqnarray}
But observe here that this conclusion can be drawn only for those states $\psi_a\vr$ ($a=2,3$) which have different mass eigenvalues \mbox{($M_2\neq M_3$)}.

Further information about the gauge field subsystem can be obtained by multiplying through the equations \mbox{\rf{310a}-\rf{310b}} by $\BSo\vr$ and $\vBS\vr$, resp., with subsequent integration over all three-space. In this way one finds from the first equation~\rf{310a}
   \begin{eqnarray}
   \idr\!\left[\frac{1}{\am^2}\,\BSo\vr\Bo\vr-\big(\vnabla\BSo\vr\big)\sdot\big(\vnabla\Bo\vr\big)\right]\hspace{-.8em}&=& \hspace{-.7em}{}-4\pi\als\idr\hSo\vr\BSo\vr\;.\nonumber\\
   \label{326}
   &&
   \end{eqnarray}
This result says that the right-hand side, i.\,e. the {\em exchange mass equivalent of electric type}\/ ($\Mh c^2$), must be a real number:
   \begin{eqnarray}
   \Mh c^2&\doteqdot&\hbar c\idr\Bo\vr\cdot\ho\vr\;=\;\hbar c\idr\BSo\vr\cdot\hSo\vr\nonumber\\
   \label{327}
   &=&\frac{\hbar c}{2}\idr\left[\Bo\vr\cdot\ho\vr+\BSo\vr\cdot\hSo\vr\right]\,.
   \end{eqnarray}
A similar conclusion can be drawn also from the second equation~\rf{310b} for the {\em exchange mass equivalent}\/ ($\Mg c^2$) {\em of the magnetic type}\/:
   \begin{eqnarray}
   \Mg c^2&\doteqdot&\hbar c\idr\vB\vr\sdot\vh\vr\;=\;\hbar c\idr\vBS\vr\sdot\vhS\vr\nonumber\\
   \label{328}
   &=&\frac{\hbar c}{2}\idr\left[\vB\vr\sdot\vh\vr+\vBS\vr\sdot\vhS\vr\right]\,,
   \end{eqnarray}
which emerges in the following form:
   \begin{eqnarray}
   \label{329}
   \frac{1}{\am^2}\idr\vBS\vr\sdot\vB\vr-\idr\big(\vnabla\times\vB\vr\big)\sdot\big(\vnabla\times\vBS\vr\big)\hspace{-6em}&&\\
   &=& -4\pi\als\idr\vhS\vr\sdot\vBS\vr\;.\nonumber
   \end{eqnarray}
On the other hand, the exchange system deals also with purely imaginary integrals, e.\,g.
   \begin{eqnarray}
   \label{330}
   \idr\vBS\vr\sdot\vX\vr&=&-\idr\vB\vr\sdot\vXS\vr\;.
   \end{eqnarray}
In order to verify this one simply substitutes here the exchange field strength of the electric type (i.\,e. $\vX\vr$) from equation~\rf{315b} and obtains through integration by parts
   \begin{eqnarray}
   \label{331}
   \idr\vBS\vr\sdot\vX\vr
   \;=\;\frac{i}{\am}\idr\left[\vBS\vr\sdot\vB\vr-\BSo\vr\cdot\Bo\vr\right]\hspace{-4em}&&\\
   &\equiv&\hspace{-.5em}-\frac{i}{\am}\idr\BSmo\Bmu\;.\nonumber
   \end{eqnarray}
Obviously, the integral on the left-hand side vanishes either for $\am\rightarrow\infty$, i.\,e. when both electrons occupy states with identical mass eigenvalues \mbox{($M_2=M_3$)}, or it vanishes also for the situation where the exchange interactions of the electric ($\leadsto\Bo$) and magnetic type ($\leadsto\vB$) are of equal magnitude
   \begin{eqnarray}
   \label{332}
   \idr\BSmo\Bmu&=&\idr\left[\BSo\vr\cdot\Bo\vr-\vBS\vr\sdot\vB\vr\right]\;\simeq\;0\;.
   \end{eqnarray}
However, the magnetic-type interactions are usually much smaller than their electric counterparts, which then implies that the requirement~\rf{332} can in general not be satisfied. One will subsequently have to return to this point of omission of the integral~\rf{332} in connection with the RST {\em principle of minimal energy}\/.

In this context, the {\em exchange identities}\/ of the electric~\rf{326} and magnetic type~\rf{329} will subsequently play the role of constraints and are for this purpose recast into a more suggestive form, namely for the electric type
   \begin{eqnarray}
   \label{333}
   \NGh&\doteqdot&\idr\Bigg\{\frac{\hbar c}{4\pi\als}\left[\big(\vnabla\BSo\vr\big)\!\!\!\:\sdot\!\!\big(\vnabla\Bo\vr\big)- \frac{1}{\am^2}\,\BSo\vr\cdot\Bo\vr\right]\hspace{4em}\\
   &&\hspace{10em}-\frac{\hbar c}{2}\left[\Bo\vr\cdot\ho\vr+\BSo\vr\cdot\hSo\vr\right]\Bigg\}\;\equiv\;0\nonumber
   \end{eqnarray}
and similarly for the magnetic type
   \begin{eqnarray}
   \label{334}
   \NGg&\doteqdot&
   \idr\Bigg\{\frac{\hbar c}{4\pi\als}\left[\big(\vnabla\times\vBS\vr\big)\!\!\!\:\sdot\!\!\big(\vnabla\times\vB\vr\big)+ \big(\vnabla\!\!\!\:\sdot\!\!\vBS\vr\big)\cdot\big(\vnabla\!\!\!\:\sdot\!\!\vB\vr\big)\right.\nonumber\\
   &&\hspace{3em}\left.{}-\frac{1}{\am^2}\,\vBS\vr\!\sdot\!\!\!\:\vB\vr\right]-\frac{\hbar c}{2}\left[\vB\vr\!\sdot\!\!\!\:\vh\vr+\vBS\vr\!\sdot\!\!\!\:\vhS\vr\right]\Bigg\}\;\equiv\;0\;.\nonumber\\
   &&
   \end{eqnarray}
It should be evident that similar identities can also be deduced from the electromagnetic Poisson equations~\rf{33} and~\rf{34}, namely the {\em electric Poisson identity}\/ \cite{c9,c11}
   \begin{eqnarray}
   \label{335}
   \NGe&\doteqdot&\idr\left\{\frac{\hbar c}{4\pi\als}\left[\big(\vnabla\eAo\big)\!\!\!\:\sdot\!\!\big(\vnabla\zAo\big)+ \big(\vnabla\zAo\big)\!\!\!\:\sdot\!\!\big(\vnabla\dAo\big)+ \big(\vnabla\dAo\big)\!\!\!\:\sdot\!\!\big(\vnabla\eAo\big)\right]\right.\nonumber\\
   &&\left.{}-\frac{\hbar c}{2}\left[\eAo\big(\zjo+\djo\big)+\zAo\big(\djo+\ejo\big)+\dAo\big(\ejo+\zjo\big)\right]\right\}\;\equiv\;0\nonumber\\
   &&
   \end{eqnarray}
and its magnetic counterpart
   \begin{eqnarray}
   \label{336}
   \NGm\hspace{-.5em}&\doteqdot&\hspace{-.5em}\idr\!\left\{\frac{\hbar c}{4\pi\als}\left[\big(\vnabla\!\!\times\!\!\!\:\vAe\big)\!\!\!\:\sdot\!\!\big(\vnabla\!\!\times\!\!\!\:\vAz\big)\!+\! \big(\vnabla\!\!\times\!\!\!\:\vAz\big)\!\!\!\:\sdot\!\!\big(\vnabla\!\!\times\!\!\!\:\vAd\big)\!+\! \big(\vnabla\!\!\times\!\!\!\:\vAd\big)\!\!\!\:\sdot\!\!\big(\vnabla\!\!\times\!\!\!\:\vAe\big)\right]\right.\nonumber\\
   &&\hspace{3em}\left.{}-\frac{\hbar c}{2}\left[\vAe\!\sdot\!\!\!\:\big(\vjz+\vjd\big)+ \vAz\!\sdot\!\!\!\:\big(\vjd+\vje\big)+ \vAd\!\sdot\!\!\!\:\big(\vje+\vjz\big)\right]\right\}\;\equiv\;0\;.\nonumber\\
   &&
   \end{eqnarray}

Finally concerning the strength of the exchange interactions, the equations \mbox{\rf{311a}-\rf{311b}} tell us that for vanishing overlap of both wave functions $\psi_2$ and $\psi_3$ the exchange density $\ho$ and exchange current $\vh$ do also vanish and hence the exchange potentials $\Bo,\,\vB$ become zero. Therefore the exchange interactions can exist only for overlapping wave functions! But there is another influence on their strength, and this refers to the magnitude of the exchange radius $\am$~\rf{37}, which roughly measures the energy difference \mbox{($M_2c^2-M_3c^2$)} of both particles. In order to get a rough impression of this effect, assume the exchange density $\ho\vr$ to be of the exponential form
   \begin{eqnarray}
   \label{337}
   \ho\vr&\simeq&\eta_0\,\exp\Big[\!-\frac{r}{\ax}\Big]\,,
   \end{eqnarray}
where the length parameter $\ax$ measures the spatial extension of the overlap region. This assumption then yields for the electric exchange potential $\Bo$~\rf{311a} at the origin \mbox{($r=0$)}
   \begin{eqnarray}
   \label{338}
   \Bo\big|_{r=0}&=&4\pi\als\,\eta_0\,\am^2\ax^2\,\frac{\am^2-\ax^2}{(\am^2+\ax^2)^2}\;.
   \end{eqnarray}
Consequently, for small enough an energy difference \mbox{($M_2\approx M_3$, $\am\gg\ax$)} the exchange potential is strongest; and conversely, for increasing energy difference \mbox{($\am\rightarrow0$)} the exchange interaction tends to zero. Clearly, when the wave functions are well separated (i.\,e. \mbox{$\ax\rightarrow0$}) then the exchange effects do also vanish. Summarizing, the exchange effects become negligible for well separated particles of sufficiently different energies.

\subsection{Matter Subsystem}

Similar to the gauge field subsystem, the matter subsystem exhibits also some peculiar features which are worth to be inspected in greater detail.

When the stationary form~\rf{39} of the wave functions $\psi_a(x)$ is inserted in the general matter dynamics \mbox{\rf{232a}-\rf{232c}}, one is led to the {\em mass eigenvalue equations}\/ \cite{c17} for the spatial parts $\psi_a\vr$ of the three wave functions. Since the first particle \mbox{($a=1$)} carries a positive charge, it is different from the other two particles (i.\,e. the electrons) and therefore does not feel the exchange force. Consequently, its mass eigenvalue equation does not contain the exchange potentials $\Bo\vr$ and $\vB\vr$ but exclusively the electromagnetic potentials $\aAo\vr$ and $\vAa\vr$ \mbox{($a=2,3$)} due to the other two (identical) particles. Furthermore, the preceding splitting of the gauge objects into their space and time components suggests a similar splitting of the spatial part $\psi_a\vr$ of the wave functions which then spoils the manifest Lorentz covariance but nevertheless preserves the relativistic character of the theory. Such a splitting of the wave functions refers to the fact that the Dirac four-spinors $\psi_a\vr$ may be conceived as the direct sum of Pauli two-spinors $\aphipm\vr$:
   \begin{eqnarray}
   \label{339}
   \psi_a\vr&=&\aphip\vr\oplus\aphim\vr\\
   &&\hspace{-1.3em}(a=1,2,3)\;,\nonumber
   \end{eqnarray}
so that the matter dynamics \mbox{\rf{232a}-\rf{232c}} transcribes to these Pauli spinors as follows \cite{c17}:
   \begin{subequations}
   \begin{align}
   \label{340a}
   i\,\vsigma\sdot\vnabla\ephipm\vr+ \Big[\zAo\vr+\dAo\vr\Big]\ephimp\vr\nonumber\\
   -\Big[\vAz\vr+\vAd\vr\Big]\!\sdot\vsigma\,\ephipm\vr&= \frac{\pm M_p-M_1}{\hbar}\,c\cdot\ephimp\vr\\
   \label{340b}
   i\,\vsigma\sdot\vnabla\zphipm\vr+ \Big[\eAo\vr+\dAo\vr\Big]\zphimp\vr+\Bo\vr\dphimp\vr\hspace{-2em}\nonumber\\
   -\Big[\vAe\vr+\vAd\vr\Big]\!\sdot\vsigma\,\zphipm\vr- \vB\vr\sdot\vsigma\,\dphipm\vr&= -\frac{M_2\pm M_e}{\hbar}\,c\cdot\zphimp\vr\\
   \label{340c}
   i\,\vsigma\sdot\vnabla\dphipm\vr+ \Big[\eAo\vr+\zAo\vr\Big]\dphimp\vr+\BSo\vr\zphimp\vr\hspace{-2em}\nonumber\\
   -\Big[\vAe\vr+\vAz\vr\Big]\!\sdot\vsigma\,\dphipm\vr- \vBS\vr\sdot\vsigma\,\zphipm\vr&= -\frac{M_3\pm M_e}{\hbar}\,c\cdot\dphimp\vr\;.
   \end{align}
   \end{subequations}

Observe here again the crucial point with this three-particle system; namely that both eigenvalue equations~\rf{340b} and~\rf{340c} for the two identical particles are {\em directly}\/ coupled via the exchange potentials $\Bo\vr$ and $\vB\vr$, so that the second particle's equation~\rf{340b} contains also the third Pauli spinor $\dphipm\vr$ and vice versa; whereas the first particle~\rf{340a} couples only {\em indirectly} to the other two (identical) particles, namely via the gauge fields \mbox{$\aAo,\,\vAa$ ($a=2,3$)} being generated by just those identical particles. Therefore it is natural to suppose that such a direct (i.\,e. exchange) coupling will cause a much more intimate connection between the identical particles than is the case with their indirect coupling to the first (different) particle. Clearly this supposition requires a more thorough discussion below.

But the coupled system of mass eigenvalue equations \mbox{\rf{340a}-\rf{340c}} yields now a first hint on how to set up an energy functional so that these mass eigenvalue equations can appear as the corresponding variational equations. It is true, the sum ($\MfT$) of the mass eigenvalues $M_a$ (to be understood as functionals $\Mfa$ of the Pauli spinors)
   \begin{eqnarray}
   \label{341}
   \MfT c^2&\doteqdot&-\Mfe c^2+\Mfz c^2+\Mfd c^2\;,
   \end{eqnarray}
can not yet be taken as the desired energy functional $\EfT$; but nevertheless the mass eigenvalue equations actually are the variational equations of this {\em total mass functional}\/ \mbox{$\MfT c^2$ \rf{341}}, under the constraint of wave function normalization \mbox{($a=1,2,3$)}
   \begin{eqnarray}
   \label{342}
   \NDa&\doteqdot&\idr\ako\vr-1\nonumber\\
   &=&\idr\Big[\aphipk\vr\sdot\aphip\vr+\aphimk\vr\sdot\aphim\vr\Big]-1\;\equiv\;0\;.\qquad
   \end{eqnarray}

In order to recognize this preliminary success more clearly, resolve the mass eigenvalue equations for the mass eigenvalues $M_a$ (or $\Mfa$, resp.) by contracting those equations with the appropriate Pauli spinors and subsequent integration. Then add up the three mass functionals $\Mfa c^2$ and thus find the total mass functional as \cite{c9}
   \begin{eqnarray}
   \label{343}
   \MfT c^2&=&\MZe^2\cdot M_pc^2+\MZz^2\cdot M_ec^2+\MZd^2\cdot M_ec^2+2\Tkin+ 2\Big[\Me c^2-\Mm c^2\Big]\nonumber\\
   &&{}-2\Big[\Mh c^2-\Mg c^2\Big]\;.
   \end{eqnarray}
Here, the {\em mass renormalization factors}\/ $\MZa$ are given by
   \begin{eqnarray}
   \label{344}
   \MZa^2&\doteqdot& \idr\Big[\aphipk\vr\sdot\aphip\vr-\aphimk\vr\sdot\aphim\vr\Big]\nonumber\\
   &\equiv&\idr\bar{\psi}_a\vr\sdot\psi_a\vr\;,
   \end{eqnarray}
and the kinetic energy $\Tkin$ is the sum of the three single-particle contributions
   \begin{eqnarray}
   \label{345}
   \Tkin&=&\Tkine+\Tkinz+\Tkind\;,
   \end{eqnarray}
i.\,e. for the first particle
   \begin{eqnarray}
   \label{346}
   \Tkine&\doteqdot& i\frac{\hbar c}{2}\idr\Big[\ephimk\vr\vsigma\sdot\vnabla\ephip\vr+\ephipk\vr\vsigma\sdot\vnabla\ephim\vr\Big]\nonumber\\
   &\equiv& i\,\frac{\hbar c}{2}\idr\bar{\psi}_1\vr\vgamma\sdot\vnabla\psi_1\vr
   \end{eqnarray}
and analogously for the two identical particles \mbox{($a=2,3$)}
   \begin{eqnarray}
   \label{347}
   \Tkina&\doteqdot& -i\frac{\hbar c}{2}\idr\Big[\aphimk\vr\vsigma\sdot\vnabla\aphip\vr+\aphipk\vr\vsigma\sdot\vnabla\aphim\vr\Big]\nonumber\\
   &\equiv& -i\,\frac{\hbar c}{2}\idr\bar{\psi}_a\vr\vgamma\sdot\vnabla\psi_a\vr\;.
   \end{eqnarray}
Furthermore, the exchange mass equivalents $\Mh c^2$ and $\Mg c^2$ of the electric and magnetic type have already been defined through equations~\rf{327} and~\rf{328}; and their electromagnetic counterparts $\Me c^2$ and $\Mm c^2$ are given by
   \begin{subequations}
   \begin{align}
   \label{348a}
   \Me c^2&\doteqdot \frac{\hbar c}{2}\idr\left\{\eko\vr\cdot\Big[\zAo\vr+\dAo\vr\Big]- \zko\vr\cdot\Big[\eAo\vr+\dAo\vr\Big]\right.\nonumber\\
   &{}\hspace{5em}\left.-\dko\vr\cdot\Big[\eAo\vr+\zAo\vr\Big]\right\}\;\equiv\; \Mefe c^2+\Mefz c^2+\Mefd c^2\nonumber\\
   &&\\
   \label{348b}
   \Mm c^2&\doteqdot \frac{\hbar c}{2}\idr\left\{\vke\vr\sdot\Big[\vAz\vr+\vAd\vr\Big]- \vkz\vr\sdot\Big[\vAe\vr+\vAd\vr\Big]\right.\nonumber\\
   &{}\hspace{5em}\left.-\vkd\vr\sdot\Big[\vAe\vr+\vAz\vr\Big]\right\}\;\equiv\; \Mmfe c^2+\Mmfz c^2+\Mmfd c^2\;.\nonumber\\
   &&
   \end{align}
   \end{subequations}

Now with these arrangements it is an easy exercise to verify that the mass eigenvalue equations \mbox{\rf{340a}-\rf{340c}} indeed do represent the variational equations due to the variational principle
   \begin{eqnarray}
   \label{349}
   \delta\big(\tMfT c^2\big)&=&0\;.
   \end{eqnarray}
Here, the functional $\tMfT c^2$ is nothing else than the original $\MfT c^2$~\rf{343}, merely to be complemented by the normalization constraints~\rf{342}:
   \begin{eqnarray}
   \label{350}
   \tMfT c^2&=&\MfT c^2+\sum_{a=1}^3\lambDa\NDa\;,
   \end{eqnarray}
where the $\lambDa$ are the usual Lagrangean multipliers. The comparison of the variational equations~\rf{349} with the mass eigenvalue equations \rf{340a}-\rf{340c} yields then immediately the equality of the Lagrangean multipliers and the mass eigenvalues, i.\,e.
   \begin{subequations}
   \begin{align}
   \label{351a}
   \lambDe&=M_1c^2\\
   \label{351b}
   \lambDz&=-M_2c^2\\
   \label{351c}
   \lambDd&=-M_3c^2\;.
   \end{align}
   \end{subequations}

But observe here that, despite of this partial success, one cannot be satisfied with adopting the mass functional $\tMfT c^2$~\rf{350} as the desired energy functional because this mass functional does {\em not}\/ admit to deduce the gauge field equations \mbox{\rf{33}-\rf{34}} and \mbox{\rf{310a}-\rf{310b}} from the variational principle~\rf{349}. The reason is that the functional $\tMfT c^2$ does not contain any derivatives of the gauge fields $\aAo$, $\vAa$, $\Bo$, $\vB$; but this is indispensable for the purpose of deducing those Poisson equations for the gauge fields. Nevertheless, a certain non-relativistic limit of the variational approach~\rf{349} is sometimes used in conventional quantum theory (i.\,e. the Hartree and Hartree-Fock approximation), but this does not in every case yield acceptable results \cite{c11}. Thus the non-relativistic limit of RST requires a more thorough inspection, especially concerning the properties of the matter subsystem.

\subsection{Polarization of Matter}

Since the total mass functional $\tMfT c^2$~\rf{350} does correctly reproduce the {\em matter}\/ part of the RST dynamics, it can be expected to lend itself to an elucidation of how the matter interacts with the gauge fields. First, consider here the positively charged particle \mbox{($a=1$)}, which formally contributes to the total mass functional $\MfT c^2$~\rf{343} exclusively through its {\em electromagnetic}\/ interactions $\Memfe c^2$
   \begin{eqnarray}
   \label{352}
   \Memfe c^2&\doteqdot& \Mefe c^2-\Mmfe c^2\nonumber\\
   &=& \frac{\hbar c}{2}\idr\eko\cdot\Big[\zAo+\dAo\Big]-\frac{\hbar c}{2}\idr\vke\sdot\Big[\vAz+\vAd\Big]\nonumber\\
   &\equiv& \frac{\hbar c}{2}\idr\keumu\,\Big[\Azomo+\Adomo\Big]\;,
   \end{eqnarray}
cf. equations \mbox{\rf{348a}-\rf{348b}}. Now observe here the Gordon decomposition \cite{c18} of the first Dirac current $\keumu$ into its convective and polarization parts, i.\,e.
   \begin{eqnarray}
   \label{353}
   \keumu&=&-\qeumu-\seumu\;.
   \end{eqnarray}
Here, the {\em convection current}\/ ($\qeumu$) is given in terms of the covariant derivative~\rf{217a} as
   \begin{eqnarray}
   \label{353'}
   \qeumu&=&\frac{i\hbar}{2M_pc}\,\Big[\bar{\psi}_1\big(\Dmu\psi_1\big)-\big(\Dmu\bar{\psi}_1\big)\psi_1\Big]
   \end{eqnarray}
while the {\em polarization current}\/ ($\seumu$) is found to be of a divergence form
   \begin{eqnarray}
   \label{354}
   \seumu&=&\pno\Seomunu\;,
   \end{eqnarray}
with the {\em electromagnetic polarization tensor}\/ \mbox{$\Seomunu$ ($=-\Seonumu$)} of the first particle being defined through
   \begin{eqnarray}
   \label{355}
   \Seomunu&\doteqdot& \frac{i\hbar}{M_pc}\,\bar{\psi}_1\Sigmunu\psi_1\\
   &&\hspace{-3.2em}(\,\Sigmunu\doteqdot\frac{1}{4}[\gamma_\mu,\gamma_\nu]\,)\;.\nonumber
   \end{eqnarray}

Now by means of this decomposition~\rf{353} of the first Dirac current $\keumu$, the electromagnetic mass equivalent~\rf{352} can also be split up into two parts:
   \begin{eqnarray}
   \label{356}
   \Memfe c^2&=& -\frac{\hbar c}{2}\idr\qeumu\,\Big[\Azomo+\Adomo\Big]-\frac{\hbar c}{2}\idr\seumu\,\Big[\Azomo+\Adomo\Big]\nonumber\\
   &\doteqdot& \Mqfe c^2+\Msfe c^2\;.
   \end{eqnarray}
Obviously, the convective part $\Mqfe c^2$ represents here the interaction of the purely translational degrees of freedom of the particle ($\leadsto$ point particle), whereas the polarization part $\Msfe c^2$ refers to the particle's dipole properties (of both the electric and magnetic type). The latter property is elucidated more clearly by observing the divergence form~\rf{354} of the polarization current. This latter form admits namely to recast the polarization part $\Msfe c^2$ of the mass equivalent~\rf{356} into the following shape:
   \begin{eqnarray}
   \label{357}
   \Msfe c^2&\doteqdot& -\frac{\hbar c}{2}\idr\seumu\,\Big[\Azomo+\Adomo\Big]\nonumber\\
   &=&\frac{\hbar c}{2}\idr\left\{\evSe\sdot\Big[\vEz+\vEd\Big]+\mvSe\sdot\Big[\vHz+\vHd\Big]\right\}\;,\quad
   \end{eqnarray}
with the electric (e) and magnetic (m) {\em polarization densities}\/ \mbox{$\evSe=\{\eSejo\}$} and \mbox{$\mvSe=\{\mSejo\}$} being related to the space and time components of the original four-tensor $\Seomunu$~\rf{355} through
   \begin{subequations}
   \begin{align}
   \label{358a}
   \eSejo\;\doteqdot\;\Seoouju&\Longleftrightarrow \evSe\;=\;-\frac{i\hbar}{2M_pc}\,\Big[\ephipk\vsigma\ephim-\ephimk\vsigma\ephip\Big]\\
   \label{358b}
   \mSejo\;\doteqdot\;\frac{1}{2}\,\varepsilon^{jkl}\Seokulu&\Longleftrightarrow \mvSe\;=\;\frac{\hbar}{2M_pc}\,\Big\{\ephipk\vsigma\ephip-\ephimk\vsigma\ephim\Big\}\;.
   \end{align}
   \end{subequations}
The time component $\eso$ of the polarization current $\seumu$~\rf{354} is then nothing else than the source of the electric dipole density $\evSe$~\rf{358a}
   \begin{eqnarray}
   \label{359}
   \eso&=&-\vnabla\sdot\evSe\;,
   \end{eqnarray}
whereas the curl of the magnetic dipole density $\mvSe$~\rf{358b} is related to the space part \mbox{$\vse=\{\seujo\}$} by
   \begin{eqnarray}
   \label{360}
   \vse&=&\vnabla\times\mvSe\;.
   \end{eqnarray}

Thus the conclusion is that a Dirac particle owns both a translational \mbox{($\sim q_\mu$)} and a rotational \mbox{($\sim s_\mu$)} handle to be acted upon by the electromagnetic fields. This will become even clearer below in the course of looking for the non-relativistic approximations.

However, the point with this polarization phenomenon is now that in RST the {\em identical}\/ Dirac particles are subjected {\em additionally}\/ to the exchange forces; and since the latter are described by the same gauge formalism as their electromagnetic counterparts, it is logical to assume that those exchange forces do also act upon both the translational and the rotational degrees of freedom of the identical particles. Indeed, one is easily convinced of this by inspecting now in more detail these exchange interactions of the second and third particle \mbox{($a=2,3$)}. Quite similarly to the electromagnetic case~\rf{352}, the exchange contribution to the total mass functional $\MfT c^2$~\rf{343} can also be rewritten in terms of a Lorentz invariant, i.\,e.
   \begin{eqnarray}
   \label{361}
   \Mhg c^2&\doteqdot&\Mh c^2-\Mg c^2\;=\;\frac{\hbar c}{2}\idr\Big[\Bmo\hmu+\BSmo\hSmu\Big]\;,
   \end{eqnarray}
cf. equations \mbox{\rf{327}-\rf{328}}.

Next, one works out the Gordon decomposition of the exchange current $\hmu$~\rf{224} in a way quite analogous to that for the electromagnetic current $\keumu$~\rf{353}, namely by substituting the wave functions $\psi_2$ and $\psi_3$ from the second and third Dirac equation \mbox{\rf{232b}-\rf{232c}}. This then yields in a straightforward manner the following form for the exchange current $\hmu$:
   \begin{eqnarray}
   \label{362}
   \hmu&=&\bmu+\zmu-i\left[\Azono-\Adono\right]\Zmunu-i\BSno\left[\Szomunu-\Sdomunu\right]\;.
   \end{eqnarray}
Of course, the first two terms are here just the exchange counterparts of the convection and spin currents $\qeumu$ and $\seumu$ of the first particle~\rf{353}, i.\,e. more concretely
   \begin{subequations}
   \begin{align}
   \label{363a}
   \bmu&=\frac{i\hbar}{2M_ec}\,\Big[\bar{\psi}_2\big(\Dmu\psi_3\big)-\big(\Dmu\bar{\psi}_2\big)\psi_3\Big]\\
   \label{363b}
   \zmu&=\pno\Zmunu\;,
   \end{align}
   \end{subequations}
with the {\em exchange polarization tensor}\/ $\Zmunu$ being given by
   \begin{eqnarray}
   \label{364}
   \Zmunu&\doteqdot&\frac{i\hbar}{M_ec}\,\bar{\psi}_2\Sigmunu\psi_3\;.
   \end{eqnarray}

When this decomposition~\rf{362} of the exchange current $\hmu$ is now substituted into the mass equivalent~\rf{361}, the latter splits up into four contributions, where the first two correspond to the translational and rotational degrees of freedom of the two identical particles; for the rotational case see the electromagnetic counterpart~\rf{357} where the electric and magnetic fields \mbox{$\vEa,\,\vHa$ ($a=2,3$)} have to be replaced by their exchange counterparts \mbox{$\vX,\,\vY$ \rf{315b} and \rf{315d}}, and similarly for the dipole densities. However, in addition to these exchange counterparts of the electromagnetic forces, there arise two new terms in the Gordon decomposition of the exchange current $\hmu$~\rf{362}, i.\,e. the terms with the polarization tensors $\Zmunu$, $\Szomunu$, and $\Sdomunu$. Therefore, if the exchange current $\hmu$~\rf{362} is inserted into the exchange mass equivalent~\rf{361}, there will emerge terms quadratic in the gauge potentials which then implies certain non-linearities in the theory (as it is to be expected for a non-Abelian gauge theory). Naturally, these nonlinearities will entail considerable computational complications so that one will omit them for the present linear approximation. Thus for our present linearized approach one can resort to the following truncated form of the exchange current $\hmu$~\rf{362}
   \begin{eqnarray}
   \label{365}
   \hmu&\Rightarrow&\bmu+\zmu\;,
   \end{eqnarray}
which appears then as the formal counterpart of the electromagnetic currents such as that for the first particle~\rf{353}. This will readily be elaborated in greater detail in connection with the non-relativistic limit.

\subsection{Non-Relativistic Limit}\label{s34}

The value of any physical theory must surely be measured by the accuracy of its predictions with respect to the observational data. But here it is important to relate this question to the various degrees of accuracy which one wishes to demand from a theoretical formalism. Indeed, one of the most frequent gradations of accuracy refers to the relativistic vs. non-relativistic approaches. Naturally, one will prefer to inspect and test first the non-relativistic limit of the theory in question, before its fully relativistic consequences are considered. In the present context of the {\em principle of minimal energy}\/, this viewpoint would entail that one first has to deduce the non-relativistic limit of this principle from its original relativistic form, namely in order to test the corresponding {\em non-relativistic}\/ predictions, e.\,g. in the field of atomic physics. But for this purpose, one will find it logically desirable that the non-relativistic form of the original RST field equations should be identical to those variational equations which may be deduced from the non-relativistic limit of the {\em principle of minimal energy}\/. Or in other words, the following commutative arrangement should be true:\bigskip\\
\begin{picture}(160,200)(0,0)
\setlength{\unitlength}{1mm}
\put(0,50){\framebox(50,18){$\displaystyle{\mbox{Relativistic} \atop \mbox{Field Equations}}$}}
\put(0.5,50.5){\framebox(49,17){}}
\put(85,50){\framebox(50,18){$\displaystyle{\mbox{Non-Relativistic} \atop \mbox{Field Equations}}$}}
\put(85.5,50.5){\framebox(49,17){}}
\put(0,0){\framebox(50,18){$\displaystyle{\mbox{Relativistic Principle} \atop \mbox{of Minimal Energy}}$}}
\put(0.5,0.5){\framebox(49,17){}}
\put(85,0){\framebox(50,18){$\displaystyle{\mbox{Non-Relativistic Principle} \atop \mbox{of Minimal Energy}}$}}
\put(85.5,0.5){\framebox(49,17){}}
\put(25,18){\vector(0,1){32}}
\put(110,18){\vector(0,1){32}}
\put(50,9){\vector(1,0){35}}
\put(50,59){\vector(1,0){35}}
\put(50,49.8){\makebox(35,18){${\mbox{\footnotesize non-relativistic} \atop \mbox{\footnotesize approximation}}$}}
\put(50,-0.2){\makebox(35,18){${\mbox{\footnotesize non-relativistic} \atop \mbox{\footnotesize approximation}}$}}
\put(16.2,13.7){\rotatebox{90}{\makebox(40,18){${\mbox{\footnotesize variational} \atop \mbox{\footnotesize equations}}$}}}
\put(101.2,13.7){\rotatebox{90}{\makebox(40,18){${\mbox{\footnotesize variational} \atop \mbox{\footnotesize equations}}$}}}
\end{picture}
\bigskip

For the verification of this demand one may first turn to the mass eigenvalue equations whose non-relativistic approximation is easily found by simply eliminating the ``negative'' Pauli components $\aphim\vr$ and thus concentrating on the remaining eigenvalue equations for the ``positive'' components $\aphip\vr$. For this purpose, one formally resolves the original mass eigenvalue equations \mbox{\rf{340a}-\rf{340c}} for the negative Pauli components $\aphim\vr$ in the following approximate way:
   \begin{subequations}
   \begin{align}
   \label{366a}
   \ephim\vr&\cong\frac{i\hbar}{2M_pc}\,\vsigma\sdot\Big[\vnabla+i\vAi\Big]\,\ephip\vr\\
   \label{366b}
   \zphim\vr&\cong-\frac{i\hbar}{2M_ec}\,\vsigma\sdot\Big[\vnabla+i\vAii\Big]\,\zphip\vr+ \frac{\hbar}{2M_ec}\,\big(\vB\sdot\vsigma\big)\dphip\\
   \label{366c}
   \dphim\vr&\cong-\frac{i\hbar}{2M_ec}\,\vsigma\sdot\Big[\vnabla+i\vAiii\Big]\,\dphip\vr+ \frac{\hbar}{2M_ec}\,\big(\vBS\sdot\vsigma\big)\zphip\;.
   \end{align}
   \end{subequations}
Here and in the following, we use certain combinations of both the magnetic three-vector potentials $\vAa\vr$ and of the electric potentials $\aAo\vr$:
   \begin{subequations}
   \begin{align}
   \label{367a}
   \vAi&\doteqdot\vAz+\vAd\hspace{4em}\iAo\;\doteqdot\;\zAo+\dAo\\
   \label{367b}
   \vAii&\doteqdot\vAe+\vAd\hspace{3.7em}\iiAo\;\doteqdot\;\eAo+\dAo\\
   \label{367c}
   \vAiii&\doteqdot\vAe+\vAz\hspace{3.4em}\iiiAo\;\doteqdot\;\eAo+\zAo\;.
   \end{align}
   \end{subequations}
The reason for this is that these combinations transform {\em inhomogeneously}\/ as
   \begin{subequations}
   \begin{align}
   \label{368a}
   \vAi&\Rightarrow\vAi'=\vAi+\vnabla\alpha_1\\
   \label{368b}
   \vAii&\Rightarrow\vAii'=\vAii+\vnabla\alpha_2\\
   \label{368c}
   \vAiii&\Rightarrow\vAiii'=\vAiii+\vnabla\alpha_3\;,
   \end{align}
   \end{subequations}
provided the wave functions $\psi_a\vr$ and the exchange potential $\vB\vr$ undergo a {\em homogeneous}\/ magnetic gauge transformation of the following kind \mbox{($a=1,2,3$)}:
   \begin{subequations}
   \begin{align}
   \label{369a}
   \psi_a\vr&\Rightarrow\psi'_a\vr={\rm e}^{-i\alpha_a}\psi_a\vr\\
   \label{369b}
   \vB\vr&\Rightarrow\vB'\vr={\rm e}^{-i(\alpha_2-\alpha_3)}\vB\vr\\
   \label{369c}
   \Bo\vr&\Rightarrow\Bo'\vr={\rm e}^{-i(\alpha_2-\alpha_3)}\Bo\vr\;.
   \end{align}
   \end{subequations}
(For a general discussion of the RST gauge transformations see ref.~\cite{c12}.) As a consequence of this gauge arrangement, the negative Pauli components $\aphim\vr$ \mbox{\rf{366a}-\rf{366c}} do inherit the same homogeneous transformation behaviour, i.\,e.
   \begin{eqnarray}
   \label{370}
   \aphim\vr&\Rightarrow\aphim'\vr={\rm e}^{-i\alpha_a}\!\cdot\!\aphim\vr\;,
   \end{eqnarray}
so that one can expect that the emerging energy eigenvalue equations for the positive Pauli spinors $\aphip\vr$ will also be found to be gauge covariant.

And indeed, substituting those negative Pauli spinors $\aphim\vr$ \rf{366a}-\rf{366c} into the {\em relativistic}\/ mass eigenvalue equations \mbox{\rf{340a}-\rf{340c}} yields the corresponding {\em non-relativistic linearized}\/ energy eigenvalue equations for the positive Pauli spinors $\aphip\vr$ in the following form
   \begin{subequations}
   \begin{align}
   \label{371a}
   \left\{-\frac{\hbar^2}{2M_p}\,\big(\vnabla+i\vAi\big)^2+\hbar c\,\iAo+ \frac{\hbar^2}{2M_p}\,\big(\vHi\sdot\vsigma\big)\right\}\ephip&=\Epe\cdot\ephip\\
   \label{371b}
   \left\{-\frac{\hbar^2}{2M_e}\,\big(\vnabla+i\vAii\big)^2-\hbar c\,\iiAo+ \frac{\hbar^2}{2M_e}\,\big(\vHii\sdot\vsigma\big)\right\}\zphip\nonumber\\
   + \left\{\frac{\hbar^2}{2M_e}\,\big(\vY\sdot\vsigma\big)-\hbar c\,\Bo-i\,\frac{\hbar^2}{2M_e}\left[\big(\vnabla\sdot\vB\big)+2\vB\sdot\vnabla\right]\right\}\dphip &=\Epz\cdot\zphip\\
   \label{371c}
   \left\{-\frac{\hbar^2}{2M_e}\,\big(\vnabla+i\vAiii\big)^2-\hbar c\,\iiiAo+ \frac{\hbar^2}{2M_e}\,\big(\vHiii\sdot\vsigma\big)\right\}\dphip\nonumber\\
   + \left\{\frac{\hbar^2}{2M_e}\,\big(\vYS\sdot\vsigma\big)-\hbar c\,\BSo-i\,\frac{\hbar^2}{2M_e}\left[\big(\vnabla\sdot\vBS\big)+2\vBS\sdot\vnabla\right]\right\}\zphip &=\Epd\cdot\dphip\;.
   \end{align}
   \end{subequations}
Here the (non-relativistic) Pauli energy eigenvalues $\Epa$ are defined in a nearby way in terms of the particle rest masses $M_p$, $M_e$ and the mass eigenvalues $M_a$ through
   \begin{subequations}
   \begin{align}
   \label{372a}
   \Epe&\doteqdot-(M_p+M_1)c^2\\
   \label{372b}
   \Epz&\doteqdot-(M_e-M_2)c^2\\
   \label{372c}
   \Epd&\doteqdot-(M_e-M_3)c^2\;.
   \end{align}
   \end{subequations}
   
Observe again the specific way in which the Pauli eigenvalue system \rf{371a}-\rf{371c} reflects the interactive structure of the three-particle system: Equation~\rf{371a} for the first (different) particle says that this particle is subjected exclusively to the {\em electromagnetic}\/ interactions with the other two (identical) particles, namely via the electric and magnetic potentials \mbox{$\iAo$, $\vAi$ \rf{367a}}, where the latter potential $\vAi$ refers to the interaction due to the translational (point particle) degrees of freedom. On the other hand, the presence of the magnetic field \mbox{$\vHi$ ($=\vnabla\times\vAi$)} in equation~\rf{371a} is due to the magnetic dipole character of the first particle. As a consequence of this ordinary interaction via the electromagnetic potentials $\iAo$ and $\vAi$, the first eigenvalue equation~\rf{371a} is of the usual Hamiltonian form
   \begin{eqnarray}
   \label{373}
   \hatHi\ephip&=&\Epe\cdot\ephip
   \end{eqnarray}
where the Hamiltonian $\hatHi$ is evidently given by
   \begin{eqnarray}
   \label{374}
   \hatHi&=&-\frac{\hbar^2}{2M_p}\,\big(\vnabla+i\vAi\big)^2+\hbar c\,\iAo+\frac{\hbar^2}{2M_p}\,\big(\vHi\sdot\vsigma\big)\;.
   \end{eqnarray}

Clearly, such a purely electromagnetic type of interaction is active also for the second and third particle, cf. equations \mbox{\rf{371b}-\rf{371c}}; but additionally these two ({\em identical}\/) particles do feel the exchange forces, where the latter subdivide again into those of the electric type ($\sim\Bo$) and those of the magnetic type \mbox{($\sim\vB$)}. Consequently, the eigenvalue equations~\rf{371b}-\rf{371c} for the two identical particles are not of that simple form as for the first particle~\rf{373} but rather do appear in the modified Hamiltonian form
   \begin{subequations}
   \begin{align}
   \label{375a}
   \hatHii\zphip+\hathii\dphip&=\Epz\cdot\zphip\\
   \label{375b}
   \hatHiii\dphip+\hathiii\zphip&=\Epd\cdot\dphip\;.
   \end{align}
   \end{subequations}
Here the ordinary Hamiltonians $\hat{H}_{\rm II,III}$ are of the usual electromagnetic type \rf{374}, i.\,e.
   \begin{eqnarray}
   \label{376}
   \hat{H}_{\rm II,III}&=&-\frac{\hbar^2}{2M_e}\,\big(\vnabla+i\vec{A}_{\rm II,III}\big)^2-\hbar c\,{{}^{\rm(\!\!\:II,III\!\!\:)\!\!\!}A_0}+\frac{\hbar^2}{2M_e}\,\big(\vec{H}_{\rm II,III}\sdot\vsigma\big)\;,\qquad
   \end{eqnarray}
and furthermore the ``exchange'' Hamiltonians $\hat{h}_{\rm II,III}$ can be read off immediately from \rf{371b}-\rf{371c} as
   \begin{subequations}
   \begin{align}
   \label{377a}
   \hathii&=\frac{\hbar^2}{2M_e}\,\big(\vY\sdot\vsigma\big)-\hbar c\,\Bo-i\,\frac{\hbar^2}{2M_e}\Big[\!\big(\vnabla\sdot\vB\big)+2\vB\sdot\vnabla\Big]\\
   \label{377b}
   \hathiii&=\frac{\hbar^2}{2M_e}\,\big(\vYS\sdot\vsigma\big)-\hbar c\,\BSo-i\,\frac{\hbar^2}{2M_e}\Big[\!\big(\vnabla\sdot\vBS\big)+2\vBS\sdot\vnabla\Big]\;.
   \end{align}
   \end{subequations}

It is true, the Pauli eigenvalue system \rf{371a}-\rf{371c} is not strictly linear because the terms quadratic in the vector potentials $\vec{A}_{\rm I,II,III}$ have been retained; but this is only in order to {\em formally}\/ ensure the gauge covariance of that Pauli system where the exchange Hamiltonians \rf{377a}-\rf{377b} must be {\em presumed}\/ to transform homogeneously, as is the case with the exchange vector potential $\vB$~\rf{369b}. It should be a matter of course that the {\em strict}\/ gauge covariance can exist only in the original non-linear theory!

\subsection{Pauli Energy Functional}

Once the non-relativistic approximation of the original relativistic mass eigenvalue equations \mbox{\rf{340a}-\rf{340c}} is now firmly established, one would like to go also the other way round by recovering just these non-relativistic equations as the variational equations due to the non-relativistic limit of the original mass functional $\tMfT c^2$~\rf{350}. However, this non-relativistic {\em Pauli energy functional}\/ ($\tEfp$, say) can easily be obtained by simply looking for the non-relativistic limit forms of each of the constituents of $\tMfT c^2$. This procedure will also further elucidate the mathematical consistency of the RST formalism. From the physical point of view, it should be self-evident that the practical usefulness of such a non-relativistic functional $\tEfp$ will consist in the development of variational techniques for obtaining approximative solutions to the non-relativistic eigenvalue system \rf{371a}-\rf{371c}, see below.

\subsubsection[Non-Identical Particle]{Non-Identical Particle\/ {\rm($a=1$)}}

As the first one of these limit forms consider the relativistic kinetic energy $\Tkin$, which is the sum of the individual single-particle contributions $\Tkina$ \mbox{\rf{346}-\rf{347}}. Eliminating here the negative Pauli spinor $\ephim\vr$ from the kinetic energy $\Tkine$ of the first particle~\rf{346} by means of the approximation~\rf{366a} yields
   \begin{eqnarray}
   \label{378}
   \Tkine\;\Rightarrow\;\Eskine&=& -\frac{\hbar^2}{2M_p}\idr\ephipk\left[\vsigma\sdot\big(\vnabla+i\vAi\big)\right]^2\ephip\nonumber\\
   &&{}\hspace{-1em}+i\, \frac{\hbar^2}{4M_p}\idr\ephipk\left\{\vsigma\sdot\big(\vnabla+i\vAi\big),\vsigma\sdot\vAi\right\}\ephip\;.\qquad\qquad
   \end{eqnarray}
But here one can resort to the operator identities
   \begin{subequations}
   \begin{align}
   \label{379a}
   \left[\vsigma\sdot\big(\vnabla+i\vAi\big)\right]^2&\equiv \big(\vnabla+i\vAi\big)^2-\vsigma\sdot\vHi\\
   \label{379b}
   \left\{\vsigma\sdot\big(\vnabla+i\vAi\big),\vsigma\sdot\vAi\right\}&\equiv \vnabla\sdot\vAi+2\vAi\sdot\big(\vnabla+i\vAi\big)+i\,\vsigma\sdot\vHi\\
   &\hspace{-2em}(\mbox{with}\ \vnabla\sdot\vAi=0)\;,\nonumber
   \end{align}
   \end{subequations}
and this recasts the non-relativistic limit $\Eskine$ of $\Tkine$~\rf{378} into its final form
   \begin{eqnarray}
   \label{380}
   \Tkine\;\Rightarrow\;\Eskine&=& -\frac{\hbar^2}{2M_p}\idr\ephipk\big(\vnabla+i\vAi\big)^2\ephip\nonumber\\ &&{}\hspace{-1em}+\frac{\hbar^2}{4M_p}\idr\ephipk\big(\vsigma\sdot\vHi\big)\ephip-\frac{\hbar c}{2}\idr\vAi\sdot\vqqe\;.\qquad\qquad
   \end{eqnarray}
The non-relativistic approximation $\vqqe$ of the convection four-current $q_{1\mu}$~\rf{353'} is easily found here as
   \begin{eqnarray}
   \label{381}
   \vqqe&=&-\frac{i\hbar}{2M_pc}\left\{\ephipk\big(\vnabla+i\vAi\big)\ephip- \left[\big(\vnabla+i\vAi\big)\ephip\right]^\dagger\ephip\right\}\,,\qquad
   \end{eqnarray}
namely by simply omitting the negative Pauli spinors $\ephim$ for its spatial part $\vec{q}_1$.

It is interesting to remark here that the non-relativistic limit $\Eskine$~\rf{380} of the kinetic energy $\Tkine$ does not agree with the expected form $\Ekine$ from conventional quantum mechanics, i.\,e.
   \begin{eqnarray}
   \label{382}
   \Ekine&\doteqdot&-\frac{\hbar^2}{2M_p}\idr\ephipk\big(\vnabla+i\vAi\big)^2\ephip\;,
   \end{eqnarray}
rather this does appear only as the first part of $\Eskine$~\rf{380}! Obviously, the relativistic kinetic energy $\Tkine$~\rf{346} does contain also the magnetic interaction energies of the polarization and convection type (\mbox{$\leadsto$ second} line of equation~\rf{380}). This is of relevance for the non-relativistic limit of the first particle's contribution $\Mfe c^2$ to the total mass functional $\MfT c^2$~\rf{343}
   \begin{eqnarray}
   \label{383}
   -\Mfe c^2&=&M_pc^2\cdot\MZe^2+2\Tkine+2\Memfe c^2
   \end{eqnarray}
because the magnetic interaction energies (on the right-hand side) must combine with the kinetic energies into a consistent non-relativistic approximation $\Efp$ of the mass functional. For instance, consider the mass renormalization factor $\MZe^2$ of the first particle which is given by equation~\rf{344} for $a=1$. Eliminating from that equation the negative Pauli spinor $\ephim\vr$ by means of~\rf{366a} with regard of the {\em relativistic}\/ normalization condition~\rf{342} yields
   \begin{eqnarray}
   \label{384}
   \MZe^2&=&1-2\idr\ephimk\ephim\nonumber\\
   &\Rightarrow&1+ \frac{1}{2}\left(\frac{\hbar}{M_pc}\right)^2\idr\ephipk\left[\vsigma\sdot\big(\vnabla+i\vAi\big)\right]^2\ephip\;,
   \end{eqnarray}
and thus one finds by use of the operator identity~\rf{379a} and the non-relativistic kinetic energy $\Ekine$~\rf{382}
   \begin{eqnarray}
   \label{385}
   \MZe^2&\Rightarrow&1-\frac{\Ekine}{M_pc^2}-\frac{\hbar^2}{2M_p^2c^2}\idr\ephipk\big(\vsigma\sdot\vHi\big)\ephip\;.
   \end{eqnarray}
Therefore the first two contributions to the first mass functional $\Mfe c^2$~\rf{383} combine to the following non-relativistic form
   \begin{eqnarray}
   \label{386}
   M_pc^2\cdot\MZe^2+2\Tkine&\Rightarrow&M_pc^2+\Ekine-\hbar c\idr\vAi\sdot\vqqe\;,
   \end{eqnarray}
which evidently contains still the first particle's magnetic interaction energy, but now exclusively of the {\em convection}\/ type ($\sim\vAi$) and no longer of the polarization type ($\sim\vHi$), as was originally the case with $\Eskine$~\rf{380}.

Finally, for the non-relativistic limit of the mass functional $\Mfe c^2$~\rf{383} it becomes necessary to consider also the non-relativistic limit of the first particle's mass equivalent $\Memfe c^2$~\rf{352}, whose splitting into a convection and a polarization part has already been specified in relativistic form through equation~\rf{356}. Turning here first to the convection part $\Mqfe c^2$
   \begin{eqnarray}
   \label{387}
   \Mqfe c^2&\doteqdot&-\frac{\hbar c}{2}\idr\qeumu\,\big(\Azomo+\Adomo\big)\;,
   \end{eqnarray}
it is immediately obvious that this object splits up in a natural way into a sum of an {\em electric}\/ ($\rightarrow$~time) part $\Mqefe c^2$ and a magnetic ($\rightarrow$~space) part $\Mqmfe c^2$, i.\,e.
   \begin{eqnarray}
   \label{388}
   \Mqfe c^2&=&\Mqefe c^2+\Mqmfe c^2\;,
   \end{eqnarray}
with the electric part being given by the time components of the four-vectors as
   \begin{eqnarray}
   \label{389}
   \Mqefe c^2&=&-\frac{\hbar c}{2}\idr\eqo\,\big(\zAo+\dAo\big)
   \end{eqnarray}
and analogously for the magnetic part in terms of the corresponding space components
   \begin{eqnarray}
   \label{390}
   \Mqmfe c^2&=&\frac{\hbar c}{2}\idr\vqe\sdot\big(\vAz+\vAd\big)\;.
   \end{eqnarray}
Clearly the non-relativistic limit of this magnetic convection part is easily written down by means of the non-relativistic convection current $\vqqe$~\rf{381} as
   \begin{eqnarray}
   \label{391}
   \Mqmfe c^2&\Rightarrow&\MMqmfe c^2\;=\;\frac{\hbar c}{2}\idr\vqqe\sdot\big(\vAz+\vAd\big)
   \end{eqnarray}
and thus compensates for the magnetic convection energy in the non-relativistic limit $\Eskine$~\rf{380} of the relativistic $\Tkine$. But also the electric convection energy $\Mqefe c^2$~\rf{389} receives a very simple approximation, namely by observing that the time component $\eqo$ of the convection current $q_{1\mu}$~\rf{353'} contains the time derivative of the first Dirac spinor $\psi_1\vr$~\rf{39} such that
   \begin{eqnarray}
   \label{392}
   \eqo\vr&=&2\left(\frac{M_1}{2M_p}-\frac{\hbar}{2M_p}\,\iAo\right)\bar{\psi}_1\vr\psi_1\vr\;\Rightarrow\; -\ephipk\vr\ephip\vr\qquad\qquad
   \end{eqnarray}
where for the non-relativistic limit one may put here $M_1\Rightarrow-M_p$. Thus the electric convection energy\rf{389} becomes in the non-relativistic limit
   \begin{eqnarray}
   \label{393}
   \Mqefe c^2&\Rightarrow&\MMqefe c^2\;=\;\frac{\hbar c}{2}\idr\iAo\vr\ephipk\vr\ephip\vr\;,
   \end{eqnarray}
which together with its magnetic counterpart $\MMqmfe c^2$ fixes then the non-relativistic limit of the total convection energy $\Mqfe c^2$~\rf{388}.

Therefore, what finally remains to be determined (for the electromagnetic mass equivalent $\Memfe c^2$~\rf{352}) is its polarization part $\Msfe c^2$~\rf{357}. But this is a simple matter when one observes that the {\em electric}\/ dipole density $\evSe$~\rf{358a} of the first Dirac field is built up by the product of positive {\em and}\/ negative Pauli spinors $\ephipm\vr$, whereas the {\em magnetic}\/ dipole density $\mvSe$~\rf{358b} contains also the product of the positive Pauli spinors $\ephip\vr$ {\em alone}\/. If now all the terms with negative Pauli spinors are dropped for the non-relativistic limit, one arrives at the following limit forms of the Dirac dipole densities \rf{358a}-\rf{358b}:
   \begin{subequations}
   \begin{align}
   \label{394a}
   \evSe&\Rightarrow0\\
   \label{394b}
   \mvSe&\Rightarrow\frac{\hbar}{2M_pc}\,\ephipk\vsigma\ephip\;,
   \end{align}
   \end{subequations}
and this then fixes the non-relativistic limit of the spin polarization energy $\Msfe c^2$~\rf{357} as
   \begin{eqnarray}
   \label{395}
   \Msfe c^2&\Rightarrow&\MMsfe c^2\;=\;\frac{\hbar^2}{4M_p}\idr\big(\vHz+\vHd\big)\sdot\big(\ephipk\vsigma\ephip\big)\;.
   \end{eqnarray}

Summarizing, the electromagnetic mass equivalent of the first particle \rf{352} becomes in the non-relativistic limit with regard of the present results \rf{391}, \rf{393} and \rf{395}:
   \begin{eqnarray}
   \label{396}
   \Memfe c^2&\equiv&\Mqefe c^2+\Mqmfe c^2+\Msfe c^2\nonumber\\
   \Rightarrow\;\MMemfe c^2&=&\frac{\hbar c}{2}\idr\iAo\vr\ephipk\vr\ephip\vr+\frac{\hbar c}{2}\idr\vqqe\sdot\vAi\nonumber\\
   &&{}+ \frac{\hbar^2}{4M_p}\idr\vHi\sdot\big(\ephipk\vsigma\ephip\big)\;.
   \end{eqnarray}
Of course, this result can be obtained also via the direct splitting of $\Memfe c^2$ into a non-relativistic electric part $\MMefe c^2$ and a non-relativistic magnetic part $\MMmfe c^2$, cf.~\rf{352}. But if this is now combined with the non-relativistic limit of the rest mass plus kinetic energy~\rf{386} into the first particle's total mass functional~\rf{383}, one finally arrives at the desired {\em Pauli energy functional}\/ $\Epfe$:
   \begin{eqnarray}
   \label{397}
   \Mfe c^2-M_pc^2&\Rightarrow&\Epfe\;=\;-\frac{\hbar^2}{2M_p}\idr\ephipk\big(\vnabla+i\vAi\big)^2\ephip\nonumber\\
   &&{}\quad+\hbar c\idr\iAo\ephipk\ephip+\frac{\hbar^2}{2M_p}\idr\vHi\sdot\big(\ephipk\vsigma\ephip\big)\;.\nonumber\\
   &&
   \end{eqnarray}
And indeed, if this functional is further modified according to the method of Lagrangean multipliers as usual to $\tEpfe$
   \begin{eqnarray}
   \label{398}
   \tEpfe&\doteqdot&\Epfe+\lambdape\NNDe\;,
   \end{eqnarray}
with the non-relativistic limit $\NNDe$ of the relativistic normalization constraint~\rf{342} being easily determined as
   \begin{eqnarray}
   \label{399}
   \NNDe&\doteqdot&\idr\ephipk\ephip-1\;\equiv\;0\;,
   \end{eqnarray}
then {\em the variational equation
   \begin{eqnarray}
   \label{3100}
   \delta\tEpfe&=&0
   \end{eqnarray}
with respect to the first Pauli spinor $\ephip$ is found to just agree with the first particle's non-relativistic equation~\rf{371a}!}\/ Of course the first Lagrange multiplier $\lambdape$ must be identified here with the first Pauli energy eigenvalue $\Epe$~\rf{372a}, i.\,e.
   \begin{eqnarray}
   \label{3101}
   \lambdape&=&-\Epe\;.
   \end{eqnarray}

It should also be a matter of course that the additivity of the relativistic mass functional $\MfT c^2$~\rf{341} will transcribe to its non-relativistic limit, i.\,e.
   \begin{eqnarray}
   \label{3102}
   \MfT c^2\hspace{-.5em}&\equiv&\hspace{-.5em}-\Mfe c^2+\Mfz c^2+\Mfd c^2\quad\Rightarrow\quad\Efp\doteqdot\Epfe+\Epfz+\Epfd\;.\nonumber\\
   &&
   \end{eqnarray}
But since the first Pauli functional $\tEpfe$~\rf{398} does depend exclusively upon the first Pauli spinor $\ephip$, but not upon the other two spinors $\zphip$ and $\dphip$, the extremalization of $\tEfp$ with respect to $\ephip$ is equivalent to the extremalization of merely the first Pauli functional $\tEpfe$~\rf{398}. Therefore it is not necessary for the variational deduction of the corresponding first non-relativistic equation~\rf{371a} to determine the other two functionals $\Epfz$ and $\Epfd$:
   \begin{eqnarray}
   \label{3103}
   \delta_1\tEfp=0&\Leftrightarrow&\delta_1\tEpfe=0\;.
   \end{eqnarray}

\subsubsection[Identical Particles]{Identical Particles\/ \rm($a=2,3$)}

However, such a variational independence cannot exist between the second and third Pauli functionals $\Epfz$ and $\Epfd$ because the corresponding eigenvalue equations \rf{371b} and \rf{371c} display a direct {\em exchange coupling}\/ between both Pauli spinors $\zphip$ and $\dphip$. Therefore one expects that the desired $\Epfz$ will depend not only upon $\zphip$ but also upon $\dphip$, and vice versa for $\Epfd$. But on the other hand, the general structure of both the original relativistic mass functionals \mbox{($a=2,3$)}
   \begin{eqnarray}
   \label{3104}
   \Mfa c^2&=&\MZa^2\cdot M_ec^2+2\Tkina+2\Memfa c^2-2\Mhgfa c^2
   \end{eqnarray}
demonstrates that they formally differ from the first functional $-\Mfe c^2$ \rf{383} merely by the presence of their exchange parts $\Mhgfa c^2$, i.\,e.
   \begin{subequations}
   \begin{align}
   \label{3105a}
   \Mhgfz c^2&=\frac{\hbar c}{2}\idr\Bo\vr\ho\vr-\frac{\hbar c}{2}\idr\vB\vr\sdot\vh\vr\;\doteqdot\;\Mhfz c^2-\Mgfz c^2\\
   \label{3105b}
   \Mhgfd c^2&=\frac{\hbar c}{2}\idr\BSo\vr\hSo\vr-\frac{\hbar c}{2}\idr\vBS\vr\sdot\vhS\vr\;\doteqdot\;\Mhfd c^2-\Mgfd c^2\;,
   \end{align}
   \end{subequations}
while the first three contributions are quite analogous to the case of the first particle \mbox{($a=1$)},cf.~\rf{383}. Therefore it is merely necessary to determine here explicitly the exchange contribution $\Mhgfa c^2$ for \mbox{$a=2,3$}.

The ``electric'' part $\Mhfa c^2$ hereof describes the exchange interaction of the electric type and is given by
   \begin{subequations}
   \begin{align}
   \label{3106a}
   \Mhfz c^2&=\frac{\hbar c}{2}\idr\Bo\vr\ho\vr\\
   \label{3106b}
   \Mhfd c^2&=\frac{\hbar c}{2}\idr\BSo\vr\hSo\vr\;,
   \end{align}
   \end{subequations}
where both contributions \mbox{($a=2,3$)} have already been proven to be identical, cf.~\rf{327}. But here it is a very simple thing to determine the corresponding non-relativistic limit, namely by recalling the general definition of the exchange current $\hmu$~\rf{224} whose time and space components \mbox{$\{\ho;-\vh\}$} read in terms of the Pauli spinors $\aphipm\vr$
   \begin{subequations}
   \begin{align}
   \label{3107a}
   \ho&=\zphipk\dphip+\zphimk\dphim\\
   \label{3107b}
   \vh&=\zphimk\vsigma\dphip+\zphipk\vsigma\dphim\;.
   \end{align}
   \end{subequations}
Thus the non-relativistic approximation $\hho$ of the time component~\rf{3107a} is immediately evident:
   \begin{eqnarray}
   \label{3108}
   \ho&\Rightarrow\hho\;\doteqdot\;\zphipk\dphip\;,
   \end{eqnarray}
which then yields for the exchange mass equivalents of the electric type \rf{3106a}-\rf{3106b}
   \begin{subequations}
   \begin{align}
   \label{3109a}
   \Mhfz c^2&\Rightarrow\MMhfz c^2\;\doteqdot\;\frac{\hbar c}{2}\idr\Bo\vr\zphipk\vr\dphip\vr\\
   \label{3109b}
   \Mhfd c^2&\Rightarrow\MMhfd c^2\;\doteqdot\;\frac{\hbar c}{2}\idr\BSo\vr\dphipk\vr\zphip\vr\;.
   \end{align}
   \end{subequations}

But somewhat more intricate is the determination of the non-relativistic limit of the exchange three-current $\vh\vr$~\rf{3107b}: Substituting here the negative Pauli spinors $\aphim\vr$ from the approximations \rf{366b}-\rf{366c} one is led to the following splitting of the exchange current $\vh$
   \begin{eqnarray}
   \label{3110}
   \vh&\Rightarrow\vhh\;=\;\vbb+\vzz\;,
   \end{eqnarray}
with the (complex-valued) convection part $\vbb$ being given by
   \begin{eqnarray}
   \label{3111}
   \vbb&=&\frac{i\hbar}{2M_ec}\left\{\big(\vnabla\zphip\big)^\dagger\dphip-\zphipk\big(\vnabla\dphip\big)\right\}\;.
   \end{eqnarray}
Clearly this is just the non-relativistic {\em linearized}\/ form of the exchange convection current $\bmu$~\rf{363a} and thus is the exchange counterpart of the electromagnetic convection currents $\vqqa$ (see, e.\,g., equation~\rf{381} for \mbox{$a=1$} or \rf{3117a}-\rf{3117b} below for \mbox{$a=2,3$}). Quite analogously to the electromagnetic polarization currents $\vsa$ (e.\,g. equation~\rf{360} for \mbox{$a=1$}), there emerges here the {\em exchange polarization current}\/ $\vzz$ as the curl of the {\em exchange polarization density}\/ $\mvZZ$
   \begin{subequations}
   \begin{align}
   \label{3112a}
   \vzz&=\vnabla\times\mvZZ\\
   \label{3112b}
   \mvZZ&=\frac{\hbar}{2M_ec}\,\zphipk\vsigma\dphip\;,
   \end{align}
   \end{subequations}
which is again the linearized non-relativistic version of \rf{363b}-\rf{364}.

Now as a consequence of that splitting of the exchange current $\vh$~\rf{3110} into a convection and polarization part, the exchange mass equivalents $\Mgfa c^2$ \rf{3105a}-\rf{3105b} do naturally split up also into a convection and a polarization part:
   \begin{subequations}
   \begin{align}
   \label{3113a}
   \Mgfz c^2&=\frac{\hbar c}{2}\idr\vB\vr\sdot\vh\vr\;\Rightarrow\;\MMgfz c^2\;\doteqdot\;\frac{\hbar c}{2}\idr\vB\vr\sdot\vbb\vr\nonumber\\
   &{}\hspace{17em}+\frac{\hbar^2}{4M_e}\idr\vY\sdot\big(\zphipk\vsigma\dphip\big)\\
   \label{3113b}
   \Mgfd c^2&=\frac{\hbar c}{2}\idr\vBS\vr\sdot\vhS\vr\;\Rightarrow\;\MMgfd c^2\;\doteqdot\;\frac{\hbar c}{2}\idr\vBS\vr\sdot\vbbS\vr\nonumber\\
   &{}\hspace{17em}+\frac{\hbar^2}{4M_e}\idr\vYS\sdot\big(\dphipk\vsigma\zphip\big)\;.
   \end{align}
   \end{subequations}
But the point with these non-relativistic limits of the magnetic exchange mass equivalents is now that they just generate the required exchange terms for the non-relativistic energy eigenvalue equations \rf{371b}-\rf{371c}, or \rf{375a}-\rf{375b}, resp., i.\,e. explicitly
   \begin{subequations}
   \begin{align}
   \label{3114a}
   -2\,\frac{\delta\big(\Mhgfz c^2\big)}{\delta\zphipk}&=\hathii\dphip\\
   \label{3114b}
   -2\,\frac{\delta\big(\Mhgfd c^2\big)}{\delta\dphipk}&=\hathiii\zphip\;.
   \end{align}
   \end{subequations}

Now that the exchange terms are clarified, we are left with the problem of writing down the kinetic and electromagnetic parts of both functionals $\Mfa c^2$, \mbox{$a=2,3$}~\rf{3104}. This problem, however, can easily be managed by simply referring to the analogous case of the first particle~\rf{383}. Namely, despite the presence of an additional exchange term \mbox{($\leadsto\vB,\vBS$)} in the approximate expressions \rf{366b}-\rf{366c} for the negative Pauli spinors $\zphim$ and $\dphim$, the rest mass and kinetic energy of both identical particles \mbox{($a=2,3$)} do combine {\em in the linear approximation}\/ to a result quite analogous to the first case~\rf{386}, i.\,e.
   \begin{subequations}
   \begin{align}
   \label{3115a}
   M_ec^2\cdot\MZz^2+2\Tkinz&\Rightarrow M_ec^2+\Ekinz-\hbar c\idr\vAii\sdot\vqqz\\
   \label{3115b}
   M_ec^2\cdot\MZd^2+2\Tkind&\Rightarrow M_ec^2+\Ekind-\hbar c\idr\vAiii\sdot\vqqd\;,
   \end{align}
   \end{subequations}
where the non-relativistic kinetic energies $\Ekina$ for both identical particles \mbox{($a=2,3$)} are of course defined quite analogously as for the first particle~\rf{382}:
   \begin{subequations}
   \begin{align}
   \label{3116a}
   \Ekinz&\doteqdot-\frac{\hbar^2}{2M_e}\idr\zphipk\big(\vnabla+i\vAii\big)^2\zphip\\
   \label{3116b}
   \Ekind&\doteqdot-\frac{\hbar^2}{2M_e}\idr\dphipk\big(\vnabla+i\vAiii\big)^2\dphip\;,
   \end{align}
   \end{subequations}
and similarly for the non-relativistic convection currents $\vqqa$ \mbox{($a=2,3$)}
   \begin{subequations}
   \begin{align}
   \label{3117a}
   \vqqz&\doteqdot-\frac{i\hbar}{2M_ec}\left\{\zphipk\big(\vnabla+i\vAii\big)\zphip- \left[\big(\vnabla+i\vAii\big)\zphip\right]^\dagger\zphip\right\}\\
   \label{3117b}
   \vqqd&\doteqdot-\frac{i\hbar}{2M_ec}\left\{\dphipk\big(\vnabla+i\vAiii\big)\dphip- \left[\big(\vnabla+i\vAiii\big)\dphip\right]^\dagger\dphip\right\}\;,
   \end{align}
   \end{subequations}
cf.~\rf{381} for the analogous case with the first particle. (However, observe here that the Gordon decomposition for the {\em identical}\/ particles \mbox{$a=2,3$} reads
   \begin{eqnarray}
   \label{3118}
   \kaumu&=&\qaumu+\saumu\;,
   \end{eqnarray}
in contrast to the case~\rf{353} with the first, {\em different}\/ particle \mbox{$a=1$}. See also below for the three-vector version~\rf{3122} hereof.)

But also the {\em electromagnetic}\/ contributions $\Memfa c^2$ of both identical particles are nothing else than copies of the first case $\Memfe c^2$~\rf{396}. However, it is very instructive to deduce this also alternatively via the original decomposition~\rf{352}, which refers to an electric ($\Mefa c^2$) and a magnetic part ($\Mmfa c^2$), rather than via the other decomposition~\rf{356} into a convection part ($\Mqfe c^2$) and a polarization part ($\Msfe c^2$). Here the electric contributions $\Mefa c^2$ appear especially simple because one merely has to approximate the Dirac densities $\ako\vr$ by the first term with the contraction of the positive Pauli spinors $\aphip\vr$, i.\,e.
   \begin{eqnarray}
   \label{3119}
   \ako&\doteqdot&\aphipk\aphip+\aphimk\aphim\;\Rightarrow\;\akko\;\doteqdot\;\aphipk\aphip\;,
   \end{eqnarray}
and this immediately yields the non-relativistic limits of the electric mass equivalents~\rf{348a} as
   \begin{subequations}
   \begin{align}
   \label{3120a}
   \Mefz c^2\;\equiv\;-\frac{\hbar c}{2}\idr\zko\vr\iiAo\vr&\Rightarrow\MMefz c^2\nonumber\\
   \MMefz&\doteqdot- \frac{\hbar c}{2}\idr\iiAo\vr\cdot\zphipk\vr\zphip\vr\nonumber\\
   &{}\\
   \label{3120b}
   \Mefd c^2\;\equiv\;-\frac{\hbar c}{2}\idr\dko\vr\iiiAo\vr&\Rightarrow\MMefd c^2\nonumber\\
   \MMefd&\doteqdot- \frac{\hbar c}{2}\idr\iiiAo\vr\cdot\dphipk\vr\dphip\vr\;.\nonumber\\
   &{}
   \end{align}
   \end{subequations}

However, the analogous treatment of the magnetic mass equivalents $\Mmfa c^2$ \rf{348b} is somewhat more complicated. First observe here that for their {\em linearized}\/ form the negative Pauli spinors $\aphim\vr$ \rf{366b}-\rf{366c} may be further approximated as follows
   \begin{subequations}
   \begin{align}
   \label{3121a}
   \zphim\vr&\Rightarrow-\frac{i\hbar}{2M_ec}\,\vsigma\sdot\big(\vnabla+i\vAii\big)\zphip\\
   \label{3121b}
   \dphim\vr&\Rightarrow-\frac{i\hbar}{2M_ec}\,\vsigma\sdot\big(\vnabla+i\vAiii\big)\dphip\;.
   \end{align}
   \end{subequations} 
Next by use of these approximations, the Dirac currents $\vka\vr$ are split up into the sum of a convection part and a polarization part
   \begin{eqnarray}
   \label{3122}
   \vka&\doteqdot&\aphimk\vsigma\aphip+\aphipk\vsigma\aphim\;\Rightarrow\;\vkka\;\doteqdot\;\vqqa+\vsa\;,
   \end{eqnarray}
where the convection parts $\vqqa$ are given by \rf{3117a}-\rf{3117b} and similarly the polarization currents are found again to be of the analogous form as for the first particle~\rf{360}, i.\,e.
   \begin{subequations}
   \begin{align}
   \label{3123a}
   \vsa&=\vnabla\times\mvSa\\
   \label{3123b}
   \mvSa&=\frac{\hbar}{2M_ec}\,\aphipk\vsigma\aphip\\
   &{}(a=2,3)\;.\nonumber
   \end{align}
   \end{subequations}
Consequently, if this splitting of the Dirac currents $\vka\vr$~\rf{3122} is substituted into the magnetic mass equivalents $\Mmfa c^2$~\rf{348b} there occurs a corresponding decomposition into two parts:
   \begin{subequations}
   \begin{align}
   \label{3124a}
   \Mmfz c^2\;=\;-\frac{\hbar c}{2}\idr\vkz\sdot\vAii&\Rightarrow\MMmfz c^2\nonumber\\
   \MMmfz&\doteqdot-\frac{\hbar c}{2}\idr\vqqz\sdot\vAii-\frac{\hbar c}{2}\idr\vHii\sdot\mvSz\nonumber\\
   &{}\\
   \label{3124b}
   \Mmfd c^2\;=\;-\frac{\hbar c}{2}\idr\vkd\sdot\vAiii&\Rightarrow\MMmfd c^2\nonumber\\
   \MMmfd c^2&\doteqdot\;-\frac{\hbar c}{2}\idr\vqqd\sdot\vAiii-\frac{\hbar c}{2}\idr\vHiii\sdot\mvSd\;.\nonumber\\
   &{}
   \end{align}
   \end{subequations}

Clearly it is not surprising that these magnetic mass equivalents of both identical particles appear very similar to the case of the first particle, cf.~\rf{396}, which then applies also to the combined electric and magnetic objects
   \begin{subequations}
   \begin{align}
   \label{3125a}
   \Memfz c^2&=\Mefz c^2-\Mmfz c^2\;\Rightarrow\;\MMemfz c^2\nonumber\\
   \MMemfz c^2&\doteqdot-\frac{\hbar c}{2}\idr\iiAo\big(\zphipk\zphip\big)+\frac{\hbar c}{2}\idr\vqqz\sdot\vAii\nonumber\\
   &{}\hspace{16em}+\frac{\hbar^2}{4M_e}\idr\vHii\sdot\big(\zphipk\vsigma\zphip\big)\\
   \label{3125b}
   \Memfd c^2&=\Mefd c^2-\Mmfd c^2\;\Rightarrow\;\MMemfd c^2\nonumber\\
   \MMemfd c^2&\doteqdot-\frac{\hbar c}{2}\idr\iiiAo\big(\dphipk\dphip\big)+\frac{\hbar c}{2}\idr\vqqd\sdot\vAiii\nonumber\\
   &{}\hspace{16em}+\frac{\hbar^2}{4M_e}\idr\vHiii\sdot\big(\dphipk\vsigma\dphip\big)\;.
   \end{align}
   \end{subequations}
This agreement of the non-relativistic limits of all {\em three}\/ electromagnetic objects $\Memfa c^2$ \mbox{($a=1,2,3$)} validates the former assertion that this limit is independent of the special decomposition of $\Memfa c^2$, i.\,e. either the decomposition with respect to an electric and a magnetic part (as shown by equation~\rf{352}) or with respect to a convection and a polarization part (as shown by equation~\rf{356}).

But the crucial point with these electromagnetic mass equivalents is now again that, through their combination with the kinetic and rest mass terms \rf{3115a}-\rf{3116b}, the convection parts ($\sim\vqqa$) do cancel so that the non-relativistic Pauli functionals $\Epfa$ of both identical particles \mbox{($a=2,3$)} as the non-relativistic limits of the mass functionals $\Mfa c^2$~\rf{3104}
   \begin{eqnarray}
   \label{3126}
   \Mfa c^2-M_ec^2&\Rightarrow&\Epfa\\
   &&\hspace{-6em}(a=2,3)\nonumber
   \end{eqnarray}
are emerging in a relatively simple shape, namely
   \begin{subequations}
   \begin{align}
   \label{3127a}
   \Epfz&=-\frac{\hbar^2}{2M_e}\idr\zphipk\big(\vnabla+i\vAii\big)^2\zphip-\hbar c\idr\iiAo\cdot\zphipk\zphip\nonumber\\
   &{}+\frac{\hbar^2}{2M_e}\idr\vHii\sdot\big(\zphipk\vsigma\zphip\big)-2\Mhgfz\\
   &{}\nonumber\\
   \label{3127b}
   \Epfd&=-\frac{\hbar^2}{2M_e}\idr\dphipk\big(\vnabla+i\vAiii\big)^2\dphip-\hbar c\idr\iiiAo\cdot\dphipk\dphip\nonumber\\
   &{}+\frac{\hbar^2}{2M_e}\idr\vHiii\sdot\big(\dphipk\vsigma\dphip\big)-2\Mhgfd\;.
   \end{align}
   \end{subequations}
Evidently, these Pauli functionals for the {\em identical}\/ particles are of the same form as that one for the {\em different}\/ particle~\rf{397}, apart from the additionally emerging exchange mass equivalents $\Mhgfa c^2$~\rf{3105a}-\rf{3105b} which can of course occur only for identical particles. But with the present Pauli energy functionals \rf{3127a}-\rf{3127b} our final goal has now actually been attained; namely, {\em the variational equations
   \begin{subequations}
   \begin{align}
   \label{3128a}
   \delta\tEpfa&=0\\
   \label{3128b}
   \tEpfa&\doteqdot\Epfa+\lambdapa\NNDa\\
   &{}\hspace{1em}(a=2,3)\nonumber
   \end{align}
   \end{subequations}
turn out to be identical to just those non-relativistic Pauli equations}\/ \rf{371b}-\rf{371c}. Here the Lagrangean multipliers $\lambdapa$ are to be identified again with the Pauli energy eigenvalues \rf{372b}-\rf{372c}:
   \begin{eqnarray}
   \label{3129}
   \lambdapa&\doteqdot&-\Epa\;,
   \end{eqnarray}
and the non-relativistic approximation to the normalization constraint~\rf{342} is of course given by
   \begin{eqnarray}
   \label{3130}
   \NNDa&\doteqdot&\idr\aphipk\aphip-1\;\equiv\;0\;.
   \end{eqnarray}

Summarizing the results, the relativistic mass eigenvalue equations \rf{340a}-\rf{340c} actually do admit the interpretation of being just the variational equations due to the mass functional $\tMfT c^2$~\rf{350}, which is the sum of the three individual mass functionals $\tMfa c^2$ \mbox{($a=1,2,3$)}. This sum is decoupled in the sense that the $a$\/th eigenvalue equation of the system \rf{340a}-\rf{340c} is the variational equation (with respect to the $a$\/th Pauli spinor $\aphipmk$) of the $a$\/th mass functional $\tMfa c^2$. The non-relativistic limit of this mass eigenvalue system is given by equations \rf{371a}-\rf{371c}, which can be obtained in two different ways: \mbox{{\bf (i)} either} one eliminates the negative Pauli spinors $\aphim\vr$ directly from the relativistic eigenvalue equations by means of the approximations \rf{366a}-\rf{366c}, or \mbox{{\bf (ii)} one} first eliminates the negative Pauli spinors from the relativistic mass functional $\tMfT c^2$ in order to obtain its non-relativistic limit, i.\,e. the Pauli energy functional $\tEfp$~\rf{3102}, and afterwards one looks for the variational equations due to this non-relativistic functional $\tEfp$. Thus the logical scheme sketched at the beginning of subsection~{\bf\ref{s34}} of the present section has actually been validated.

But though this coincidence of the results (obtained along different routes) is very satisfying, one nevertheless cannot be contented with this situation since it is obvious that the gauge field equations (e.\,g. the Poisson equations \rf{33}-\rf{34}) cannot be deduced via a variational procedure from that relativistic mass functional $\tMfT c^2$ nor from its non-relativistic approximation $\tEfp$! The reason is here that both the mass functional $\tMfT c^2$ and its non-relativistic approximation $\tEfp$ do not contain any {\em derivative}\/ of the gauge fields! Therefore, if we wish to possess a (relativistic or non-relativistic) energy functional, whose variational equations are required to embrace {\em both}\/ the matter {\em and}\/ gauge field equations, then we have to resort now to a new construction: the {\em RST energy functional}\/ ($\tEfT$). However, the mass and Pauli functionals discussed so far will thereby not become obsolete, but on the contrary will be of valuable help for constructing the desired RST functional $\tEfT$.

\section{RST Energy Functional}\label{s4}
\indent 

\textrm After it has become sufficiently clear now that the mass functional $\tMfT c^2$ cannot directly serve as the desired energy functional $\tEfT$, one is motivated to turn back to the total RST energy $\ET$~\rf{251}. Indeed, this object contains both the matter energy $\ED$~\rf{253a} {\em and}\/ the gauge field energy $\EG$~\rf{253b}, which is a necessary condition for the variational deduction of both the matter {\em plus}\/ gauge field equations from the corresonding functional $\EfT$. Nevertheless, this total energy functional $\EfT$ cannot be immediately adopted as the desired goal. The reason is that the matter and gauge fields are coupled in such a way that the mass eigenvalue equations plus Poisson equations do {\em not}\/ appear as the corresponding variational equations! On the other hand, the definition of the field energy $\ET$~\rf{251} as the integral over the energy density $\TToo\vr$ appears to be very natural and absolutely consistent with all that one knows about the logical structure of the successful field theories in theoretical physics. Consequently, when one wishes to modify the original RST functional $\EfT$ in order to meet with the complete set of field equations, then one must try to admit only those modifications which do preserve the numerical value of the original $\ET$. A special class of such modifications consists obviously in adding to the original functional $\EfT$ some constraint which is {\em automatically}\/ obeyed by all solutions of the RST eigenvalue problem. Indeed, we will readily demonstrate that constraints of this kind are provided by just those {\em Poisson identities}\/, cf. equations \rf{335}-\rf{336}, but not by the {\em exchange identities}\/ \rf{333}-\rf{334}, so that the inclusion of the exchange forces requires a slight modification of the original proposal $\ET$~\rf{251}.

\subsection{Physical Meaning of the Poisson Identities}

The role played by the Poisson identities becomes elucidated by a closer inspection of the splitting of the total field energy $\EfT$~\rf{252}. Its matter part $\EfD$~\rf{253a} has already been shown to be the sum of three single-particle contributions $\EDfa$ \mbox{($a=1,2,3$)}
   \begin{eqnarray}
   \label{41}
   \EfD&=&\EDfe+\EDfz+\EDfd\;,
   \end{eqnarray}
which then adds up to the total mass functional $\MfT c^2$ minus the mass equivalents of the electric and exchange type~\cite{c8,c11}
   \begin{eqnarray}
   \label{42}
   \EfD&=&\MfT c^2-2\big(\Me c^2-\Mh c^2\big)\;.
   \end{eqnarray}
Recall here that the total mass functional has already been specified by equation~\rf{343} so that the matter energy~\rf{42} appears now in the following (relativistic) form
   \begin{eqnarray}
   \label{43}
   \EfD&=&\MZe^2\cdot M_pc^2+\MZz^2\cdot M_ec^2+\MZd^2\cdot M_ec^2\nonumber\\
   &&{}+2\Tkin-2\big(\Mm c^2-\Mg c^2\big)\;.
   \end{eqnarray}
This is a very plausible result because it says that the energy of matter consists of the (renormalized) rest mass energies of the three particles (first line on the right-hand side of~\rf{43}) plus the proper kinetic energy of motion $2\Tkin$ minus the {\em magnetic}\/ mass equivalents of the electromagnetic \mbox{($\sim\Mm$)} and exchange \mbox{($\sim\Mg$)} type. Observe here that the {\em electric}\/ mass equivalents \mbox{($\sim M^{\rm(\!\!\:e,h\!\!\:)\!}$)} do not contribute to the matter energy $\ED$~\rf{43}, quite in contrast to the mass equivalents of the magnetic type \mbox{($\sim M^{\rm(\!\!\:m,g\!\!\:)\!}$)} which represent here the field-theoretic analogue of the well-known ``minimal substitution'' \mbox{($\vec{p}\rightarrow\vec{p}+\hbar\vec{A}$)} for the coupling of a spinless point particle of momentum $\vec{p}$ \mbox{($\Rightarrow\frac{\hbar}{i}\vnabla$)} to a three-vector potential $\vec{A}$, see the kinetic energies $\Ekina$~\rf{382} and \rf{3116a}-\rf{3116b}.

But concerning the Poisson identities, they are rather connected with the gauge field contribution $\EG$~\rf{253b} of $\ET$. This is better realized by splitting up $\EG$ into its electromagnetic and exchange parts, i.\,e.
   \begin{eqnarray}
   \label{44}
   \EG&=&\ER-\EC\;,
   \end{eqnarray}
where the energy content $\ER$ of the real modes \mbox{($\sim\Aaomu$)} is built up again by the electric (e) and magnetic (m) contributions~\cite{c8}
   \begin{subequations}
   \begin{align}
   \label{45a}
   \ER&=\ERe+\ERm\\
   \label{45b}
   \ERe&=\frac{\hbar c}{4\pi\als}\idr\left[\vEe\sdot\vEz+\vEz\sdot\vEd+\vEd\sdot\vEe\right]\\
   \label{45c}
   \ERm&=\frac{\hbar c}{4\pi\als}\idr\left[\vHe\sdot\vHz+\vHz\sdot\vHd+\vHd\sdot\vHe\right]\;,
   \end{align}
   \end{subequations}
and similarly for the complex modes \mbox{($\sim\Bmu$)}
   \begin{subequations}
   \begin{align}
   \label{46a}
   \EC&=\ECh+\ECg\\
   \label{46b}
   \ECh&=\frac{\hbar c}{4\pi\als}\idr\vXS\sdot\vX\\
   \label{46c}
   \ECg&=\frac{\hbar c}{4\pi\als}\idr\vYS\sdot\vY\;.
   \end{align}
   \end{subequations}
Observe here the important fact that the exchange energy $\EC$~\rf{46a}-\rf{46c} is always positive \mbox{($\EC\geq0$)} but enters the gauge field energy $\EG$~\rf{44} with a negative sign. This implies that it becomes energetically more favourable to undergo the {\em exchange interactions}\/ which, on the other hand, is possible only for {\em identical}\/ particles if the basic conservation laws are to be respected, see below!

The relationship between the Poisson identities and the various contributions \rf{45a}-\rf{46c} becomes now immediately evident through simply recalling the definitions of the field strengths in terms of the potentials, cf.~\rf{315a}-\rf{315d}. For instance, consider the electric Poisson identity~\rf{335} and find that this identity expresses nothing else than just the identity of the electric gauge field energy $\ERe$~\rf{45b} and the electric mass equivalent $\Me c^2$~\rf{348a}, i.\,e.
   \begin{eqnarray}
   \label{47}
   \NGe&=&\ERe-\Me c^2\;\equiv\;0\;.
   \end{eqnarray}
Or in other words, the electrostatic gauge field energy $\ERe$ equals the electrostatic interaction energy $\Me c^2$ of the matter field modes with the gauge field modes. In a similar way, it is easily seen that the physical content of the magnetostatic Poisson identity~\rf{336} consists in the identity of the magnetic gauge field energy $\ERm$~\rf{45c} and the magnetic interaction energy $\Mm c^2$~\rf{348b} of the matter and gauge field modes, i.\,e.
   \begin{eqnarray}
   \label{48}
   \NGm&=&\ERm-\Mm c^2\;\equiv\;0\;.
   \end{eqnarray}

However, the situation with the analogous exchange identities is somewhat more intricate. Namely, the field equations~\rf{310a}-\rf{310b} for the exchange potentials $\Bo$, $\vB$ are not of the simple Poisson form but rather contain a ``mass term'' \mbox{($\sim\am^{-2}$)} and therefore do rather resemble the static form of the Klein-Gordon equations. When this fact is combined with the former integral relations~\rf{331} one arrives at the exchange counterparts of the electromagnetic identities~\rf{47}-\rf{48} in the following form:
   \begin{subequations}
   \begin{align}
   \label{49a}
   \NGh&=\ECh+\frac{\hbar c}{4\pi\als}\cdot\frac{1}{\am^2}\idr\BSmo\Bmu-\Mh c^2\;\equiv\;0\\
   \label{49b}
   \NGg&=\ECg+\frac{\hbar c}{4\pi\als}\cdot\frac{1}{\am^2}\idr\BSmo\Bmu-\Mg c^2\;\equiv\;0\;.
   \end{align}
   \end{subequations}
This result says that the energy content of the exchange field modes (of the electric (h) and magnetic (g) types) differs from the corresponding exchange mass equivalents $\Mh c^2$~\rf{327} and $\Mg c^2$~\rf{328} by the spatial integral of the Lorentz invariant $\BSmo\Bmu$. Thus it is only for states with infinite exchange length $\am$~\rf{37} that the exchange energies $\ECh$ and $\ECg$~\rf{46b}-\rf{46c} can coincide with the corresponding mass equivalents $\Mh c^2$ and $\Mg c^2$.

This non-identity (for finite $\am$) of the exchange energies and their mass equivalents will readily force us to modify the RST field energy $\ET$~\rf{251} in order to obtain a physically reasonable energy functional $\tEfT$. 

\subsection{Relativistic Variational Principle}

The construction of the desired relativistic energy functional $\tEfT$ has already been described in great detail for a system of two {\em different}\/ particles (i.\,e. positronium, ref.\,\cite{c9}). Therefore it may be sufficient here to briefly comment on the analogous construction for the present three-particle system (of two identical and one different particle). Namely, the principal idea refers again to the modification of the original RST energy $\ET$~\rf{251} by means of the aforementioned Poisson and exchange {\em identities}\/ so that the numerical value of the total energy is preserved (as far as possible).

In the first step, one puts together the total energy $\EfT$~\rf{252}-\rf{253b} from the matter $\EfD$~\rf{42} and the gauge field energy $\EfG$~\rf{44}-\rf{46c} in order to find the first tentative proposal $\EifT$ in the following form:
   \begin{eqnarray}
   \label{410}
   \EifT&=&\MfT c^2-2\big(\Me c^2-\Mh c^2\big)+\ERe+\ERm-\ECh-\ECg\;.\hspace{3em}
   \end{eqnarray}
Next, the mass equivalents of the electric type are eliminated from this first proposal just by means of both identities~\rf{47} and \rf{49a} of the electric type, which then yields the second tentative form as
   \begin{eqnarray}
   \label{411}
   \EiifT&=&\MfT c^2-\ERe+\ERm+\ECh-\ECg+\frac{\hbar c}{4\pi\als}\cdot\frac{2}{\am^2}\idr\BSmo\Bmu\,.\hspace{4em}
   \end{eqnarray}
From this intermediate result a third proposal is obtained now by substituting here the mass functional $\MfT c^2$ from equation~\rf{343}, which brings the total RST energy functional $\EfT$ to its final form $\EiiifT$
   \begin{eqnarray}
   \label{412}
   \EiiifT&=&\MZe^2\cdot M_pc^2+\MZz^2\cdot M_ec^2+\MZd^2\cdot M_ec^2+2\Tkin\nonumber\\
   &&{}+\ERe-\ERm-\ECh+\ECg+\frac{\hbar c}{4\pi\als}\cdot\frac{2}{\am^2}\idr\BSmo\Bmu\;.\hspace{2em}
   \end{eqnarray}

Observe here that the numerical value of the corresponding functional $\EiiifT$ upon the solutions of the RST eigenvalue problem will {\em exactly}\/ reproduce the RST energy $\ET$ in its original form~\rf{251}! The reason is that all those manipulations leading us from that initial form~\rf{251} to the present $\EiiifT$~\rf{412} are {\em exactly}\/ valid for the RST solutions. Furthermore, the latter result~\rf{412} appears to be very plausible because the first line refers to the matter subsystem and thus specifies its rest mass and kinetic energy while the second line obviously consists in the energy of the gauge field subsystem.

But now comes a subtle point which just refers to this numerical equivalence of the modified functional $\EiiifT$~\rf{412} and its original predecessor $\ET$~\rf{251}. Namely, a very natural expectation is that the gauge field subsystem should contribute its part to the total energy $\ET$ via the field strengths (i.\,e. $\vE,\vH;\vX,\vY$), but not via the potentials (i.\,e. $\aAo,\vAa;\Bo,\vB$). But obviously, this expectation is violated by the last term on the right-hand side of~\rf{412}, which is built up exclusively by the exchange potentials $\Bmu$. On the other hand, the presence of that questionable term is indispensable from the mathematical viewpoint because it is a consequence of the rigorous use of the Poisson and exchange identities, which ensure the numerical equivalence of $\ET$~\rf{251} and $\EiiifT$~\rf{412}. {\em Thus the mathematical rigour runs into conflict with the physical intuition.}\/ In order to escape from this dilemma, one needs some convincing argument which settles the conflict either in favour of physics or in favour of mathematics. Here it seems wise to take the pragmatic viewpoint; and this means that one lets settle the question through the circumstance wether the stationary gauge field equations~\rf{33}-\rf{34} and \rf{310a}-\rf{310b} can be deduced from the functional $\EiiifT$~\rf{412}, either {\em with}\/ the presence {\em or}\/ the omission of the questionable potential term. Anticipating the result, physics gets the better of both and the potential term must be omitted, i.\,e. we adopt the desired energy functional ($\EivfT$, say) to be of the following form
   \begin{eqnarray}
   \label{413}
   \EivfT\hspace{-.7em}&=&\hspace{-.7em}\MZe^2\cdot M_pc^2+\MZz^2\cdot M_ec^2+\MZd^2\cdot M_ec^2+2\Tkin+\ERe-\ERm-\ECh+\ECg\;,\nonumber\\
   &&
   \end{eqnarray}
otherwise the exchange equations~\rf{310a}-\rf{310b} could not be identified as the variational equations due to the corresponding energy functional.

Nevertheless, the present result~\rf{413}, if interpreted as a functional $\EivfT$ acting over the configuration space of matter and gauge fields $\{\aphipm\vr$, $\Aaomu\vr$, $\Bmu\vr\}$, will not yet (but almost) reproduce the RST eigenvalue system. Recall here that the latter system consists of the Poisson equations~\rf{33}-\rf{34} for the real gauge field modes $\Aaomu$, the exchange equations~\rf{310a}-\rf{310b} for the complex modes $\Bmu$, and also of the mass eigenvalue equations~\rf{340a}-\rf{340c} for the Pauli spinors $\aphipm\vr$. However, in order that the members of this coupled system of matter and gauge fields do actually emerge as the variational equations of an RST energy functional ($\tEfT$, say) one merely has to complement the present result $\EivfT$~\rf{413} by certain constraints via the method of Lagrangean multipliers. Clearly, the first one of these constraints must refer to the relativistic {\em normalization condition}~\rf{342}. But also the {\em exchange identities}\/~\rf{333}-\rf{334} and {\em Poisson identities}\/~\rf{335}-\rf{336} must be adopted as constraints for the variational procedure. Thus one ultimately arrives at the desired energy functional $\tEfT$ in the following form:
   \begin{eqnarray}
   \label{414}
   \tEfT&=& \EivfT+\sum_{a=1}^3\lambDa\cdot\NDa+\lambdaGe\cdot\NGe+\lambdaGm\cdot\NGm+\lambdaGh\cdot\NGh+\lambdaGg\cdot\NGg\;.\nonumber\\
   &&
   \end{eqnarray}

And indeed, it is now a very instructive and enlightening exercise to identify the whole set of RST field equations as just the variational equations due to that functional $\tEfT$:
   \begin{enumerate}
   \item the relativistic {\em mass eigenvalue equations}\/~\rf{340a}-\rf{340c} appear as the variational equations with respect to the Pauli spinors $\aphipm\vr$ (or their Hermitean conjugates $\aphipmk\vr$, resp.)
      \begin{eqnarray}
      \label{415}
      \frac{\delta\tEfT}{\delta\aphipm}&=&0\;.
      \end{eqnarray}
   Here the Lagrangean multipliers $\lambDa$ are due to the normalization constraints and are to be identified again with the mass eigenvalues $M_a$~\rf{39}, quite analogously to the case of the mass functional~\rf{351a}-\rf{351c}:
      \begin{subequations}
      \begin{align}
      \label{416a}
      \lambDe&=M_1c^2\\
      \label{416b}
      \lambDz&=-M_2c^2\\
      \label{416c}
      \lambDd&=-M_3c^2\;.
      \end{align}
      \end{subequations}
   \item the {\em electromagnetic Poisson equations}\/~\rf{33}-\rf{34} turn out as the variational equations with respect to the gauge potentials $\aAo$, $\vAa$. For the link between the Maxwellian charge and current densities $\ajo$, $\vja$ and the Dirac densities $\ako$, $\vka$ see equations~\rf{240a}-\rf{240e}. The Lagrangean multipliers due to the Poisson constraints must adopt the following values:
      \begin{subequations}
      \begin{align}
      \label{417a}
      \lambdaGe&=-2\\
      \label{417b}
      \lambdaGm&=2\;.
      \end{align}
      \end{subequations}   
   \item the {\em exchange equation}\/~\rf{310a} of the electric type is found to play the role of the variational equation of $\tEfT$ with respect to the variation of the exchange potential $\Bo$. Here the Lagrangean multipliers $\lambdaGh$ and $\lambdaGg$ due to the exchange constraints~\rf{333}-\rf{334} must be chosen in the following way:
      \begin{subequations}
      \begin{align}
      \label{418a}
      \lambdaGh&=2\\
      \label{418b}
      \lambdaGg&=-2\;,
      \end{align}
      \end{subequations}
   which completes the set of Lagrangean multipliers.
   \item the {\em exchange equation}~\rf{310b} of the magnetic type is found to appear as the variational equation of $\tEfT$ with respect to the variation of the magnetic exchange field $\vB$. Both exchange equations~\rf{310a}-\rf{310b} represent the crucial point for the preceding discussion of the potential term \mbox{($\sim\BSmo\Bmu$)} in the rejected energy functional $\EiiifT$~\rf{412}. Indeed, it is easy to see that both the electromagnetic Poisson equations~\rf{33}-\rf{34} and the mass eigenvalue equations~\rf{340a}-\rf{340c} are insensitive to the presence or omission of that questionable potential term. But surely it is rather satisfying that one has to {\em omit}\/ this term on two mutually supporting grounds: namely, (i) physical plausibility and (ii) variational deducibility of the exchange equations~\rf{310a}-\rf{310b}. However unfortunately, the original definition~\rf{251} of the RST energy $\ET$ must be abandoned to a certain extent under this compulsion of making a decision in favour of physical plausibility.    
   \end{enumerate}

Observe here that the energy functional $\tEfT$~\rf{414} consists of two rather different contributions, namely of the {\em physical}\/ contribution $\EivfT$~\rf{413}, which is the sum of all physical energies (i.\,e. rest mass, kinetic, electromagnetic, exchange energy) and of the {\em constraints}\/ (i.\,e. wave function normalization, Poisson and exchange identities). For the practical applications, where one may resort to appropriate trial functions for the RST fields (wave functions and gauge potentials), one will therefore select the trial functions in such a way that all the constraints are obeyed automatically so that it is sufficient to extremalize the physical energy $\EivfT$~\rf{413} rather than the constrained energy $\tEfT$~\rf{414}.

\subsection{Non-Relativistic Principle}

For the concrete applications, e.\,g. in the field of atomic and molecular physics, one can frequently be satisfied with the non-relativistic approximation of the desired predictions. Thus if the {\em RST principle of minimal energy}\/ is selected as the preferred variational method, one will be interested also in its non-relativistic limit. Clearly, one expects here that the non-relativistic form of the variational technique can be handled more easily than its relativistic version. In this sense one wishes to see now the non-relativistic approximation of the present energy functional $\tEfT$~\rf{414}. But naturally for this purpose it is merely necessary to look for the non-relativistic forms of the individual contributions of $\tEfT$.

Turning first to the physical contribution $\EivfT$~\rf{413} of $\tEfT$~\rf{414}, one recalls that the particle rest masses ($M_pc^2$ and $M_ec^2$) are to be neglected for the non-relativistic approach and thus the matter contribution to $\EiiifT$ (first line on the right-hand side of~\rf{412}) is simply replaced by the non-relativistic kinetic energy $\Ekin$ as the sum of the three single-particle energies $\Ekina$
   \begin{eqnarray}
   \label{419}
   \Ekin&\doteqdot&\Ekine+\Ekinz+\Ekind
   \end{eqnarray}
as defined by equations~\rf{382} and \rf{3116a}-\rf{3116b}. Next, consider the energy content of the gauge field subsystem, which is given by the second line on the right-hand side of equation~\rf{412}. Since the gauge field energy $\EG$~\rf{44} is defined exclusively in terms of the field strengths, see equations~\rf{45a}-\rf{46c}, this energy contribution remains formally invariant when passing over to the non-relativistic limit since this limit concerns essentially the motion of matter, but not directly the mechanism of generation of the gauge fields (Maxwell equations). Therefore the non-relativistic approximations do refer preferably to the matter subsystem, which (apart from the kinetic energies) enters the energy functional $\tEfT$~\rf{414} via the constraints of wave function normalization $\NDa$~\rf{342}, Poisson identities $\NGe$ and $\NGm$~\rf{335}-\rf{336}, and exchange identities $\NGh$ and $\NGg$~\rf{333}-\rf{334}. Here the non-relativistic form $\NNDa$ of the original wave function normalizations $\NDa$ has already been specified through equations~\rf{3130} so that we are left now with the task to present also the non-relativistic shape of the Poisson and exchange identities.

Since, however, these identities are built up by the gauge field energies $\{\ERe,\ERm,\ECh,\ECg\}$ and their mass equivalents $\{\Me c^2,\Mm c^2,\Mh c^2,\Mg c^2\}$, see equations~\rf{47}-\rf{49b}, it becomes merely necessary to look for the non-relativistic forms of these mass equivalents:
   \begin{subequations}
   \begin{align}
   \label{420a}
   \NGe&\Rightarrow\NNGe\;\doteqdot\;\ERe-\MMe c^2\;\equiv\;0\\
   \label{420b}
   \NGm&\Rightarrow\NNGm\;\doteqdot\;\ERm-\MMm c^2\;\equiv\;0\\
   \label{420c}
   \NGh&\Rightarrow\NNGh\;\doteqdot\;\frac{\hbar c}{4\pi\als}\idr\Big[\big(\vnabla\BSo\big)\sdot\big(\vnabla\Bo\big)-\frac{1}{\am^2}\,\BSo\Bo\Big]-\MMh c^2\;\equiv\;0\\
   \label{420d}
   \NGg&\Rightarrow\NNGg\;\doteqdot\;\ECg+\frac{\hbar c}{4\pi\als}\cdot\frac{1}{\am^2}\idr\BSmo\Bmu-\MMg c^2\;\equiv\;0\;.
   \end{align}
   \end{subequations}
Here the desired non-relativistic expressions have already been specified by equations~\rf{3120a}-\rf{3120b} and \rf{3124a}-\rf{3124b} for the electromagnetic mass equivalents $\MMefa c^2$ and $\MMmfa c^2$; and similarly the exchange mass equivalents $\MMhfa c^2$ and $\MMgfa c^2$ are given by equations~\rf{3109a}-\rf{3109b} and \rf{3113a}-\rf{3113b}. All these results can now be put together for the non-relativistic version ($\tEEfT$, say) of the relativistic energy functional $\tEfT$~\rf{414}, which then ultimately appears in the following form:
   \begin{eqnarray}
   \label{421}
   \tEEfT&=&\Ekin+\ERe+\ERm-\ECh+\ECg\nonumber\\
   &&{}+\hbar c\idr\!\left[\vHi\sdot\mvSe+\vHii\sdot\mvSz+\vHiii\sdot\mvSd\right]\nonumber\\
   &&{}+\sum_{a=1}^3\lambdapa\cdot\NNDa+\lambdaGe\cdot\NNGe+\lambdaGh\cdot\NNGh+\lambdaGg\cdot\NNGg\;.\hspace{3em}
   \end{eqnarray}

Surely this is a plausible result from the physical point of view because it says that the non-relativistic energy $\EET$ of any RST field configuration is composed of essentially three constituents: \mbox{{\bf (i)} the} first line represents the sum of the kinetic energy $\Ekin$ of matter plus all four kinds of the gauge field energy, \mbox{{\bf (ii)} the} second line is the interaction energy of the individual magnetic dipole densities $\mvSa$ and the magnetic fields $\vHiiiiii$ of the {\em other}\/ particles \mbox{($\leadsto$ no} magnetic self-interactions); and finally, \mbox{{\bf (iii)} the} third line is the collection of constraints. Observe here that the magnetic constraint $\NNGm$~\rf{420b} does not explicitly emerge in the third (i.\,e. constraint) line because it has been eliminated in favour of the dipole interaction term (second line). As a consequence, the magnetic field energy $\ERm$ appears now with a positive sign in the non-relativistic functional $\tEEfT$~\rf{421}, in contrast to the relativistic functional~\rf{413}.

Now it is again a nice consistency check to deduce the non-relativistic RST equations from that functional $\tEEfT$~\rf{421}:
   \begin{enumerate}
   \item the non-relativistic {\em energy eigenvalue equations}\/~\rf{371a}-\rf{371c} are obtained through the extremalization of $\tEEfT$ with respect to the positive Pauli spinors $\aphip$ \mbox{($a=1,2,3$)};
   \item the {\em electric Poisson equations}\/~\rf{33} are obtained through extremalization of $\tEEfT$ with respect to the electric gauge potentials $\aAo$, where merely the relativistic charge densities $\ajo\vr$ (or $\ako$, resp.) are to be replaced by their non-relativistic forms $\akko\vr$, i.\,e. concretely
      \begin{eqnarray}
      \label{422}
      \ako&\doteqdot&\aphipk\aphip+\aphimk\aphim\;\Rightarrow\;\akko\;\doteqdot\;\aphipk\aphip\;;
      \end{eqnarray}
   \item the {\em magnetic Poisson equations}\/~\rf{34} emerge through extremalization of $\tEEfT$ with respect to the magnetic gauge potentials $\vAa$, where the relativistic current densities $\vja\vr$ (or $\vka$, resp.) have merely to be replaced by their non-relativistic forms $\vkka$, see equation~\rf{3122} for \mbox{$a=2,3$} ( and similarly for \mbox{$a=1$)};
   \item the {\em electric exchange equations}\/~\rf{310a} arise through the extremalization of $\tEEfT$ with respect to the electric exchange potential $\Bo$ (or $\BSo$, resp.), where the exchange density $\ho\vr$ is to be replaced by its non-relativistic version $\hho\vr$, see equation~\rf{3108};
   \item the {\em magnetic exchange equation}~\rf{310b} is obtained through extremalization of $\tEEfT$ with respect to the magnetic exchange potential $\vB$ (or $\vBS$, resp.). Clearly, the exchange current $\vh$ is to be replaced here again by its non-relativistic form $\vhh$~\rf{3110}.
   \end{enumerate}

Summarizing, one actually has achieved the original goal, namely to construct a relativistic energy functional $\tEfT$ (together with its non-relativistic approximation $\tEEfT$) from the intrinsic logic of RST, such that the solutions of the coupled set of (non-)relativistic RST field equations do just represent the stationary points of those functionals!

\section{Fermionic and Bosonic States}\label{s5}
\indent

Naturally, the establishment of a new variational principle must provoke intense endeavours in order to either falsificate the new theoretical structure or to validate it by testing it for various applications and thus elaborating its logical consistency and practical usefulness. Especially one first wishes to become convinced that the predictions of the new formalism do coincide with those of the conventional theory for some of the standard situations where the results are already well known and in perfect agreement with the observations. In this sense, we select the hydrogen state $\zpdh$ for a comparison of the conventional and RST treatment. This one-particle example can be solved exactly in both the conventional theory \cite{c18} and in RST; and it should not come as a surprise that the results of both approaches do exactly coincide. For instance, the energy ($\Econv$) of the quantum state is identified with the mass eigenvalue ($M_2c^2$), see equation~\rf{535} below. On the other hand, the proposed RST energy $\tEfT$~\rf{414} is composed of various contributions due to the matter and gauge fields to be complemented by the constraints. But despite this different physical status of both competing quantities $\Econv$ and $\tEfT$, their values upon the selected state $\zpdh$ do exactly coincide, see equation~\rf{553} below. Moreover, the RST functional $\tEfT$ meets also with the plausible expectation that for the non-relativistic limit there should occur some kind of {\em spin degeneracy}\/. The reason for this is that the spin-orbit coupling must disappear in the lowest-order approximation where the spin effect upon the energy levels is neglected. Subsequently it is demonstrated that the RST formalism does correctly account for this non-relativistic spin degeneracy.

But clearly, such a coincidence of the conventional and RST predictions for the one-particle systems can not yet ensure the general confidence in the claimed {\em RST principle of minimal energy}\/; but rather one would like to see this principle working well also in the field of the few-particle systems (or even many-particle systems). Therefore, in order to present a less trivial situation, we consider positronium as a typical but sufficiently simple few-particle problem. More concretely, we demonstrate the working of the {\em principle of minimal energy}\/ as an approximative technique for calculating the energy of the bound few-particle systems (albeit only in the non-relativistic regime). The treatment of such a two-body problem is surely a more rigorous test of the {\em principle of minimal energy}\/ since in RST the two positronium constituents (i.\,e. electron and positron) must occupy {\em bosonic states}\/ \cite{c11}; and if any physical relevance is claimed for that principle, it must be shown to produce meaningful results also for such exotic quantum states with non-zero boson number. However, in order to not plunge here too deeply in the tedious calculations, we are satisfied with a treatment of the state $\zePe$ in the non-relativistic limit to be combined with the spherically symmetric approximation. It turns out that already the choice of a simple trial function with only two variational parameters $\beta$ and $\nu$ (see equation~\rf{5125} below) is sufficient in order to predict the (non-relativistic) energy of the state $\zePe$ up to an accuracy of 10\% (see the figure on p.\,\pageref{f1}). This is the same magnitude of deviation as for the analogous groundstate predictions in ref.~\cite{c11} and thus hints on the hypothesis that the non-relativistic RST predictions may come very close to their conventional counterparts (or even coincide with them) if more subtle approximation techniques were applied. 

\subsection{Selection of Spinor Basis}

The Pauli spinors $\aphipm\vr$~\rf{339} are elements of a two-dimensional complex linear vector space which may be conceived as a representation space of the orthogonal group $SO(3)$ (or its covering group $SU(2)$, resp.). According to the specific transformation law under this group, those elements are called ``spinors''. Any such spinor can be described in terms of its components relative to some spinor basis whose selection is arbitrary on principle, but an intelligent choice mostly facilitates the managing of the concrete problems. Since the Dirac wave functions $\psi_a$~\rf{339} appear as the direct sum of two Pauli spinors $\aphipm$, one can select a separate 2-basis for any Pauli 2-space. Especially for the description of the present groundstate situation one first starts with the {\em $\zeta$ basis}\/ $\{\zopp,\zopm\}$ for the ``positive'' Pauli component and its concomitant $\{\zepp,\zepm\}$ for the ``negative'' component, see ref.\,\cite{c11}. Each of these two basis systems consists of eigenspinors for total angular momentum $\hvJ$ \mbox{($\doteqdot\hvL+\hvS$)} such that
   \begin{subequations}
   \begin{align}
   \label{51a}
   \hvJ^{\;2}\,\zjml&=\hbar^2\,j(j+1)\,\zjml\\
   \label{51b}
   \hJz\,\zjml&=m\hbar\,\zjml\\
   \label{51c}
   \hvL^{\,2}\,\zjml&=\hbar^2\,l(l+1)\,\zjml\\
   \label{51d}
   \hvS^{\,2}\,\zjml&=\hbar^2\,s(s+1)\,\zjml\;.
   \end{align}
   \end{subequations}

For the positronium groundstate to be treated here one adopts the quantum numbers $j=\frac{1}{2}$, $l=0$, $m=\pm\frac{1}{2}$ for the positive Pauli component and $j=\frac{1}{2}$, $l=1$, $m=\pm\frac{1}{2}$ for the negative Pauli component ($s=\frac{1}{2}$ in any case). However, we do not use directly this basis system~\rf{51a}-\rf{51d} but we equip it with a further phase degree of freedom by using the {\em ``\/$\omega$ basis''}\/ which is related to the above mentioned {\em ``$\zeta$ basis''}\/ through \cite{c11}
   \begin{subequations}
   \begin{align}
   \label{52a}
   \wop&={\rm e}^{-i\bbar\phi}\cdot\zopp\\
   \label{52b}
   \wom&={\rm e}^{i\bbar\phi}\cdot\zopm\\
   \label{52c}
   \wep&={\rm e}^{-i\bbar\phi}\cdot\zepp\\
   \label{52d}
   \wem&={\rm e}^{i\bbar\phi}\cdot\zepm\;.
   \end{align}
   \end{subequations}
Here $\bbar$ is a real constant to be fixed later on by means of additional requirements. This constant $\bbar$ may (or may not) be different for different particles (i.\,e. $\bbar\rightarrow \bbar_p$ for the positively charged particle ($a=1$) and $\bbar\rightarrow \bbar_e$ for the electron ($a=2$)). The new basis elements are still eigenspinors of the $z$ component ($\hJz$) of angular momentum, i.\,e.
   \begin{subequations}
   \begin{align}
   \label{53a}
   \hJz\,\wop&=-\big(\bbar-\frac{1}{2}\big)\hbar\cdot\wop\\
   \label{53b}
   \hJz\,\wom&=\big(\bbar-\frac{1}{2}\big)\hbar\cdot\wom\\
   \label{53c}
   \hJz\,\wep&=-\big(\bbar-\frac{1}{2}\big)\hbar\cdot\wep\\
   \label{53d}
   \hJz\,\wem&=\big(\bbar-\frac{1}{2}\big)\hbar\cdot\wem\;,
   \end{align}
   \end{subequations}
and thus coincide with the old basis system~\rf{51a}-\rf{51d} for $\bbar=0$. But for arbitrary $\bbar$ one has a new situation with a {\em non-unique}\/ basis system over three-space, i.\,e.
   \begin{eqnarray}
   \label{54}
   \woepm(\phi+2\pi)&=&{\rm e}^{\mp i2\pi \bbar}\woepm(\phi)\;,
   \end{eqnarray}
where $\{r,\vartheta,\phi\}$ are the spherical polar coordinates as usual. Especially for $\bbar=\frac{1}{2}$ one has vanishing eigenvalues of angular momentum
   \begin{eqnarray}
   \label{55}
   \hJz\,\woepm&=&0\\
   &&\hspace{-4em}(\bbar=\frac{1}{2})\;.\nonumber
   \end{eqnarray}
This basis system ($\bbar=\frac{1}{2}$) is double-valued over the two-sphere $S^2$ and has been used for the treatment of the positronium groundstate in the preceding papers~\cite{c9,c10,c11}; but for the sake of generality we leave the constant $\bbar$ unspecified for the time being.

A first hint on how to fix the value of $\bbar$ arises now from the specific shape of the Dirac three-currents $\vka$~\rf{3122} which decompose in spherical polar coordinates as
   \begin{eqnarray}
   \label{56}
   \vka&=&\akr\cdot\vec{e}_r+\akt\cdot\vec{e}_\vartheta+\akp\cdot\vec{e}_\phi\;.
   \end{eqnarray}
Decomposing here also the (relativistic) Pauli spinors $\aphipm\vr$~\rf{339} with respect to the selected $\omega$ basis~\rf{52a}-\rf{52d} as ($a=1,2$)
   \begin{subequations}
   \begin{align}
   \label{57a}
   \aphip\vr&=\aMRp\cdot\wop+\aMSp\cdot\wom\\
   \label{57b}
   \aphim\vr&=-i\left\{\aMRm\cdot\wep+\aMSm\cdot\wem\right\}
   \end{align}
   \end{subequations}
lets appear the Dirac (relativistic) densities $\ako\vr$~\rf{3119} in the following form
   \begin{eqnarray}
   \label{58}
   \ako\vr&=&\frac{\aMRpS\cdot\aMRp+\aMSpS\cdot\aMSp+\aMRmS\cdot\aMRm+\aMSmS\cdot\aMSm}{4\pi}\;.\qquad\quad
   \end{eqnarray}
Evidently, these densities do not yet provide an immediate handle for fixing the parameter $\bbar$.

This situation changes now when one considers also the Dirac currents $\vka$~\rf{56}, which by their very definitions \rf{243a}-\rf{243c} are always real-valued objects:
   \begin{subequations}
   \begin{align}
   \label{59a}
   \akr&=\frac{i}{4\pi}\left\{\aMRpS\cdot\aMRm+\aMSpS\cdot\aMSm-\aMRmS\cdot\aMRp-\aMSmS\cdot\aMSp\right\}\\
   \label{59b}
   \akt&=-\frac{i}{4\pi}\left\{{\rm e}^{2i(\bbar-\frac{1}{2})\phi}\cdot\MCa-{\rm e}^{-2i(\bbar-\frac{1}{2})\phi}\cdot\MCaS\right\}\\
   &{}(\MCa\doteqdot\aMRpS\cdot\aMSm+\aMRmS\cdot\aMSp)\nonumber\\
   \label{59c}
   \akp&= \frac{\sin\vartheta}{4\pi}\left\{\aMRpS\cdot\aMRm+\aMRmS\cdot\aMRp-\aMSpS\cdot\aMSm-\aMSmS\cdot\aMSp\right\}\nonumber\\
   &\hspace{1.5em}-\frac{\cos\vartheta}{4\pi}\left\{{\rm e}^{2i(\bbar-\frac{1}{2})\phi}\cdot\MCa+{\rm e}^{-2i(\bbar-\frac{1}{2})\phi}\cdot\MCaS\right\}\;.
   \end{align}
   \end{subequations}
But here a nearby restriction upon the parameter $\bbar$ suggests itself, namely through the plausible demand that the Dirac currents $\vka$~\rf{56}, with their components being specified by~\rf{59a}-\rf{59c}, must be {\em unique}\/~(!) vector fields over three-space (albeit only apart from the origin $r=0$ and the $z$ axis $\vartheta=0,\pi$). Evidently this demand of uniqueness reads in terms of the spherical polar coordinates $\{r,\vartheta,\phi\}$ 
   \begin{eqnarray}
   \label{510}
   \vka(r,\vartheta,\phi+2\pi)&=&\vka(r,\vartheta,\phi)\;,
   \end{eqnarray}
and thus the values of $\bbar$ become restricted to the range
   \begin{eqnarray}
   \label{511}
   \bbar&=&\frac{1}{2}(n+1)\\
   &&\hspace{-2em}(n=0,\pm1,\pm2,\pm3,...)\nonumber
   \end{eqnarray}
which then entails also {\em (half-)integer}\/ quantum numbers for the $z$ component of angular momentum~\rf{53a}-\rf{53d}:
   \begin{eqnarray}
   \label{512}
   j_z&=&\pm\big(\bbar-\frac{1}{2}\big)\;=\;\pm\frac{n}{2}\;.
   \end{eqnarray}

Notice here that this {\em (half-)integrity}\/ arises as a consequence of the demand of uniqueness with respect to certain physical {\em densities}\/ (i.\,e. Dirac current), whereas the corresponding {\em integral}\/ quantum numbers of conventional non-relativistic quantum mechanics are mostly traced back in the textbooks to the uniqueness requirement for the {\em wave functions}\/ themselves (not the densities). The lowest values of $j_z$~\rf{512} are $j_z=\pm\frac{1}{2}$ for $\bbar=0$ and $j_z=0,\pm1$ for $\bbar=\pm\frac{1}{2}$. Thus for the first case ($\bbar=0$) we have a {\em fermionic basis}\/ and for the second case ($\bbar=\pm\frac{1}{2}$) one deals with a {\em bosonic basis}\/. In this sense, a Dirac particle is said to occupy a fermionic quantum state $\psi$ if the {\em ``boson number''}\/ $\bbar$ of its spinor basis is zero ($\bbar=0$), and a bosonic quantum state if the boson number $\bbar$ equals $\pm\frac{1}{2}$. Observe that through this arrangement the fermionic or bosonic character of the quantum state of a Dirac particle is defined by reference to the corresponding spinor basis.

Naturally, it must appear very tempting to interprete the results~\rf{511}-\rf{512} in the sense that {\em angular momentum}\/ $J_z$ be quantized in units of $\frac{n}{2}$. However, it is important to see that this ``quantization'' occurs in connection with the basis system~\rf{53a}-\rf{53d}, not in connection with the {\em unique wave amplitudes}\/ $\aMRpm,\,\aMSpm$ or the wave functions $\psi_a$, resp., which are normally thought to specify the angular momentum of the considered quantum state. Therefore one conceives here the ansatz parameter $\bbar$ to be a new quantum number ({\em boson number}\/) which does influence the magnitude of angular momentum but does not fix it completely (i.\,e. the ultimate fixation, if possible, of angular momentum is assumed to occur through cooperation of $\bbar$ and the (unique) wave amplitudes $\aMRpm\vr,\,\aMSpm\vr$).

\subsection{Two-Particle Eigenvalue System}

For the {\em two}\/-particle systems, to be considered now, all fields which are due to the third particle must naturally vanish, and therefore the original eigenvalue system for three particles becomes cut down to an eigenvalue system for only two particles. Thus the abstract three-particle equations~\rf{340a}-\rf{340c} become converted to the corresponding eigenvalue system in terms of the wave amplitudes $\aMRpm,\,\aMSpm$ ($a=1,2$) which are due to the ansatz~\rf{57a}-\rf{57b} for the Pauli spinors. Each of both Dirac wave functions $\psi_a$ \mbox{($a=1,2$)} becomes thus parametrized by four complex-valued fields $\aMRpm,\,\aMSpm$; and each of the two Dirac equations~\rf{232a}-\rf{232b}, or Pauli equations~\rf{340a}-\rf{340b}, resp., yields then a coupled system of four equations for these wave amplitudes. For the sake of brevity, it may suffice here to explicitly reproduce only the case of the first particle ($a=1$):
   \begin{subequations}
   \begin{align}
   \label{513a}
   \frac{\partial\,\eMRp}{\partial r} +\frac{i}{r}\cdot\frac{\partial\,\eMRp}{\partial\phi}+\frac{\bbar_p}{r}\cdot\eMRp+\zAo\cdot\eMRm +\big[i\,\zAr-\sin\vartheta\cdot\zAp\big]\cdot\eMRp\hspace{-9em}&\hspace{9em}\nonumber\\
   {}+{\rm e}^{2i(\bbar_p-\frac{1}{2})\phi}\cdot\Bigg\{\frac{1}{r}\cdot\frac{\partial\,\eMSp}{\partial\vartheta} +\frac{\cot\vartheta}{r}\,\big[\bbar_p\cdot\eMSp-i\,\frac{\partial\,\eMSp}{\partial\phi}\big]\hspace{-1em}&\nonumber\\
   +\big[i\,\zAt+\cos\vartheta\cdot\zAp\big]\cdot\eMSp\Bigg\}&=\frac{M_p-M_1}{\hbar}\,c\cdot\eMRm
   \\
   \label{513b}
   \frac{\partial\,\eMSp}{\partial r} -\frac{i}{r}\cdot\frac{\partial\,\eMSp}{\partial\phi}+\frac{\bbar_p}{r}\cdot\eMSp+\zAo\cdot\eMSm +\big[i\,\zAr+\sin\vartheta\cdot\zAp\big]\cdot\eMSp\hspace{-7em}&\hspace{7em}\nonumber\\
   {}-{\rm e}^{-2i(\bbar_p-\frac{1}{2})\phi}\cdot\Bigg\{\frac{1}{r}\cdot\frac{\partial\,\eMRp}{\partial\vartheta} +\frac{\cot\vartheta}{r}\,\big[\bbar_p\cdot\eMRp+i\,\frac{\partial\,\eMRp}{\partial\phi}\big]\hspace{-2.8em}&\nonumber\\
   +\big[i\,\zAt-\cos\vartheta\cdot\zAp\big]\cdot\eMRp\Bigg\}&=\frac{M_p-M_1}{\hbar}\,c\cdot\eMSm\\
   \label{513c}
   \frac{\partial\,\eMRm}{\partial r} -\frac{i}{r}\cdot\frac{\partial\,\eMRm}{\partial\phi}+\frac{2-\bbar_p}{r}\cdot\eMRm-\zAo\cdot\eMRp -\big[i\,\zAr-\sin\vartheta\cdot\zAp\big]\cdot\eMRm\hspace{-11em}&\hspace{11em}\nonumber\\
   {}-{\rm e}^{2i(\bbar_p-\frac{1}{2})\phi}\cdot\Bigg\{\frac{1}{r}\cdot\frac{\partial\,\eMSm}{\partial\vartheta} +\frac{\cot\vartheta}{r}\,\big[\bbar_p\cdot\eMSm-i\,\frac{\partial\,\eMSm}{\partial\phi}\big]\hspace{-1em}&\nonumber\\
   +\big[i\,\zAt+\cos\vartheta\cdot\zAp\big]\cdot\eMSm\Bigg\}&=\frac{M_p+M_1}{\hbar}\,c\cdot\eMRp\\
   \label{513d}
   \frac{\partial\,\eMSm}{\partial r} +\frac{i}{r}\cdot\frac{\partial\,\eMSm}{\partial\phi}+\frac{2-\bbar_p}{r}\cdot\eMSm-\zAo\cdot\eMSp -\big[i\,\zAr+\sin\vartheta\cdot\zAp\big]\cdot\eMSm\hspace{-9.5em}&\hspace{9.5em}\nonumber\\
   {}+{\rm e}^{-2i(\bbar_p-\frac{1}{2})\phi}\cdot\Bigg\{\frac{1}{r}\cdot\frac{\partial\,\eMRm}{\partial\vartheta} +\frac{\cot\vartheta}{r}\,\big[\bbar_p\cdot\eMRm+i\,\frac{\partial\,\eMRm}{\partial\phi}\big]\hspace{-2.5em}&\nonumber\\ +\big[i\,\zAt-\cos\vartheta\cdot\zAp\big]\cdot\eMRm\Bigg\}&=\frac{M_p+M_1}{\hbar}\,c\cdot\eMSp\;.
   \end{align}
   \end{subequations}

Since this (relativistic) system of mass eigenvalue equations for the wave amplitudes looks somewhat intricate, it will be highly instructive to spend some effort on its detailed analysis. First let us mention that a similar system of eigenvalue equations does hold also for the second particle ($a=2$) which, however, can easily be obtained by means of the particle permutation symmetry:
   \begin{subequations}
   \begin{align}
   \label{514a}
   \eMRpm\Leftrightarrow\zMRpm\qquad&\qquad\quad\eMSpm\Leftrightarrow\zMSpm\\
   \label{514b}
   \eAmo=\{\eAo;\eAr,\eAt,\eAp\}\quad&\Leftrightarrow\quad\zAmo=\{\zAo;\zAr,\zAt,\zAp\}\\
   \label{514c}
   M_1\Leftrightarrow M_2\qquad\quad M_p&\Leftrightarrow -M_e\qquad\quad \bbar_p\Leftrightarrow \bbar_e\;.
   \end{align}
   \end{subequations}
Next, recall that this system~\rf{513a}-\rf{514c} of eigenvalue equations is mathematically consistent only if the {\em boson numbers}\/ $\{\bbar_p,\,\bbar_e\}$ do obey again the former restriction~\rf{511} because it is only for this case that the exponential factors become {\em unique}\/ functions over three-space, i.\,e.
   \begin{eqnarray}
   \label{515}
   {\rm e}^{\pm2i(\bbar_p-\frac{1}{2})\phi}&\Rightarrow&{\rm e}^{\pm in\phi}\\
   &&\hspace{-7em}(n=0,\pm1,\pm2,...)\nonumber\;.
   \end{eqnarray}

But what is most important for the present discussion, this refers to the fact that the eigenvalue system~\rf{513a}-\rf{514c} can be viewed also as the set of variational equations due to the RST energy functional $\tEfT$~\rf{414}. In order to become convinced of this claim one merely has to represent the individual contributions of the energy functional in terms of the wave amplitudes $\aMRpm,\,\aMSpm$ and carry through the variational procedure, which itself again requires the uniqueness condition~\rf{515}, namely in order that the necessary partial integrations can be carried out in an unambiguous way. Observe also that, in the most general case, all three components $\aAr,\,\aAt,\,\aAp$ of the vector potential $\vAa\vr$ will be non-zero; and this requires that in such a general situation the Pauli spinors $\aphipm\vr$~\rf{57a}-\rf{57b} necessarily must appear as a superposition of those simple ``spin-up'' ($\sim\mathcal{R}$) and ``spin-down'' ($\sim\mathcal{S}$) components. It is only in very special situations that one can deal exclusively with {\em one}\/ spin direction (either ``up'' or ``down''), see below.

\subsection{One-Particle State $\zpdh$}

Surely, a single Dirac particle, moving in an external field $\exAmu$, will occupy a {\em fermionic state}\/, i.\,e. the boson number $\bbar_e$ of a single electron must be adopted to be zero ($\bbar_e=0$). For instance, one encounters such a situation for the hydrogen atom where, for the sake of simplicity, the proton mass $M_p$ will be presumed to be infinite. The Dirac density $\eko\vr$ of such an infinitely heavy particle will be found to be pointlike, i.\,e. one puts
   \begin{eqnarray}
   \label{516}
   \eko\vr&\equiv&\ejo\vr\;\Rightarrow\;\delta^3\vr\;,
   \end{eqnarray}
and consequently the electric potential $\eAo\vr$~\rf{35a} of such a point charge must degenerate to the Coulomb potential
   \begin{eqnarray}
   \label{517}
   \eAo&\Rightarrow&\frac{\als}{r}\;.
   \end{eqnarray}
The relativistic spectrum of this one-particle eigenvalue problem is exactly known (see e.\,g. ref.~\cite{c18}); and if we take this physical situation as the simplest test case for our RST energy functional $\tEfT$~\rf{414} one must expect to recover just that conventional spectrum~\cite{c18}. Clearly, this will then entail some confidence into the proposed RST energy functional $\tEfT$~\rf{414}. But sometimes the application of a new formalism admits also a new view upon the considered standard situation; and in the present case this alternative view provides us with the emergence of the {\em spin degeneracy}\/ when passing over to the non-relativistic limit. This effect is mostly omitted in the text books (see, e.\,g., ref.~\cite{c18}).

We will treat here this phenomenon for the excited state $\zpdh$ of a Dirac electron in the {\em electrostatic field}\/~\rf{517} of a pointlike proton. Observe that all the magnetic fields are presumed to be zero, so that there exists locally no preferred direction in space for the orientation of the electronic spin. Therefore a ``spin-up'' solution must be expected to carry the same energy $\ET$ as a ``spin-down'' solution or as any linear combination of both. However, this supposition is not true for the relativistic case, and the reason for this is the spin-orbit coupling of the angular momenta. In order to recognize this more clearly, one first recalls the conventional treatment of the relativistic hydrogen atom~\cite{c18}. Here the stationary solution $\psi_2\vr$~\rf{339} of the Dirac equation~\rf{232b} for the electron (i.\,e. our second particle, $a=2$) is presented in terms of the angular momentum eigenvectors $\zjml$ \rf{51a}-\rf{51d} as
   \begin{subequations}
   \begin{align}
   \label{518a}
   \zphip\vr&\Rightarrow\Phi(r)\cdot\zdee\\
   \label{518b}
   \zphim\vr&\Rightarrow-iW\Phi(r)\cdot\zdez\;,
   \end{align}
   \end{subequations}
where the radial function $\Phi(r)$ is given by
   \begin{eqnarray}
   \label{519}
   \Phi(r)&=&N_c\,r^\nu{\rm e}^{-\beta r}\;.
   \end{eqnarray}
The normalization constant $N_c$ for this solution is
   \begin{eqnarray}
   \label{520}
   N_c&=&\left[(1+W^2)\cdot\frac{\Gamma(3+2\nu)}{(2\beta)^{3+2\nu}}\right]^{-\frac{1}{2}}\,,
   \end{eqnarray}
the decay parameter $\beta$ of the exponential function is found as
   \begin{eqnarray}
   \label{521}
   \beta&=&\frac{M_ec}{\hbar}\cdot\frac{2W}{1+W^2}\\
   &&\hspace{-5em}\big(\,W\doteqdot\frac{2}{\als}\left[1-\sqrt{1-\left(\frac{\als}{2}\right)^2}\right]\big)\,,\nonumber
   \end{eqnarray}
and finally the power $\nu$ is given by
   \begin{eqnarray}
   \label{522}
   \nu&=&-1+2\sqrt{1-\left(\frac{\als}{2}\right)^2}\;\,.
   \end{eqnarray}
Concerning the corresponding energy eigenvalue ($\Econv$, say) of this quantum state $\zpdh$, it may be sufficient here to quote its non-relativistic approximation up to order $\als^4$~\cite{c18}:
   \begin{eqnarray}
   \label{523}
   \frac{\Econv-M_ec^2}{M_ec^2}&\cong&-\als^2\left\{\frac{1}{2n^2}+\frac{\als^2}{2n^3}\left(\frac{1}{j+\frac{1}{2 }}-\frac{3}{4n}\right)\right\}\,,
   \end{eqnarray}
i.\,e. for the present state $\zpdh$ with principal quantum number $n=2$ and $j=\frac{3}{2}$
   \begin{eqnarray}
   \label{524}
   \frac{\Econv-M_ec^2}{M_ec^2}&\Rightarrow&-\frac{\als^2}{8}-\frac{\als^4}{128}\;.
   \end{eqnarray}
Of course the first term of the non-relativistic expansion~\rf{523} yields the usual Schr\"{o}dinger eigenvalue ($\Es$, say) of the non-relativistic hydrogen atom
   \begin{eqnarray}
   \label{525}
   \Es&=&-\frac{1}{2n^2}\,\als^2M_ec^2\;=\;-\frac{e^2}{2\ab}\cdot\frac{1}{n^2}\;,
   \end{eqnarray}
with the atomic energy unit (a.\,u.) given in terms of the elementary charge ($e$) and Bohr radius ($\ab$) as
   \begin{eqnarray}
   \label{526}
   \frac{e^2}{\ab}&\cong&27.2116...\ [{\rm eV}]\;.
   \end{eqnarray}

But now it is especially instructive to oppose this standard treatment of the idealized hydrogen atom \rf{516}-\rf{517} to the corresponding RST treatment. Here it is first necessary to specify the general eigenvalue equations \rf{514a}-\rf{514c} for the second particle ($a=2$) which is assumed to occupy (as a single particle) a {\em fermionic}\/ state, i.\,e. one puts the boson number $\bbar_e$ to zero ($\bbar_e=0$). Moreover, all the magnetic fields are also assumed to vanish (i.\,e. $\eAr=\eAt=\eAp\equiv0$); and finally one introduces the real-valued wave amplitudes $\zRpm,\,\zSpm$ by putting
   \begin{subequations}
   \begin{align}
   \label{527a}
   \zMRpm&\equiv\zMRpmS\;\doteqdot\;\zRpm\\
   \label{527b}
   \zMSpm&\doteqdot{\rm e}^{i\phi}\cdot\zSpm\;.
   \end{align}
   \end{subequations}
For such a simplified situation the system of eigenvalue equations for the second particle becomes cut down to the following form:
   \begin{subequations}
   \begin{align}
   \label{528a}
   \frac{\partial\zRp}{\partial r}+\eAo\cdot\zRm+\frac{1}{r}\left[\frac{\partial\zSp}{\partial\vartheta}+\cot\vartheta\cdot\zSp\right] &=&\hspace{-.5em}-\frac{M_e+M_2}{\hbar}\,c\cdot\zRm\\
   \label{528b}
   \frac{\partial\zSp}{\partial r}+\frac{1}{r}\,\zSp+\eAo\cdot\zSm-\frac{1}{r}\frac{\partial\zRp}{\partial\vartheta} &=&\hspace{-.5em}-\frac{M_e+M_2}{\hbar}\,c\cdot\zSm\\
   \label{528c}
   \frac{\partial\zRm}{\partial r}+\frac{2}{r}\,\zRm-\eAo\cdot\zRp-\frac{1}{r}\left[\frac{\partial\zSm}{\partial\vartheta}+\cot\vartheta\cdot\zSm\right] &=&\hspace{-.5em}\frac{M_2-M_e}{\hbar}\,c\cdot\zRp\\
   \label{528d}
   \frac{\partial\zSm}{\partial r}+\frac{1}{r}\,\zSm-\eAo\cdot\zSp+\frac{1}{r}\frac{\partial\zRm}{\partial\vartheta} &=&\hspace{-.5em}\frac{M_2-M_e}{\hbar}\,c\cdot\zSp\;.
   \end{align}
   \end{subequations}

This simplified system of eigenvalue equations for the electronic (real-valued) wave amplitudes suggests to try the following product ansatz
   \begin{subequations}
   \begin{align}
   \label{529a}
   \zRpm(r,\vartheta)&=\cos\vartheta\cdot\nRpm(r)\\
   \label{529b}
   \zSpm(r,\vartheta)&=\sin\vartheta\cdot\nSpm(r)\;,
   \end{align}
   \end{subequations}
and this recasts the original eigenvalue system \rf{528a}-\rf{528d} into the following equations for the spherically symmetric functions
   \begin{subequations}
   \begin{align}
   \label{530a}
   \frac{\dif{\nRp(r)}}{\dif{r}}+\eAo\cdot\nRm(r)+\frac{2}{r}\cdot\nSp(r) &=-\frac{M_e+M_2}{\hbar}\,c\cdot\nRm(r)\\
   \label{530b}
   \frac{\dif{\nSp(r)}}{\dif{r}}+\frac{1}{r}\,\nSp(r)+\eAo\cdot\nSm(r)+\frac{1}{r}\cdot\nRp(r) &=-\frac{M_e+M_2}{\hbar}\,c\cdot\nSm(r)\\
   \label{530c}
   \frac{\dif{\nRm(r)}}{\dif{r}}+\frac{2}{r}\,\nRm(r)-\eAo\cdot\nRp(r)-\frac{2}{r}\cdot\nSm(r) &=\frac{M_2-M_e}{\hbar}\,c\cdot\nRp(r)\\
   \label{530d}
   \frac{\dif{\nSm(r)}}{\dif{r}}+\frac{1}{r}\,\nSm(r)-\eAo\cdot\nSp(r)-\frac{1}{r}\cdot\nRm(r) &=\frac{M_2-M_e}{\hbar}\,c\cdot\nSp(r)\;.
   \end{align}
   \end{subequations}
For the solution of this set of ordinary differential equations one tries the nearby ansatz
   \begin{subequations}
   \begin{align}
   \label{531a}
   \nRpm(r)&=p_\pm\cdot r^\nu\exp[-\beta r]\\
   \label{531b}
   \nSpm(r)&=f_\pm\cdot r^\nu\exp[-\beta r]
   \end{align}
   \end{subequations}
which then yields two sets of four algebraic equations for the determination of the four constants $p_\pm,\,f_\pm$ and for the ansatz parameters $\nu$ and $\beta$. The first set contains only the power $\nu$
   \begin{subequations}
   \begin{align}
   \label{532a}
   \nu\cdot p_++2\cdot f_++\als\cdot p_-&=0\\
   \label{532b}
   p_++(\nu+1)\cdot f_++\als\cdot f_-&=0\\
   \label{532c}
   -\als\cdot p_++(\nu+2)\cdot p_--2\cdot f_-&=0\\
   \label{532d}
   -\als\cdot f_++(\nu+1)\cdot f_--p_-&=0\;,
   \end{align}
   \end{subequations}
whereas the second set relates the decay parameter $\beta$ to the mass eigenvalue $M_2$
   \begin{subequations}
   \begin{align}
   \label{533a}
   \beta\cdot p_+&=\frac{M_e+M_2}{\hbar}\,c\cdot p_-\\
   \label{533b}
   \beta\cdot f_+&=\frac{M_e+M_2}{\hbar}\,c\cdot f_-\\
   \label{533c}
   \beta\cdot p_-&=\frac{M_e-M_2}{\hbar}\,c\cdot p_+\\
   \label{533d}
   \beta\cdot f_-&=\frac{M_e-M_2}{\hbar}\,c\cdot f_+\;.
   \end{align}
   \end{subequations}
Naturally, the first set \rf{532a}-\rf{532d} says that the power $\nu$ of the RST ansatz \rf{531a}-\rf{531b} is identical to the conventional value~\rf{522}; and similarly the second set \rf{533a}-\rf{533d} fixes the RST value of the decay parameter $\beta$ also to its conventional counterpart~\rf{521}, where the mass eigenvalue ($M_2$) of the electron turns out as
   \begin{eqnarray}
   \label{534}
   M_2&=&M_e\cdot\frac{1-W^2}{1+W^2}\;=\;M_e\sqrt{1-\left(\frac{\als}{2}\right)^2}
   \end{eqnarray}
with the relativistic parameter $W$ being defined below equation~\rf{521}.

This exact coincidence of the RST results and their conventional counterparts mentioned above seems to signal that RST is merely a formal one-to-one transcription of the conventional formalism; but actually RST owns a richer structure. This becomes more obvious when one faces now the question of the ``energy'' carried by the considered wave function $\psi\vr$. In the conventional theory, this energy is quite generally identified for the presently considered one-particle systems with the mass eigenvalue $M_2c^2$:
   \begin{eqnarray}
   \label{535}
   \Econv&=&M_2c^2\;.
   \end{eqnarray}
This is most easily seen in the non-relativistic limit by expanding the present RST result $M_2c^2$~\rf{534} with respect to the fine-structure constant $\als$:
   \begin{eqnarray}
   \label{536}
   M_2c^2&\cong&M_ec^2\left\{1-\frac{\als^2}{8}-\frac{\als^4}{128}\:\ldots\right\}
   \end{eqnarray}
and comparing this to the conventional result~\rf{524}. (For a general proof, see ref.s~\cite{c15,c16}.) However, in RST the total energy $\tET$~\rf{414} will in general not be found to agree with some mass eigenvalue! It is only for the present oversimplified situation \rf{516}-\rf{517} that the RST energy $\tET$ can agree with the mass eigenvalue $M_2c^2$. This is easily seen by explicit calculation of the various contributions to the RST energy $\tET$.

First, one makes oneself sure of the relativistic normalization condition \rf{342} for the electron by imposing upon the constants $p_\pm,\,f_\pm$ the following condition:
   \begin{subequations}
   \begin{align}
   \label{537a}
   p_+^2+2\cdot f_+^2+p_-^2+2\cdot f_-^2&=\nN^2\\
   \label{537b}
   \nN^2&=\frac{1}{3}\cdot\frac{\Gamma(3+2\nu)}{(2\beta)^{3+2\nu}}\;.
   \end{align}
   \end{subequations}
Next, the Pauli spinors $\zphipm\vr$ \rf{57a}-\rf{57b} are inserted into the second mass renormalization factor $\MZz$~\rf{344} which yields quite generally
   \begin{eqnarray}
   \MZz^2&=&\int\!\frac{{\rm d}^3\vec{r}}{4\pi}\left\{\zMRpS\cdot\zMRp+\zMSpS\cdot\zMSp-\zMRmS\cdot\zMRm-\zMSmS\cdot\zMSm\right\}\,,\nonumber\\
   \label{538}
   &&
   \end{eqnarray}
i.\,e. in particuar for the presently considered wave amplitudes \rf{529a}-\rf{529b} and \rf{531a}-\rf{531b}, resp.,
   \begin{eqnarray}
   \label{539}
   M_ec^2\cdot\MZz^2&=&M_ec^2-2M_ec^2\left[\frac{p_-^2}{\nN^2}+2\frac{f_-^2}{\nN^2}\right]\,.
   \end{eqnarray}
Here the normalization condition~\rf{537a} has also been used.

The kinetic energy $\Tkinz$~\rf{347} represents a more complicated problem; inserting here the former ansatz \rf{57a}-\rf{57b} for the Pauli spinors $\zphipm\vr$ with the wave amplitudes $\zMRpm,\,\zMSpm$ being further specified by equations \rf{527a}-\rf{527b}, \rf{529a}-\rf{529b}, and \rf{531a}-\rf{531b} yields by straightforward integration
   \begin{eqnarray}
   \label{540}
   \Tkinz&=&\hbar c\left(\frac{\beta}{1+\nu}\right)\frac{1}{\nN^2}\,\Big\{p_+\cdot p_--2\,\big[p_+\cdot f_-+p_-\cdot f_+\big]\Big\}\;.
   \end{eqnarray}
Obviously, both the kinetic energy $\Tkinz$ and the mass renormalization term \rf{539} are originally built up by the integration constants of both the ``positive'' ($p_+,\,f_+$) and the ``negative'' ($p_-,\,f_-$) type. But the relations \rf{533a}-\rf{533d} fix their relative magnitudes as
   \begin{subequations}
   \begin{align}
   \label{541a}
   p_-&=W\cdot p_+\\
   \label{541b}
   f_-&=W\cdot f_+\;.
   \end{align}
   \end{subequations} 
Here the relativistic parameter $W$ has already been defined below equation~\rf{521}, but on the other hand it can be written also in terms of the mass eigenvalue $M_2$ as
   \begin{eqnarray}
   \label{542}
   W&=&\sqrt{\frac{M_e-M_2}{M_e+M_2}}\;.
   \end{eqnarray}
This provides us with the possibility to express both quantities by means of the positive-type constants alone, i.\,e. the mass renormalization term~\rf{539} and the kinetic energy~\rf{540} appear then in the following form
   \begin{subequations}
   \begin{align}
   \label{543a}
   M_ec^2\cdot\MZz^2&=M_ec^2-2M_ec^2\frac{W^2}{\nN^2}\,\Big\{p_+^2+2f_+^2\Big\}\\
   \label{543b}
   \Tkinz&=\frac{M_ec^2}{1+\nu}\frac{(2W)^2}{1+W^2}\frac{1}{\nN^2}\,\Big\{p_+^2+2f_+^2\Big\}\;.
   \end{align}
   \end{subequations}
This form of the results will become important for the deduction of the corresponding non-relativistic limit; but for the present relativistic case one may use the values of the constants $p_+,\,f_+$ following from the algebraic equations \rf{533a}-\rf{533d} together with the normalization condition~\rf{537a}
   \begin{subequations}
   \begin{align}
   \label{544a}
   p_+&=\nN\cdot\sqrt{\frac{2}{3(1+W^2)}}\\
   \label{544b}
   f_+&=-\nN\cdot\frac{1}{\sqrt{6(1+W^2)}}
   \end{align}
   \end{subequations}
so that the energy contributions \rf{543a}-\rf{543b} adopt their final shape as
   \begin{subequations}
   \begin{align}
   \label{545a}
   M_ec^2\cdot\MZz^2&=M_ec^2-\frac{2W^2}{1+W^2}\cdot M_ec^2\\
   \label{545b}
   \Tkinz&=\frac{4}{1+\nu}\frac{W^2}{(1+W^2)^2}\cdot M_ec^2\;.
   \end{align}
   \end{subequations}

Thus we are left with the problem of calculating the last physical contribution to the energy $\tEfT$~\rf{414}, i.\,e. the gauge field energy $\ERe$~\rf{45b} which of course reduces to a simpler form for the two-particle systems:
   \begin{eqnarray}
   \label{546}
   \ERe&=&\frac{\hbar c}{4\pi\als}\idr\vEe\sdot\vEz\;=\;\frac{\hbar c}{4\pi\als}\idr\big(\vnabla\eAo\big)\sdot\big(\vnabla\zAo\big)\;.\qquad
   \end{eqnarray}
Observe again that we need consider only the physical part $\EivfT$~\rf{413} of the proper functional $\tEfT$~\rf{414} if all the constraints are obeyed by our ansatz for the matter and gauge fields. But here the relativistic normalization constraint~\rf{342} has already been demonstrated to hold in the form~\rf{537a}; and furthermore it is also very instructive and satisfying to see in what way the electric Poisson identity~\rf{47} is validated. To this end, one first computes the electric mass equivalent $\Mefz c^2$~\rf{348a} by use of the first potential $\eAo\vr$~\rf{517} and the second Dirac density $\zko\vr$~\rf{58}
  \begin{eqnarray}
  \Mefz c^2&=&-\frac{\hbar c}{2}\idr\zko\vr\cdot\eAo\vr\nonumber\\
  \label{547}
  &=&\frac{\hbar c}{2}\:\als\int\!\frac{{\rm d}^3\vec{r}}{4\pi r}\,\left\{\big(\zRp\big)^2+\big(\zSp\big)^2+\big(\zRm\big)^2+\big(\zSm\big)^2\right\}\,.\qquad\quad
  \end{eqnarray}
The straightforward calculation of this integral under use of the normalization constraint~\rf{537a} yields the following result
   \begin{eqnarray}
   \label{548}
   \Mefz c^2&=&-\frac{e^2}{2}\frac{\beta}{1+\nu}\;=\;-\left(\frac{\als}{1+\nu}\right)\frac{W}{1+W^2}\cdot M_ec^2\;.
   \end{eqnarray}

This result is also found for the first mass equivalent, i.\,e.
   \begin{eqnarray}
   \label{549}
   \Mefe c^2&=&-\frac{\hbar c}{2}\idr\eko\vr\cdot\zAo\vr\;=\;\Mefz c^2\;,
   \end{eqnarray}
which is most easily verified by simply observing the fact that the first Dirac density is pointlike, cf.~\rf{516}, so that the considered mass equivalent is essentially the value of the second potential $\zAo$ at the origin ($r=0$)
   \begin{eqnarray}
   \label{550}
   \Mefe c^2&=&\frac{\hbar c}{2}\cdot\zAo(\vec{0})\;.
   \end{eqnarray}
However, this required value of the second potential $\zAo$ can easily be deduced from the formal solution~\rf{35a} of the corresponding Poisson equation~\rf{33} as
   \begin{eqnarray}
   \zAo(\vec{0})&=&-\als\int\!\frac{{\rm d}^3\vec{r}\!\;'}{r'}\,\zko\vrs\nonumber\\
   &=&-\als\int\!\frac{{\rm d}^3\vec{r}}{4\pi r}\left\{\big(\zRp\big)^2+\big(\zSp\big)^2+\big(\zRm\big)^2+\big(\zSm\big)^2\right\}\nonumber\\
   \label{551}
   &=&-\als\frac{\beta}{1+\nu}\;,
   \end{eqnarray}
cf. \rf{547} and \rf{548}, so that the claimed equality~\rf{549} of both mass equivalents is immediately verified by use of equation~\rf{550}. Thus the electric Poisson identity~\rf{47} admits now to specify the desired gauge field energy $\ERe$~\rf{546} as
   \begin{eqnarray}
   \label{552}
   \ERe&=&\Me c^2\;=\;\Mefe c^2+\Mefz c^2\;=\;-\frac{2\als}{1+\nu}\frac{W}{1+W^2}\,M_ec^2\;.\quad
   \end{eqnarray}

But now that all three physical contributions to the total energy $\tEfT$ \rf{414} are explicitly known, see \rf{545a}-\rf{545b} and \rf{552}, one actually gets for their sum the expected numerical coincidence of the RST energy $\tET$ and the mass eigenvalue $M_2c^2$:
   \begin{eqnarray}
   \label{553}
   \tET&=&\EivT\;=\;M_ec^2\cdot\MZz^2+2\Tkinz+\ERe\;=\;M_2c^2\;=\;\Econv\;.\qquad
   \end{eqnarray}
Clearly, this is a very pleasant result in favour of our RST energy functional $\tEfT$ because it associates to the considered hydrogen state $\zpdh$ exactly the same energy as does the conventional theory, cf.~\rf{535}! On the other hand, the present explicit calculation of the RST energy $\tET$ in terms of the cooperation and interaction of the various atomic fields yields a much more detailed picture (of the intra-atomic situation) than is possible in the conventional theory where the atomic energy is simply identified with the mass eigenvalue, without discussion of the specific interplay of matter and gauge fields. This feature of RST (namely to provide a more detailed picture of the intra-atomic situation) becomes obvious also by considering now the non-relativistic approximation.

\subsubsection{Non-Relativistic Spin Degeneracy}

The point of departure for the non-relativistic approach is again the relativistic eigenvalue system \rf{528a}-\rf{528d} where one eliminates the wave amplitudes $\zRm,\,\zSm$ in close analogy to the elimination of the Pauli spinors $\aphim$ \rf{366a}-\rf{366c}, i.\,e. one puts
   \begin{subequations}
   \begin{align}
   \label{554a}
   \zRm&\cong-\frac{\hbar}{2M_ec}\left\{\frac{\partial\zRp}{\partial r}+\frac{1}{r\sin\vartheta}\frac{\partial}{\partial\vartheta}\big(\sin\vartheta\cdot\zSp\big)\right\}\\
   \label{554b}
   \zSm&\cong-\frac{\hbar}{2M_ec}\left\{\frac{\partial\zSp}{\partial r}+\frac{1}{r}\,\zSp-\frac{1}{r}\frac{\partial\zRp}{\partial\vartheta}\right\}\,.
   \end{align}
   \end{subequations}
Indeed, these approximative relations may immediately be deduced from their exact counterparts \rf{528a}-\rf{528b}; and if this is substituted in the remaining equations \rf{528c}-\rf{528d} one ends up with the desired non-relativistic eigenvalue equations for $\zSp$ ($\Rightarrow S_+$) and $\zRp$ ($\Rightarrow R_+$):
   \begin{subequations}
   \begin{align}
   -\frac{\hbar^2}{2M_e}\left\{\frac{\partial^2R_+}{\partial r^2}+\frac{2}{r}\frac{\partial R_+}{\partial r}+\frac{1}{r^2}\left[\frac{\partial^2R_+}{\partial\vartheta^2}+\cot\vartheta\,\frac{\partial R_+}{\partial\vartheta}\right]\right\}-\hbar c\eAo\cdot R_+&\nonumber\\
   \label{555a}
   &\hspace{-3em}=\Esz\cdot R_+\\
   -\frac{\hbar^2}{2M_e}\left\{\frac{\partial^2S_+}{\partial r^2}+\frac{2}{r}\frac{\partial S_+}{\partial r}+\frac{1}{r^2}\frac{\partial}{\partial\vartheta}\left[\frac{1}{\sin\vartheta}\frac{\partial}{\partial\vartheta}\big(\sin\vartheta\cdot S_+\big)\right]\right\}-\hbar c\eAo\cdot S_+&\nonumber\\
   \label{555b}
   &\hspace{-3em}=\Esz\cdot S_+\;.
   \end{align}
   \end{subequations}
Here the non-relativistic counterparts of the relativistic wave amplitudes $\zRp,\,\zSp$ are denoted simply by $R_+,\,S_+$; and the non-relativistic energy eigenvalue $\Esz$ is formally the same as the previous $\Epz$~\rf{372b}.

Observe also that through the elimination of the ``negative'' wave amplitudes $\zRm,\,\zSm$ the ``positive'' wave amplitudes $R_+,\,S_+$ have been {\em decoupled}\/, which is equivalent to the breaking of the spin-orbit coupling. It is true, both eigenvalue equations \rf{555a}-\rf{555b} look very similar, but they are not identical because of their different angular derivative parts. But here one may now suppose that the angular dependence \rf{529a}-\rf{529b} of the relativistic amplitudes $\zRp,\,\zSm$ is left unchanged on their way to their non-relativistic approximations $R_+,\,S_+$, so that the former relativistic factorization will apply also to the present non-relativistic case. And indeed, if that factorized ansatz \rf{529a}-\rf{529b} is inserted in the present non-relativistic equations \rf{555a}-\rf{555b} one finds that both non-relativistic amplitudes obey the {\em same}\/ eigenvalue equation, e.\,g. for the non-relativistic approximation $\oRp(r)$ of the relativistic amplitude $\nRp(r)$
   \begin{eqnarray}
   -\frac{\hbar^2}{2M_e}\left\{\frac{{\rm d}^2\oRp(r)}{{\rm d}r^2}+\frac{2}{r}\frac{{\rm d}\oRp(r)}{{\rm d}r}-\frac{2}{r^2}\,\oRp(r)\right\}-\hbar c\eAo\cdot\oRp(r)&=&\Esz\cdot\oRp(r)\;.\nonumber\\
   \label{556}
   &&
   \end{eqnarray}

Naturally in view of the successful relativistic ansatz \rf{531a}-\rf{531b}, one tries here for the non-relativistic situation quite analogously
   \begin{subequations}
   \begin{align}
   \label{557a}
   \oRp(r)&=\po\cdot r^\nuo\exp[-\beto r]\\
   \label{557b}
   \oSp(r)&=\fo\cdot r^\nuo\exp[-\beto r]\;.
   \end{align}
   \end{subequations}
The ansatz parameters $\nuo$ and $\beto$ are then determined from equations~\rf{556} as
   \begin{subequations}
   \begin{align}
   \label{558a}
   \beto&=\frac{e^2M_e}{2\hbar^2}\;=\;\frac{1}{2\ab}\\
   \label{558b}
   \nuo&=1
   \end{align}
   \end{subequations}
which of course is nothing else than the non-relativistic limit of their original counterparts $\beta$~\rf{521} and $\nu$~\rf{522}. Furthermore, the non-relativistic eigenvalue $\Esz$ is found as
   \begin{eqnarray}
   \label{559}
   \Esz&=&-\frac{1}{8}\,\als^2\cdot M_ec^2\;=\;-\frac{e^2}{8\ab}
   \end{eqnarray}
which again is the non-relativistic limit of the mass eigenvalue $M_2c^2$~\rf{534} or the convential energy $\Econv$~\rf{524}. (Recall here that the {\em non-relativistic}\/ limits are deduced from their {\em relativistic}\/ counterparts through expanding the latter with respect to the fine-structure constant $\als$.)

But concerning the non-relativistic form of the wave function itself, one is referred back to the original ansatz~\rf{57a} for the ``positive'' Pauli spinor $\zphip\vr$ and observes that the corresponding ``negative'' spinor $\zphim\vr$~\rf{57b} is neglected for the non-relativistic approach. Thus the non-relativistic spinor solution for the present fermionic ($\bbar_e=0$) state $\zpdh$ reads
   \begin{eqnarray}
   \zphip\vr&\Rightarrow&\cos\vartheta\,\oRp(r)\cdot\zopp+{\rm e}^{i\phi}\sin\vartheta\,\oSp(r)\cdot\zopm\nonumber\\
   \label{560}
   &=&r\,{\rm e}^{-\beto r}\left\{\po\cos\vartheta\,\zopp+\fo\,{\rm e}^{i\phi}\sin\vartheta\,\zopm\right\}\,.
   \end{eqnarray}
Imposing now upon this solution the non-relativistic normalization condition analogous to~\rf{399} yields the following constraint for the non-relativistic normalization constants $\po,\,\fo$:
   \begin{eqnarray}
   \label{561}
   \po^2+2\fo^2&=&\frac{1}{8\ab^5}\;.
   \end{eqnarray}
In order to satisfy this condition trivially, one introduces the {\em degeneracy parameter}\/ $\Theta$ through putting
   \begin{subequations}
   \begin{align}
   \label{562a}
   \po&=\frac{1}{\sqrt{8\ab^5}}\cos\Theta\\
   \label{562b}
   \fo&=\frac{1}{\sqrt{16\ab^5}}\sin\Theta\;,
   \end{align}
   \end{subequations}
which then recasts the non-relativistic solution~\rf{560} into its final form
   \begin{eqnarray}
   \label{563}
   \zphip\vr&\Rightarrow&\frac{r\,{\rm e}^{-\frac{r}{2\ab}}}{\sqrt{8\ab^5}}\left\{\cos\Theta\cos\vartheta\cdot\zopp+\frac{{\rm e}^{i\phi}}{\sqrt{2}}\sin\Theta\sin\vartheta\cdot\zopm\right\}\,.\qquad\quad
   \end{eqnarray}

This non-relativistic RST result should now be compared to the non-relativistic limit of the conventional form~\rf{518a} of the relativistic $\zpdh$ state. To this end, one first expands the relativistic normalization factor $N_c$~\rf{520} with respect to the fine-structure constant $\als$ which yields in the lowest order of approximation
   \begin{eqnarray}
   \label{564}
   N_c&\cong&\frac{1}{\sqrt{24\ab^5}}\;,
   \end{eqnarray}
and if furthermore the decomposition of the basis spinor $\zdee$ with respect to our {\em fermionic}\/ ($\bbar_e=0$) basis $\{\zopp,\zopm\}$ is applied, i.\,e.
   \begin{eqnarray}
   \label{565}
   \zdee&=&\sqrt{2}\left\{\cos\vartheta\cdot\zopp-\frac{1}{2}\,{\rm e}^{i\phi}\sin\vartheta\cdot\zopm\right\}\,,
   \end{eqnarray}
then the non-relativistic limit of the conventional Pauli spinor~\rf{518a} appears as
   \begin{eqnarray}
   \label{566}
   \zphip\vr&\Rightarrow&\frac{r\,{\rm e}^{-\frac{r}{2\ab}}}{\sqrt{12\ab^5}}\left\{\cos\vartheta\cdot\zopp-\frac{1}{2}\,{\rm e}^{i\phi}\sin\vartheta\cdot\zopm\right\}\,.
   \end{eqnarray}
Thus the comparison of this conventional result to the degenerate RST state~\rf{563} fixes the degeneracy parameter $\Theta$ to
   \begin{subequations}
   \begin{align}
   \label{567a}
   \cos\Theta&=\sqrt{\frac{2}{3}}\\
   \label{567b}
   \sin\Theta&=-\sqrt{\frac{1}{3}}\;.
   \end{align}
   \end{subequations}
So we see that the conventional approach provides a somewhat too simple transition~\rf{566} to the non-relativistic limit by unnecessarily fixing the degeneracy parameter $\Theta$ which, however, by the RST approach~\rf{563} is correctly left undetermined!

Finally, we have to reassure that the non-relativistic RST functional $\tEEfT$~\rf{421} actually does confirm the degeneracy effect, i.\,e. its value upon the non-relativistic RST solution~\rf{563} must be independent of the degeneracy parameter $\Theta$ introduced by equations \rf{562a}-\rf{562b}. In order to verify this by a few brief arguments, one first observes that through the present neglection of the exchange effects and of the magnetism that non-relativistic functional $\tEEfT$~\rf{421} becomes cut down to the following form
   \begin{eqnarray}
   \label{568}
   \tEEfT&\Rightarrow&\tEEfTo\;=\;\Ekin+\ERe+\lambdapz\cdot\NNDz+\lambdaGe\cdot\NNGe\;.
   \end{eqnarray}
But here the non-relativistic normalization condition~\rf{3130} is obviously satisfied by our present result~\rf{563}, and the same is true also for the non-relativistic Poisson identity~\rf{420a}. This is easily seen by explicitly calculating both mass equivalents $\MMefe c^2$ and $\MMefz c^2$~\rf{348a} which yields by simple integration
   \begin{eqnarray}
   \label{569}
   \MMefe&=&\MMefz\;=\;-\frac{e^2}{8\ab}\;,
   \end{eqnarray}
so that the gauge field energy $\ERe$ becomes
   \begin{eqnarray}
   \label{570}
   \ERe&=&\MMefe c^2+\MMefz c^2\;=\;-\frac{e^2}{4\ab}\;.
   \end{eqnarray}
Observe here that even this gauge field energy alone is independent of the degeneracy parameter $\Theta$. But this then must hold also for the kinetic energy $\Ekinz$ of the second particle. Indeed, the explicit integration for $\Ekinz$~\rf{3116a} with reference to the present solution $\zphip\vr$~\rf{563} immediately yields
   \begin{eqnarray}
   \label{571}
   \Ekinz&=&\frac{e^2}{8\ab}\;.
   \end{eqnarray} 
Thus adding up both contributions \rf{570} and \rf{571} to the total non-relativistic energy $\tEEfTo$~\rf{568} yields
   \begin{eqnarray}
   \label{572}
   \tEEfTo&=&-\frac{e^2}{8\ab}\;=\;-\frac{1}{8}\,\als^2M_ec^2
   \end{eqnarray}
in perfect agreement with the short-hand expansion of the relativistic result \rf{535}-\rf{536}. Observe also that in the non-relativistic limit the total energy $\tEEfTo$~\rf{572} coincides again with the non-relativistic eigenvalue $\Esz$~\rf{559}!

Summarizing, the constructed RST energy functional for the fermionic one-particle systems meets with all the logical demands which one can reasonably impose on such a functional. Surely, this yields sufficient motivation to test now also its viability for the more complicated physical situations, such as, e.\,g., a non-trivial two-particle system where the Dirac particles occupy {\em bosonic}\/ states!

\subsection{Simplest Bosonic State \bf($\bbar=\frac{1}{2}$)}

A bosonic state naturally occurs for the positronium groundstate~\cite{c11}. For such a bound system of a positron ($a=1$) and an electron ($a=2$), the former two-particle eigenvalue system \rf{513a}-\rf{514c} does also apply, however with identical rest masses and mass eigenvalues, i.\,e.
   \begin{subequations}
   \begin{align}
   \label{573a}
   M_e&=M_p\,\doteqdot\,M\\
   \label{573b}
   M_1&=-M_2\,\doteqdot\,-M_*\\
   \label{573c}
   \bbar_e&=\bbar_p\,=\,\frac{1}{2}\;.
   \end{align}
   \end{subequations}
Furthermore, it is well known that positronium emits a one-particle spectrum, i.\,e. the two oppositely charged constituents must occupy (almost) identical quantum states. This requires that the {\em Dirac charge densities}\/ $\ako\vr$ are identical
   \begin{eqnarray}
   \label{574}
   \eko\vr&\equiv&\zko\vr\;\doteqdot\;\pko\vr
   \end{eqnarray}
so that the {\em Maxwell densities}\/ $\ajo\vr$ \rf{243a}-\rf{243b} differ in sign:
   \begin{eqnarray}
   \label{575}
   \ejo\vr&\equiv&-\zjo\vr\;\doteqdot\;\pjo\vr\;,
   \end{eqnarray}
which then also holds for the electric potentials $\aAo\vr$ generated by these densities according to the Poisson equations~\rf{33}:
   \begin{eqnarray}
   \label{576}
   \eAo\vr&\equiv&-\zAo\vr\;\doteqdot\;\pAo\vr\;.
   \end{eqnarray}
Concerning now the magnetic fields $\vHa\vr$, one encounters two possibilities: either the magnetic fields are antiparallel ($\vHe\vr\equiv-\vHz\vr$, \/{\em para-positronium}\/) or they are parallel (\/{\em ortho-positronium}\/~\cite{c11}):
   \begin{subequations}
   \begin{align}
   \label{577a}
   \vAe\vr&\equiv\vAz\vr\;\doteqdot\;\vAp\vr\\
   \label{577b}
   \vHe\vr&\equiv\vHz\vr\;\doteqdot\;\vHp\vr\;.
   \end{align}
   \end{subequations}
(In the conventional theory, these are the para-states $\eSo$.) Clearly, such a coincidence of the electromagnetic fields requires a similar coincidence of the corresponding Dirac three-currents $\vka\vr$ (or Maxwell currents $\vja\vr$, resp.). Indeed, it is easy to see by reference to the Poisson equations~\rf{34} that the magnetic demands \rf{577a}-\rf{577b} imply the identity of the Maxwell currents $\vja\vr$ or antiparallelity of the Dirac currents $\vka\vr$, resp.
   \begin{subequations}
   \begin{align}
   \label{578a}
   \vje\vr&\equiv\vjz\vr\;\doteqdot\;\vjp\vr\\
   \label{578b}
   \vke\vr&\equiv-\vkz\vr\;\doteqdot\;\vkp\vr\;.
   \end{align}
   \end{subequations}

Of course, such a far-reaching coincidence of the matter densities and gauge fields of both particles must be induced by the corresponding relationship of the wave functions. But here it is a simple thing to satisfy both density requirements \rf{574} and \rf{578b}; indeed, resorting to the original definitions of those densities in terms of the Pauli spinors $\aphipm\vr$, cf.~\rf{3122} and \rf{422}, one puts
   \begin{subequations}
   \begin{align}
   \label{579a}
   \zphip\vr&\equiv\ephip\vr\;\doteqdot\;\pphip\vr\\
   \label{579b}
   \zphim\vr&\equiv-\ephim\vr\;\doteqdot\;-\pphim\vr\;,
   \end{align}
   \end{subequations}
which then reads in terms of the wave amplitudes \rf{57a}-\rf{57b}
   \begin{subequations}
   \begin{align}
   \label{580a}
   \eMRp&\equiv\zMRp\;\doteqdot\;\pRp\\
   \label{580b}
   \eMSp&\equiv\zMSp\;\doteqdot\;\pSp\\
   \label{580c}
   \eMRm&\equiv-\zMRm\;\doteqdot\;\pRm\\
   \label{580d}
   \eMSm&\equiv-\zMSm\;\doteqdot\;\pSm\;.
   \end{align}
   \end{subequations}
Accordingly, the relativistic two-particle eigenvalue equations \rf{340a}-\rf{340b} must collapse to the following one-particle eigenvalue problem:
   \begin{eqnarray}
   \label{581}
   i\vsigma\sdot\vnabla\pphipm-\pAo\cdot\pphimp-\big(\vAp\sdot\vsigma\big)\pphipm &=&\frac{M_*\pm M}{\hbar}\,c\cdot\pphimp\;.\qquad
   \end{eqnarray}
This abstract system may be rewritten also in terms of the (real-valued) wave amplitudes $\pRpm,\,\pSpm$, namely by simply introducing those identifications \rf{580a}-\rf{580d} into the two-particle system \rf{513a}-\rf{514c} which then collapses to the following one-particle system
   \begin{subequations}
   \begin{align}
   \label{582a}
   \frac{\partial\tRp}{\partial r}+\frac{1}{r}\frac{\partial\tSp}{\partial\vartheta}-\pAo\cdot\tRm &=\frac{M+M_*}{\hbar}\,c\cdot\tRm\\
   \label{582b}
   \frac{\partial\tSp}{\partial r}-\frac{1}{r}\frac{\partial\tRp}{\partial\vartheta}-\pAo\cdot\tSm &=\frac{M+M_*}{\hbar}\,c\cdot\tSm\\
   \label{582c}
   \frac{1}{r}\frac{\partial(r\cdot\tRm)}{\partial r}-\frac{1}{r}\frac{\partial\tSm}{\partial\vartheta}+\pAo\cdot\tRp &=\frac{M-M_*}{\hbar}\,c\cdot\tRp\\
   \label{582d}
   \frac{1}{r}\frac{\partial(r\cdot\tSm)}{\partial r}+\frac{1}{r}\frac{\partial\tRm}{\partial\vartheta}+\pAo\cdot\tSp &=\frac{M-M_*}{\hbar}\,c\cdot\tSp\;.
   \end{align}
   \end{subequations}
Here the magnetic field is neglected ($\vAp\Rightarrow0$), see ref.~\cite{c11} for its inclusion, and the modified wave amplitudes $\tRpm,\,\tSpm$ are defined through
   \begin{subequations}
   \begin{align}
   \label{583a}
   \tRpm&\doteqdot\sqrt{r\sin\vartheta}\cdot\pRpm\\
   \label{583b}
   \tSpm&\doteqdot\sqrt{r\sin\vartheta}\cdot\pSpm\;.
   \end{align}
   \end{subequations}
Naturally, this one-particle eigenvalue system \rf{582a}-\rf{582d} couples to merely a single gauge potential $\pAo$, cf.~\rf{576}; and consequently the two Poisson equations~\rf{33} become contracted to a single one for this residual potential $\pAo$
   \begin{eqnarray}
   \label{584}
   \Delta\pAo&=&-4\pi\als\,\pko\;=\;-\als\,\frac{\tRp^2+\tSp^2+\tRm^2+\tSm^2}{r\sin\vartheta}
   \end{eqnarray}
whose formal standard solution is as usual
   \begin{eqnarray}
   \pAo(r,\vartheta)&=&\frac{\als}{4\pi}\int\frac{{\rm d}^3\vec{r}\!\;'}{r'\sin\vartheta'}\,\frac{\tRp^2(r',\vartheta')+\tSp^2(r',\vartheta')+\tRm^2(r',\vartheta')+\tSm^2(r',\vartheta')}{\|\vec{r}-\vec{r}\!\;'\|}\;.\nonumber\\
   \label{585}
   &&
   \end{eqnarray}

Comparing the present bosonic system \rf{582a}-\rf{582d} together with the Poisson equation~\rf{584} to its fermionic counterpart \rf{528a}-\rf{528d}, the crucial difference must refer of course to the electric potential: whereas the Coulomb potential $\eAo$~\rf{517} is fixed from the outside for the fermionic case (so that one could obtain an exact solution), the present interaction potential $\pAo$~\rf{585} is part of the dynamics and therefore must be determined simultaneously with the wave amplitudes. Of course, this circumstance spoils the possibility of finding an exact solution for the present bosonic case; and thus one has to resort to some approximative method, e.\,g. the variational method based upon just the energy functional $\tEfT$.

In this sense, it becomes necessary to specialize also the general result for $\tEfT$~\rf{414} to the present situation where the magnetic and exchange fields are neglected. Observing here first the circumstance that the identifications \rf{580a}-\rf{580d} ultimately leave us with the one-particle fields $\tRpm,\,\tSpm$ \rf{583a}-\rf{583b}, the energy functional $\tEfT$ is found to be simplified to the corresponding one-particle contributions, i.\,e.
   \begin{eqnarray}
   \label{586}
   \tEfT&\Rightarrow&2Mc^2\cdot\tMZ^2+4\!\:\tTkin+\ERe+(\lambDe+\lambDz)\cdot\tND+\lambdaGe\cdot\tNGe\;,\qquad\qquad
   \end{eqnarray}
with the one-particle objects being given in a self-evident way through (${\rm d}^2\vec{r}\doteqdot\mbox{$r\,{\rm d}r{\rm d}\vartheta$}$)
   \begin{subequations}
   \begin{align}
   \label{587a}
   \MZe^2&=\MZz^2\,\doteqdot\,\tMZ^2\,=\,\frac{1}{2}\int\!\!{\rm d}^2\vec{r}\;\left\{\tRp^2+\tSp^2-\tRm^2-\tSm^2\right\}\\
   \Tkine&=\Tkinz\,\doteqdot\,\tTkin\nonumber\\
   &\hspace{3em}=\,\frac{\hbar c}{4}\int\!\!{\rm d}^2\vec{r}\;\left\{\tRm\frac{\partial\tRp}{\partial r}-\frac{\tRp}{r}\frac{\partial(r\tRm)}{\partial r}+\tSm\frac{\partial\tSp}{\partial r}-\frac{\tSp}{r}\frac{\partial(r\tSm)}{\partial r}\right.\nonumber\qquad\\
   &{}\hspace{8.5em}\left.  {}+\frac{1}{r}\left[\tRp\frac{\partial\tSm}{\partial\vartheta}-\tSm\frac{\partial\tRp}{\partial\vartheta}-\tSp\frac{\partial\tRm}{\partial\vartheta}+\tRm\frac{\partial\tSp}{\partial\vartheta}\right]\right\}\nonumber\\
   \label{587b}
   &{}\\
   \label{587c}
   \NDe&=\NDz\,\doteqdot\,\tND\,=\,\frac{1}{2}\int\!\!{\rm d}^2\vec{r}\;\left\{\tRp^2+\tSp^2+\tRm^2+\tSm^2\right\}-1\,\equiv\,0\\
   \label{587d}
   \ERe&=-\frac{\hbar c}{4\pi\als}\idr\|\vnabla\pAo\|^2\\
   \label{587e}
   \tMe c^2&=-\frac{\hbar c}{2}\int\!\!{\rm d}^2\vec{r}\;\pAo\cdot\left\{\tRp^2+\tSp^2+\tRm^2+\tSm^2\right\}\\
   \label{587f}
   \tNGe&\doteqdot\ERe-\tMe c^2\,\equiv\,0\;.
   \end{align}
   \end{subequations}
But with these arrangements it becomes an easy exercise to deduce {\em both}\/ the one-particle eigenvalue equations \rf{582a}-\rf{582d} {\em and}\/ the Poisson equation~\rf{584} as the variational equations due to the truncated RST energy functional $\tEfT$~\rf{586}.

\subsection{Positronium Groundstate \bf($\eeSo$)}

The bosonic variational principle~\rf{586} with the corresponding variational equations \rf{582a}-\rf{585} describes that part of the positronium spectrum which is due to the boson number $\bbar_e=\bbar_p=\frac{1}{2}$ and which additionally admits the use of real-valued wave amplitudes $\pRpm,\,\pSpm$ \rf{580a}-\rf{580d}. The question of completeness of the positronium spectrum will require an extra discussion, but the groundstate (as the state of highest symmetry) will surely be a member of the present subset defined by the energy functional~\rf{586}. As a test of this supposition one may assume that the groundstate wave amplitudes $\tRpm(r,\vartheta),\,\tSpm(r,\vartheta)$ are approximately spherically symmetric, i.\,e. we put for the groundstate
   \begin{subequations}
   \begin{align}
   \label{588a}
   \tRpm(r,\vartheta)&\Rightarrow\dRpm(r)\\
   \label{588b}
   \tSpm(r,\vartheta)&\Rightarrow\dSpm(r)\;,
   \end{align}
   \end{subequations}
and simultaneously we pass over to the non-relativistic approximation. If RST has something to do with reality, one must recover (at least approximately) for this situation the non-relativistic positronium groundstate together with the excited states, because of the non-relativistic angular momentum degeneracy. Now in order to verify this expectation, one first looks for the non-relativistic version ($\tEEfTo$) of the original energy functional $\tEfT$~\rf{586} in the spherically symmmetric approximation \rf{588a}-\rf{588b}, i.\,e.
   \begin{eqnarray}
   \label{589}
   \tEEfTo&=&2\dEkin+\dERe+2\dlambdaS\dNND+\lambdaGe\dNNGe\;,
   \end{eqnarray}
with the non-relativistic version $\dEkin$ of the one-particle kinetic energy $\tTkin$ \rf{587b} being given by
   \begin{eqnarray}
   \label{590}
   \tTkin\;\Rightarrow\;\dEkin&=& \frac{\hbar^2}{2M}\cdot\frac{\pi}{2}\int\limits_0^\infty\!\!\dif{r}\,r\,\left(\frac{\dif{\dRp}(r)}{\dif{r}}\right)^2\,,
   \end{eqnarray}
furthermore the spherically symmetric form $\dERe$ of the gauge field energy $\ERe$ \rf{587d} is ($\pAo(r,\vartheta)\Rightarrow\dotAo(r)$)
   \begin{eqnarray}
   \label{591}
   \ERe\;\Rightarrow\;\dERe&=&-\frac{\hbar c}{\als}\int\limits_0^\infty\!\!\dif{r}\,r^2\,\left(\frac{\dif{\dotAo}(r)}{\dif{r}}\right)^2\,,
   \end{eqnarray}
the mass equivalent $\tMe c^2$~\rf{587e} becomes
   \begin{eqnarray}
   \label{592}
   \tMe c^2\;\Rightarrow\;\dMMe c^2&=&-\hbar c\cdot\frac{\pi}{2}\int\limits_0^\infty\!\!\dif{r}\,r\,\dotAo(r)\dRp^2(r)\;,
   \end{eqnarray}
and finally the normalization condition \rf{587c} appears in the spherically symmetric approximations as
   \begin{eqnarray}
   \label{593}
   \tNND\;\Rightarrow\;\dNND&\doteqdot&\frac{\pi}{2}\int\limits_0^\infty\!\!\dif{r}\,r\,\dRp^2(r)-1\;\equiv\;0\;.
   \end{eqnarray}

The non-relativistic form of the eigenvalue equations \rf{582a}-\rf{582d} reads now for the spherically symmetric wave amplitude $\dRp(r)$
   \begin{eqnarray}
   \label{594}
   -\frac{\hbar^2}{2M}\frac{1}{r}\frac{\rm d}{\dif{r}}\left(r\cdot\frac{\dif{\dRp(r)}}{\dif{r}}\right)-\hbar c\dotAo(r)&=&\ES\cdot\dRp(r)\;.
   \end{eqnarray}
Of course, this equation can also be deduced by extremalizing the energy functional $\tEEfTo$~\rf{589} with respect to the wave amplitude $\dRp(r)$ ($\delta\tEEfTo=0$). For the sake of simplicity, the non-relativistic spin degeneracy is suppressed here ($\leadsto\,\dSp(r)\Rightarrow0$). Finally, the Poisson equation for the spherically symmetric gauge potential $\dotAo(r)$ appears as the corresponding variational equation of $\tEEfTo$ as
   \begin{eqnarray}
   \label{595}
   \frac{1}{r^2}\frac{\rm d}{\dif{r}}\left(r^2\cdot\frac{\dif{\dotAo(r)}}{\dif{r}}\right)\;\equiv\;\Delta_{(r)}\dotAo&=&-\frac{\pi}{2}\,\als\,\frac{\dRp^2(r)}{r}\;.
   \end{eqnarray}

The present positronium groundstate problem \rf{589}-\rf{595} has been treated in great detail in a preceding paper, see equations \mbox{(III.27\hspace{-.5pt})}-\mbox{(III.34\hspace{-.5pt})} of ref.\,\cite{c10}; and it was found that the corresponding RST predictions for the non-relativistic positronium spectrum are in acceptable agreement with the conventional spectrum
   \begin{eqnarray}
   \label{596}
   \Econv&=&-\frac{e^2}{4\ab}\frac{1}{(n_p+1)^2}\;\cong\;-\frac{6.8029...}{(n_p+1)^2}\;[{\rm eV}]\;,\\
   \nonumber&&\hspace{3em}{{}\atop\d(n_p=0,1,2,...)}
   \end{eqnarray}
see fig.\,2 of ref.\,\cite{c10}. It is true, there are certain deviations ($\sim10\%$) of the RST predictions from the conventional results~\rf{596} but these deviations can be attributed to just the use of the spherically symmetric approximation. This logical situation lends itself now to a further test of the RST energy functional~\rf{586}; namely one can apply the (non-relativistic) variational principle $\delta\stEEfT=0$ (due to the non-relativistic approximation $\stEEfT$ of $\tEET$~\rf{586}) to that excited positronium state (i.\,e.~$\zePe$) which has in the spherically symmetric approximation the same orbital symmetry as the fermionic one-particle state $\zpdh$ treated in the preceding subsection. Because of the non-relativistic angular momentum degeneracy, the non-relativistic RST energy $\stEET$ of that excited positronium state~$\zePe$ (conventional classification) should turn out according to \rf{596} as ($n_p=1$)
   \begin{eqnarray}
   \label{597}
   \stEET\Big|_{\zePe}&=&-\frac{e^2}{4\ab}\frac{1}{(1+1)^2}\;=\;-\frac{e^2}{16\ab}\;=\;-1.70...\ [{\rm eV}]\;.
   \end{eqnarray}
Of course, we are not able here to verify this claim~\rf{597} exactly, because the coupled RST system of eigenvalue and Poisson equations is not exactly solvable, not even in the non-relativistic limit; but one could accept the verification of \rf{597} as a support of the RST energy functional if the corresponding RST prediction turns out to meet with the requirement~\rf{597} up to that magnitude of inaccuracy ($\sim10\%$) which showed up also for the groundstate (see fig.\,2 of ref.\,\cite{c10}) and is to be attributed to the use of the spherically symmetric approximation.

\subsection{Excited Positronium State \bf($\zePe$)}

Concerning the excited positronium states, it must appear as a matter of course that exact (relativistic or non-relativistic) solutions of the coupled matter and gauge field system \rf{582a}-\rf{585} will not be attainable. Therefore we have to resort again to some approximation which we adopt here in a two-fold way, namely in form of a combination of the non-relativistic limit and the spherically symmetric approximation.

Turning first to the latter kind of approximation, one can of course not take over the former assumptions \rf{588a}-\rf{588b} for the {\em groundstate}\/ to the presently considered {\em excited}\/ state; but rather for the wave amplitudes $\tRpm,\,\tSpm$ we resort to that angular dependence \rf{529a}-\rf{529b} of the fermionic one-particle state $\zpdh$, i.\,e. we put
   \begin{subequations}
   \begin{align}
   \label{598a}
   \tRpm(r,\vartheta)&\Rightarrow\sRpm(r)\cdot\cos\vartheta\\
   \label{598b}
   \tSpm(r,\vartheta)&\Rightarrow\sSpm(r)\cdot\sin\vartheta\;.   
   \end{align}
   \end{subequations}
This assumption for the angular dependence of the wave amplitudes is further complemented by the assumption of strict $SO(3)$ symmetry for the gauge potential $\pAo(r,\vartheta)$, i.\,e. we put
   \begin{eqnarray}
   \label{599}
   \pAo(r,\vartheta)&\Rightarrow&\sAo(r)\;.
   \end{eqnarray}
These symmetry assumptions recast the exact positronium eigenvalue problem \rf{582a}-\rf{584} to the following approximate (but still relativistic) form:
   {\setlength{\lineskip}{2.5ex}
   \begin{subequations}
   \begin{align}
   \label{5100a}
   \frac{\dif{\sRp(r)}}{\dif{r}}+\frac{1}{r}\cdot\sSp(r)-\sAo(r)\cdot\sRm(r)&=\frac{M+M_*}{\hbar}\,c\cdot\sRm(r)\\
   \label{5100b}
   \frac{\dif{\sSp(r)}}{\dif{r}}+\frac{1}{r}\cdot\sRp(r)-\sAo(r)\cdot\sSm(r)&=\frac{M+M_*}{\hbar}\,c\cdot\sSm(r)\\
   \label{5100c}
   \frac{1}{r}\frac{\rm d}{\dif{r}}\,\big(r\sRm(r)\big)-\frac{1}{r}\cdot\sSm(r)+\sAo(r)\cdot\sRp(r) &=\frac{M-M_*}{\hbar}\,c\cdot\sRp(r)\\
   \label{5100d}
   \frac{1}{r}\frac{\rm d}{\dif{r}}\,\big(r\sSm(r)\big)-\frac{1}{r}\cdot\sRm(r)+\sAo(r)\cdot\sSp(r) &=\frac{M-M_*}{\hbar}\,c\cdot\sSp(r)\\
   \label{5100e}
   \frac{1}{r^2}\frac{\rm d}{\dif{r}}\,\big(r^2\sAo(r)\big)\,\equiv\,\Delta_{(r)}\sAo(r)&=-\frac{\pi}{4}\,\als\,\frac{\sRp^2+\sSp^2+\sRm^2+\sSm^2}{r}\;.
   \end{align}
   \end{subequations}}
Here it is also interesting to point at the differences of the present {\em two-particle}\/ eigenvalue system \rf{5100a}-\rf{5100d} compared to the former {\em one-particle}\/ eigenvalue problem \rf{530a}-\rf{530d}. Clearly these differences concern the interaction potential $\sAo(r)$ and are responsible for the different energy spectra of the one- and two-particle cases, resp. Recall here the fact that in the non-relativistic conventional theory these differences simply consist in replacing the rest mass $M$ of the one-particle problem (hydrogen) by $\frac{M}{2}$ in order to obtain the two-particle spectrum~\rf{596}!

It is also important to observe that the assumption of angular dependence \rf{598a}-\rf{598b} together with the strict spherical symmetry of the gauge potential $\sAo(r)$~\rf{599} does not spoil the principle of minimal energy. Indeed it is an easy exercise to demonstrate that the present approximate eigenvalue equations \rf{5100a}-\rf{5100e} are the variational equations due to the functional $\stEfT$ which arises from the original functional $\tEfT$~\rf{586} through applying the assumptions \rf{598a}-\rf{599} of spherical symmetry:
   \begin{eqnarray}
   \nonumber
   \tEfT\;\Rightarrow\;\stEfT&=&2Mc^2\cdot{\sMZ}^2+4\sTkin+\sERe+(\lambDe+\lambDz)\cdot\sND+\lambdaGe\cdot\sNGe\;.\\
   \label{5101}
   &&
   \end{eqnarray}
Here the relativistic mass renormalization ${\sMZ}^2$ is deduced from the original $\tilde{\MZ}^2$~\rf{587a} as
   \begin{eqnarray}
   \label{5102}
   {\sMZ}^2&=&\frac{\pi}{4}\int\limits_0^\infty\!\!\dif{r}\,r \left\{\big(\sRp(r)\big)^2+\big(\sSp(r)\big)^2-\big(\sRm(r)\big)^2-\big(\sSm(r)\big)^2\right\}\,,\qquad\qquad
   \end{eqnarray}
next the original kinetic energy $\tTkin$~\rf{587b} transcribes to $\sTkin$ as
   \begin{eqnarray}
   \nonumber
   \tTkin\;\Rightarrow\;\sTkin&=&\frac{\pi}{8}\,\hbar c\int\limits_0^\infty\!\!\dif{r}\,r \left\{\sRm(r)\cdot\frac{\dif{\sRp(r)}}{\dif{r}}-\frac{\sRp(r)}{r}\cdot\frac{\dif{(r\sRm(r))}}{\dif{r}}\right.\\
   \nonumber &&\hspace{-8em}\left.+\sSm(r)\cdot\frac{\dif{\sSp(r)}}{\dif{r}}-\frac{\sSp(r)}{r}\cdot\frac{\dif{(r\sSm(r))}}{\dif{r}}+ \frac{2}{r}\left[\sRp(r)\cdot\sSm(r)+\sSp(r)\cdot\sRm(r)\right]\right\}\,,\\
   \label{5103}
   &&
   \end{eqnarray}
furthermore the gauge field energy $\ERe$~\rf{587d} becomes a purely radial integral due to the symmetry assumption~\rf{599}
   \begin{eqnarray}
   \label{5104}
   \ERe\;\Rightarrow\;\sERe&=&-\frac{\hbar c}{\als}\int\limits_0^\infty\!\!\dif{r}\,r^2 \left(\frac{\dif{\sAo(r)}}{\dif{r}}\right)^2\,,
   \end{eqnarray}
the normalization constraint $\tND$~\rf{587c} appears now as
   \begin{eqnarray}
   \nonumber
   \tND\;\Rightarrow\;\sND&=&\frac{\pi}{4}\int\limits_0^\infty\!\!\dif{r}\,r \left\{\big(\sRp(r)\big)^2+\big(\sSp(r)\big)^2+\big(\sRm(r)\big)^2+\big(\sSm(r)\big)^2\right\}-1\\
   \nonumber
   &&\hspace{24em}\equiv0\;,\\
   \label{5105}
   &&
   \end{eqnarray}
and finally the electric Poisson identity~\rf{587f} reads
   \begin{eqnarray}
   \label{5106}
   \tNGe\;\Rightarrow\;\sNGe&=&\sERe-\sMe c^2
   \end{eqnarray}
with the electric mass equivalent $\tMe c^2$~\rf{587e} adopting its new form as
   \begin{eqnarray}
   \nonumber
   \tMe c^2\;\Rightarrow\;\sMe c^2&=&-\frac{\pi}{4}\,\hbar c\int\limits_0^\infty\!\!\dif{r}\,r\,\sAo(r) \left\{\big(\sRp(r)\big)^2+\big(\sSp(r)\big)^2+\big(\sRm(r)\big)^2\right.\\
   \nonumber
   &&\hspace{17em}\left.{}+\big(\sSm(r)\big)^2\right\}\,.\\
   \label{5107}
   &&
   \end{eqnarray}

The next step of approximation concerns again the neglection of the relativistic effects. Here, the procedure is as usual: Calculating approximately the negative Pauli amplitudes $\sRm(r)$ and $\sSm(r)$ from the equations \rf{5100a}-\rf{5100b} as
   \begin{subequations}
   \begin{align}
   \label{5108a}
   \sRm(r)&\cong\frac{\hbar}{2Mc}\left\{\frac{\dif{\sRp(r)}}{\dif{r}}+\frac{1}{r}\cdot\sSp(r)\right\}\\
   \label{5108b}
   \sSm(r)&\cong\frac{\hbar}{2Mc}\left\{\frac{\dif{\sSp(r)}}{\dif{r}}+\frac{1}{r}\cdot\sRp(r)\right\}
   \end{align}
   \end{subequations} 
and substituting this into the residual equations \rf{5100c}-\rf{5100d} yields again the non-relativistic eigenvalue equations for the positive Pauli amplitudes $\sRp,\,\sSp$, e.\,g. for $\ddRp(r)$ as the non-relativistic version of $\sRp(r)$
   \begin{eqnarray}
   \nonumber
   -\frac{\hbar^2}{2M}\left\{\frac{{\rm d}^2\ddRp(r)}{\dif{r^2}}+\frac{1}{r}\frac{\dif{\ddRp(r)}}{\dif{r}}-\frac{\ddRp(r)}{r^2}\right\}-\hbar c\ddotAo(r)\cdot\ddRp(r) &=&E_*\cdot\ddRp(r)\;.\\
   \label{5109}
   &&
   \end{eqnarray}
The same equation is also found for the spin-down component $\ddSp(r)$, but one may renounce on this if one is willing to omit the corresponding spin degeneracy. The non-relativistic eigenvalue equation~\rf{5109} must be complemented by the non-relativistic version of the Poisson equation~\rf{584}, i.\,e. by neglection of the negative Pauli amplitudes $\sRm,\,\sSm$ and by omission of the positive spin-down amplitude $\ddSp$ (due to the spin degeneracy):
   \begin{eqnarray}
   \label{5110}
   \frac{1}{r^2}\frac{\rm d}{\dif{r}}\left(r^2\cdot\frac{\dif{\ddotAo(r)}}{\dif{r}}\right)\;\equiv\;\Delta_{(r)}\ddotAo(r) &=&-\frac{\pi}{4}\,\als\,\frac{\ddRp^2(r)}{r}\;.
   \end{eqnarray}
Clearly, the coupled non-relativistic system \rf{5109}-\rf{5110} can again be understood to represent the variational equations due to an associated energy functional ($\tEEfTo$, say) which deduces from its relativistic predecessor $\stEfT$~\rf{5101} just through the approximations \rf{5108a}-\rf{5108b}, namely
   \begin{eqnarray}
   \label{5111}
   \stEfT\;\Rightarrow\;\tEEfTo&=&2\ddEkin+\sERe+2\ddlambdaS\cdot\ddNND+\lambdaGe\cdot\ddNNGe
   \end{eqnarray}
with
   \begin{subequations}
   \begin{align}
   \label{5112a}
   \ddEkin&=\frac{\hbar^2}{2M}\frac{\pi}{4}\int\limits_0^\infty\!\!\dif{r}\,r \left\{\left(\frac{\dif{\ddRp(r)}}{\dif{r}}\right)^2+\frac{\ddRp^2(r)}{r^2}\right\}\\
   \label{5112b}
   \ddNND&=\frac{\pi}{4}\int\limits_0^\infty\!\!\dif{r}\,r\,\ddRp^2(r)-1\,\equiv\,0\\
   \label{5112c}
   \ddNNGe&=\sERe-\ddMMe c^2\,\equiv\,0\\
   \label{5112d}
   \ddMMe c^2&=-\hbar c\,\frac{\pi}{4}\int\limits_0^\infty\!\!\dif{r}\,r\,\ddotAo(r)\big(\ddRp(r)\big)^2\;.  
   \end{align}
   \end{subequations}
The Lagrangean multiplier $\ddlambdaS$ is half the non-relativistic limit of the sum $\lambDe+\lambDz$, namely
   \begin{eqnarray}
   \nonumber
   \lambDe+\lambDz=(M_1-M_2)c^2&\Rightarrow&(-Mc^2-E_*)-(Mc^2+E_*)\Rightarrow-2E_*\doteqdot2\ddlambdaS\;,\\
   \label{5113}
   &&
   \end{eqnarray}
cf. \rf{351a}-\rf{351b}.

The non-relativistic form $\tEEfTo$~\rf{5111} of our {\em RST principle of minimal energy}\/ is now just the right point of departure for calculating approximately the (non-relativistic) energy of the considered positronium state $\zePe$. However, concerning the intra-atomic situation from the viewpoint of various energy contributions, this variational principle $\delta\tEEfTo$ owns some interesting features to be first elaborated in some detail.

\subsubsection{Virial Theorem}

The total energy $\tEETo$~\rf{5111} of any solution $\ddRp(r),\,\ddotAo(r)$ of the eigenvalue problem \rf{5109}-\rf{5110} is obviously built up exclusively by the sum of the kinetic energy $\ddEkin$ and the gauge field energy $\sERe$, provided the considered solutions satisfy the normalization constraint~\rf{5112b} and the Poisson identity~\rf{5112c}. Thus one first has two equations for the four energy contributions $E_*,\,\ddEkin,\,\sERe,\,\ddMMe c^2$; namely
   \begin{subequations}
   \begin{align}
   \label{5114a}
   \tEETo&=2\ddEkin+\sERe\\
   \label{5114b}
   \sERe&=\ddMMe c^2\;.
   \end{align}
   \end{subequations}
Additionally, a third relation can be obtained from the eigenvalue equation~\rf{5109} by multiplying through that equation with $\ddRp(r)$ and integrating over, using the normalization constraint~\rf{5112b}:
   \begin{eqnarray}
   \label{5115}
   E_*&=&\ddEkin+\ddMMe c^2\;.
   \end{eqnarray}
But the interesting point here is now that a fourth relation does exist, i.\,e.
   \begin{eqnarray}
   \label{5116}
   \sERe&=&4\cdot\big(\ddMMe c^2-E_*\big)\;,
   \end{eqnarray}
and this fact then admits to express all energy contributions in terms of the electric mass equivalent $\ddMMe c^2$
   \begin{subequations}
   \begin{align}
   \label{5117a}
   E_*&=\frac{3}{4}\,\ddMMe c^2\\
   \label{5117b}
   \tEETo&=\frac{1}{2}\,\ddMMe c^2\,=\,\frac{2}{3}\,E_*\\
   \label{5117c}
   \ddEkin&=-\frac{1}{4}\,\ddMMe c^2\;.
   \end{align}
   \end{subequations}
The total kinetic energy of the two-particle system is thus revealed as just half of its electric interaction energy (up to sign), i.\,e.
   \begin{eqnarray}
   \label{5118}
   2\ddEkin&=&-\frac{1}{2}\,\ddMMe c^2\;=\;-\frac{1}{2}\,\sERe\;,
   \end{eqnarray}
and furthermore the total energy $\tEETo$ equals minus the total kinetic energy (i.\,e. twice the kinetic energy per particle)
   \begin{eqnarray}
   \label{5119}
   \tEETo&=&-2\ddEkin\;=\;\frac{1}{2}\,\ddMMe c^2\;.
   \end{eqnarray}
This result is the well-known {\em virial theorem}\/ of classical and quantum mechanics~\cite{c19,c20} which in those conventional theories does apply to systems of particles being bound together via the Coulomb force.

In order to verify the validity of this virial theorem for the present RST situation, one merely has to supply the proof of the fourth energy relation~\rf{5116}. For this purpose, one reconsiders the present variational principle $\delta\tEEfTo=0$~\rf{5111} and subjects the corresponding extremal field configurations to a global spatial dilatation ($r\Rightarrow\lambda\cdot r,\,\lambda=\mbox{const}$), i.\,e.
   \begin{subequations}
   \begin{align}
   \label{5120a}
   \ddRp(r)&\Rightarrow\ddRp(\lambda r)\\
   \label{5120b}
   \ddotAo(r)&\Rightarrow\ddotAo(\lambda r)\;.
   \end{align}
   \end{subequations}
According to this common modification of the matter and gauge fields, the various contributions to the energy functional $\tEEfTo$~\rf{5111} undergo the following changes:
   \begin{subequations}
   \begin{align}
   \label{5121a}
   \ddEkin&\Rightarrow\ddEkin\,,\;\mbox{no change}\\
   \label{5121b}
   \sERe&\Rightarrow\frac{1}{\lambda}\cdot\sERe\\
   \label{5121c}
   \ddNND&\Rightarrow\frac{1}{\lambda^2}\,(\ddNND+1)-1\\
   \label{5121d}
   \ddMMe c^2&\Rightarrow\frac{1}{\lambda^2}\cdot\ddMMe c^2\;.
   \end{align}
   \end{subequations}
Putting these individual changes together, the energy functional $\tEETo$ reacts as follows to the dilatational transformation \rf{5120a}-\rf{5120b}
   \begin{eqnarray}
   \nonumber
   \tEETo\Rightarrow\tEETo(\lambda)&=& 2\ddEkin+\frac{1}{\lambda}\cdot\sERe+\ddlambdaS\left[\frac{1}{\lambda^2}\,(\ddNND+1)-1\right]\\
   \nonumber
   &&\hspace{12em}{}+\lambdaGe\left[\frac{1}{\lambda}\,\sERe-\frac{1}{\lambda^2}\,\ddMMe c^2\right]\,.\\
   \label{5122}
   &&
   \end{eqnarray}
But since we started from an extremal field configuration ($\delta\tEEfTo=0$), we obviously must have
   \begin{eqnarray}
   \label{5123}
   \left.\frac{\dif{\tEETo(\lambda)}}{\dif{\lambda}}\right|_{\lambda=1}&=&0\;,
   \end{eqnarray}
and this just reproduces the desired fourth relation~\rf{5116}, since the Lagrangean multiplier $\ddlambdaS$ is to be identified with the non-relativistic energy eigenvalue $E_*$, see equation~\rf{5113}.

In the present context of this RST virial theorem, it is important to remark that the extremal field configuration must not necessarily be an exact solution of the associated eigenvalue problem \rf{5109}-\rf{5110}. Rather, it is merely necessary that the dilatated fields \rf{5120a}-\rf{5120b} remain members of that subspace of trial functions which is considered for extremalizing the energy functional $\tEETo$\rf{5111}! This then means that the RST virial theorem does apply also to the {\em approximate}\/ variational solutions of the original eigenvalue problem \rf{5109}-\rf{5110}. In order to present an example of this type, consider the set of hydrogen-like wave functions which appear as the product of some polynomial $P_\nu(r)$ times an exponential function
   \begin{subequations}
   \begin{align}
   \label{5124a}
   \ddRp(r)&\Rightarrow\ddot{N}\cdot P_\nu(r)\cdot\exp[-\beta r]\\
   \label{5124b}
   P_\nu(r)&=\sum_{n=0}^\nu a_nr^n\\
   \nonumber
   &{}\hspace{-2.5em}(\ddot{N}\,=\,\mbox{normalization constant})\;.
   \end{align}
   \end{subequations}
Since the polynomial coefficients $a_n$ and the decay parameter $\beta$ are understood to work as the variational parameters for extremalizing the functional $\tEETo$~\rf{5111}, the dilatational transformations~\rf{5120a} obviously do not change the hydrogen-like character of the trial functions, since those transformations merely induce some irrelevant change of the variational parameters $\{a_n,\beta\}$. The virial theorem \rf{5118}-\rf{5119} must then also hold for the extremal configurations within the subset of hydrogen-like wave functions, see the treatment of the groundstate in ref.~\cite{c10}.

\subsubsection{Energy of Excited State $\zePe$}

Those trial functions of the type \rf{5124a}-\rf{5124b} are well-suited in order to estimate the (non-relativistic) excitation energy of the considered singlet state $\zePe$ (this is the conventional classification; the corresponding RST classification requires an extra discussion). However, in order to keep the calculations as simple as possible, we resort to the simplest possible form of trial functions by regarding only one term of the polynomial $P_\nu$~\rf{5124b}, i.\,e. we try
   \begin{eqnarray}
   \label{5125}
   \ddRp(r)&=&\ddot{N}\,r^\nu\,{\rm e}^{-\beta r}\;.
   \end{eqnarray}
It is true, this is the same form as for the the precedent one-particle situation \rf{557a}-\rf{557b}; but of course the variational parameters $\nu$ and $\beta$ must adopt here other values than those being specified by equations \rf{558a}-\rf{558b}. Despite such a simple form of the trial functions, the subsequent calculations represent a brief demonstration of how to make use of the {\em RST principle of minimal energy}\/.

In the first step, one calculates the normalization constant $\ddot{N}$ in terms of the variational parameters $\nu$ and $\beta$, which yields by reference to the normalization constraint~\rf{5112b}
   \begin{eqnarray}
   \label{5126}
   \ddot{N}^2&=&\frac{4}{\pi}\cdot\frac{(2\beta)^{2\nu+2}}{\Gamma(2\nu+2)}\;.
   \end{eqnarray}
Using always this value for the normalization constant $\ddot{N}$, one can omit the constraint~\rf{5112b} in the energy functional~\rf{5111}. And once a normalized trial function $\ddRp(r)$ is at hand now, one can immediately compute the kinetic energy $\ddEkin$~\rf{5112a} in terms of the variational parameters. The general form looks as follows
   \begin{eqnarray}
   \label{5127}
   \ddEkin&=&\frac{\hbar^2}{2M}\,(2\beta)^2\cdot\epskin(\nu)
   \end{eqnarray}
where the dimensionless function $\epskin(\nu)$ of the variational parameter $\nu$ is given for that simple trial function~\rf{5125} by
   \begin{eqnarray}
   \label{5128}
   \epskin(\nu)&=&\frac{1}{2\nu+1}\left(\frac{1}{2\nu}+\frac{1}{4}\right)\,,
   \end{eqnarray}
provided $\nu$ is a (half-)integer: $\nu=0,\frac{1}{2},1,\frac{3}{2},2,\frac{5}{2},...$ (for the case of more complicated functions $\epskin$ see equation~\mbox{(IV.22\hspace{-.5pt})} of ref.~\cite{c10}).

In the next step we have to determine the gauge potential $\ddotAo(r)$ from the Poisson equation~\rf{5110}, because only in this case we can omit the Poisson constraint~\rf{5112c} in the energy functional $\tEEfTo$~\rf{5111}. Strictly speaking, it is not even necessary to determine the gauge potential $\ddotAo(r)$ itself but rather it suffices to determine its field strength $\frac{\dif{\ddotAo(r)}}{\dif{r}}$, since only this enters the physical part of the energy functional $\tEEfTo$ (provided the Poisson identity is satisfied). The general form of the field strength is chosen to look as follows \cite{c10}
   \begin{eqnarray}
   \label{5129}
   \frac{\dif{\ddotAo(r)}}{\dif{r}}&=&-\frac{\als}{r^2}\left\{1-{\rm e}^{-2\beta r}\big[1+\ddot{f}(r)\big]\right\}\,,
   \end{eqnarray}
and if this is substituted into the Poisson equation~\rf{5110} one gets as the differential equation for the ansatz function $\ddot{f}(r)$
   \begin{eqnarray}
   \nonumber
   \frac{\dif{\ddot{f}(r)}}{\dif{r}}-2\beta\,\big[1+\ddot{f}(r)\big]&=&-\frac{\pi}{4}\,\ddot{N}^2r^{2\nu+1}\\
   \label{5130}
   &=&-\frac{2\beta}{\Gamma(2\nu+2)}\,(2\beta r)^{2\nu+1}\;.
   \end{eqnarray}

The solution of such a simple differential equation can easily be determined by first introducing the dimensionless variable $y$ through
   \begin{eqnarray}
   \label{5131}
   y&\doteqdot&2\beta r
   \end{eqnarray}
which recasts the original differential equation into the following form
   \begin{eqnarray}
   \label{5132}
   \frac{\dif{\ddot{f}(y)}}{\dif{y}}-\big[1+\ddot{f}(y)\big]&=&-\frac{y^{2\nu+1}}{\Gamma(2\nu+2)}
   \end{eqnarray}
with the corresponding solution
   \begin{eqnarray}
   \label{5133}
   \ddot{f}(y)&=&\sum_{n=1}^{2\nu+1}\frac{y^n}{n!}\;,
   \end{eqnarray}
provided the variational parameter $\nu$ is a (half-)integer: $\nu=0,\frac{1}{2},1,\frac{3}{2},...$ Since furthermore the field strength~\rf{5129} reads in terms of the dimensionless variable $y$
   \begin{eqnarray}
   \label{5134}
   \left(\frac{\dif{\ddotAo(y)}}{\dif{y}}\right)^2&=&(2\beta\als)^2\,\frac{\left\{1-{\rm e}^{-y}[1+\ddot{f}(y)]\right\}^2}{y^4}\;,
   \end{eqnarray}
 the gauge field energy $\sERe$~\rf{5104} adopts the following form
    \begin{eqnarray}
    \label{5135}
    \sERe&=&-\als\hbar c\cdot2\beta\cdot\epspot(\nu)\;.
    \end{eqnarray}
Here the dimensionless function $\epspot(\nu)$ of the variational parameter $\nu$ is given by
   \begin{eqnarray}
   \label{5136}
   \epspot(\nu)&=&\int\limits_0^\infty\!\!\dif{y}\,\left\{\frac{1-{\rm e}^{-y}[1+\ddot{f}(y)]}{y}\right\}^2\,,
   \end{eqnarray}
i.\,e. for our special case~\rf{5133}
   \begin{eqnarray}
   \nonumber
   \epspot(\nu)&=&\sum_{n,m=1}^{2\nu+1}\frac{(n+m-2)!}{n!\,m!}\cdot\frac{1}{2^{n+m-1}} -\sum_{n=2}^{2\nu+1}\frac{2}{n(n-1)}\left(1-\frac{1}{2^{n-1}}\right)\,.\\
   \label{5137}
   &&
   \end{eqnarray}
Observe here that both functions $\epskin(\nu)$~\rf{5128} and $\epspot(\nu)$~\rf{5137} are defined here exclusively for (half-)integer values of the variational parameter $\nu=0,\frac{1}{2},1,\frac{3}{2},...$, since for general $\nu$ such integrals as showing up in equation~\rf{5136} could not be given in a simple analytic form! But on principle, $\nu$ is to be considered as a continuous variable for which both functions $\epskin(\nu)$ and $\epspot(\nu)$ do surely exist.

But now that all constraints are satisfied and both the kinetic energy $\ddEkin$~\rf{5127} and the gauge field energy $\sERe$~\rf{5135} are explicitly known in terms of the variational parameters $\nu$ and $\beta$, one can build up by means of them the total energy $\tEETo$~\rf{5111}
   \begin{eqnarray}
   \label{5138}
   \tEETo(\beta,\nu)&=& 2\ddEkin+\sERe\;=\;e^2\,\ab\,(2\beta)^2\cdot\epskin(\nu)-e^2\,(2\beta)\cdot\epspot(\nu)\qquad\qquad\\
   \nonumber
   &&\hspace{2em}(\ab\;\doteqdot\;\frac{\hbar^2}{Me^2}\ \ldots\ \mbox{Bohr radius})\;.
   \end{eqnarray}
According to the {\em principle of minimal energy}\/, the minimal value of this function $\tEETo(\beta,\nu)$ yields the energy of the considered state $\zePe$ in {\em both}\/ the non-relativistic limit {\em and}\/ in the spherically symmetric approximation. Turning first to the minimalization process with respect to the decay parameter $\beta$, one has
   \begin{eqnarray}
   \label{5139}
   0\;=\;\frac{\partial\tEETo(\beta,\nu)}{\partial\beta}&=&2e^2\,\big[4\ab\cdot\epskin(\nu)\cdot\beta-\epspot(\nu)\big]\;,
   \end{eqnarray}
which fixes the minimalizing value $\beta_*$ of the decay constant $\beta$ to
   \begin{eqnarray}
   \label{5140}
   \beta_*&=&\frac{1}{4\ab}\cdot\frac{\epspot(\nu)}{\epskin(\nu)}\;.
   \end{eqnarray}
Now substituting this back into the kinetic and gauge field energies $\ddEkin$~\rf{5127} and $\sERe$~\rf{5135} one arrives at
   \begin{subequations}
   \begin{align}
   \label{5141a}
   \ddEkin\Big|_{\beta_*}&=\frac{\hbar^2}{2M}\,(2\beta_*)^2\cdot\epskin(\nu) \,=\,\frac{e^2}{8\ab}\cdot\frac{\epspot^2(\nu)}{\epskin(\nu)}\\
   \label{5141b}
   \sERe\Big|_{\beta_*}&=-e^2\,(2\beta_*)\cdot\epspot(\nu) \,=\,-\frac{e^2}{2\ab}\cdot\frac{\epspot^2(\nu)}{\epskin(\nu)}\;,
   \end{align}
   \end{subequations}
and this result just realizes the two-particle {\em virial theorem}\/~\rf{5118}! (Clearly, the one-particle virial theorem is also satisfied, cf. equations \rf{570}-\rf{572}). Accordingly, the total energy $\tEETo$~\rf{5119} becomes
   \begin{eqnarray}
   \label{5142}
   \tEETo&=&-\frac{e^2}{16\ab}\frac{[2\epspot(\nu)]^2}{\epskin(\nu)}\;\equiv\;\Econv\cdot S(\nu)
   \end{eqnarray}
where $\Econv$ is the prediction~\rf{596} of standard quantum mechanics for principal quantum number $n_p=1$; and the function $S(\nu)$
   \begin{eqnarray}
   \label{5143}
   S(\nu)&\doteqdot&\frac{[2\epspot(\nu)]^2}{\epskin(\nu)}
   \end{eqnarray}
contributes to $\tEETo$ by its maximally possible value $S_*$ (according to the {\em principle of minimal energy}\/).

If this maximal value $S_*$ of the function $S(\nu)$ turned out to be unity (i.\,e. $S_*=1$), the RST prediction $\tEETo$~\rf{5142} would exactly agree with the conventional prediction $\Econv\big|_{n_p=1}$~\rf{597}. Perhaps this hypothesis of the ``exact'' coincidence of the RST and conventional predictions is correct for the {\em true}\/ non-relativistic limit of RST, which does not rely on the spherically symmetric approximation \rf{598a}-\rf{599} of the gauge potential $\pAo(r,\vartheta)$. But as the discussion of the groundstate $\eeSo$ in ref.s~\cite{c10,c11} has shown, the latter type of approximation induces a deviation of magnitude $\sim10\%$, which is then to be expected also for the present case of $\zePe$. Indeed, approximating the properly continuous function $S(\nu)$~\rf{5143} (with the discrete values as shown in the figure below) by an interpolating polynomial of 11th order yields a maximum $S_*=0.9119...$ at $\nu_*=1.7942...$ and thus realizes the 10\% expectations (see fig.).

\newpage
\subsubsection{Energy Level Diagram}

\begin{center}
\begin{figure}[h]
\label{f1}
\includegraphics[bb=3cm 23.5cm 0cm 0cm, scale=0.8]{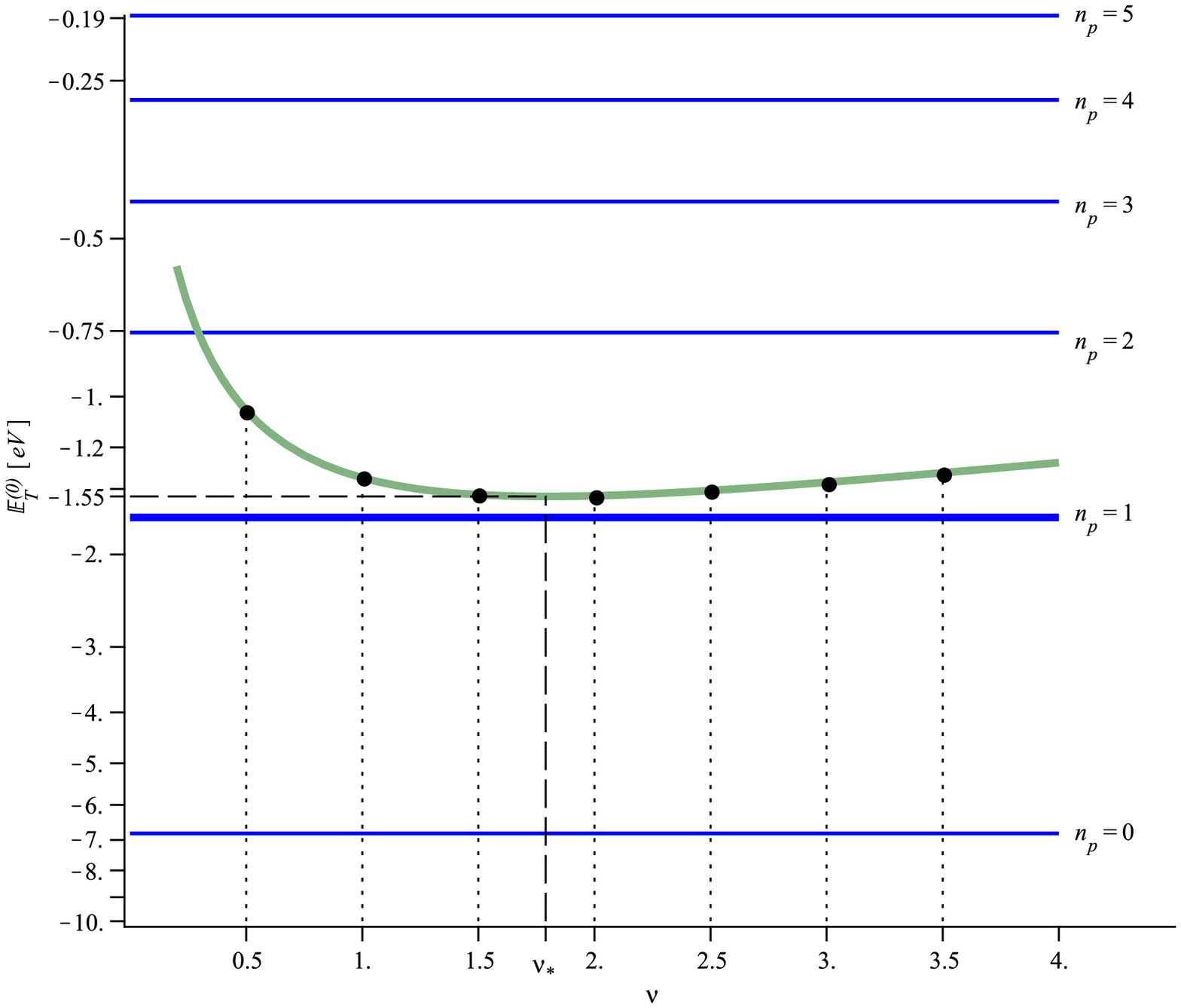}
\end{figure}
\end{center}
\vspace*{13cm}
\noindent {\bf\itshape Minimal value of the energy\/ $\tEETo$~\rf{5142}}\medskip\\
\indent The coincidence of the RST prediction $\tEETo$~\rf{5142} with the conventional prediction $\Econv\big|_{n_p=1}$~\rf{597} would occur if the maximal value of the function $S(\nu)$~\rf{5143} were unity ($S(\nu)\Rightarrow1$). Since, however, the maximal value $S_*$ due to the trial function \rf{5125} is found as $S_*=0.9119...$, one ends up with a deviation of (roughly) 10\%, i.\,e. the RST prediction for the $\zePe$ energy is $-1.55...\,[{\rm eV}]$ as compared to the conventional prediction $\Econv\big|_{n_p=1}=1.70...\,[{\rm eV}]$. This deviation is to be ascribed partly to the use of the spherically symmetric approximation and partly to the use of an oversimplified trial function~\rf{5125}. Thus it remains to be clarified to what extent the RST and conventional predictions would agree if better approximation techniques for the RST calculations could be applied.

\end{document}